\documentclass[amsmath,amssymb,aps,prd,11pt,tightenlines,superscriptaddress,nofootinbib,preprintnumbers,notitlepage]{revtex4-1}

\input{header}

\usepackage{lineno}
\usepackage{mathtools}
\usepackage{booktabs}
\usepackage{xcolor}
\usepackage{colortbl}
\usepackage{pifont}
\usepackage{amssymb}
\usepackage{amsmath}
\usepackage{subcaption}
\usepackage{bm}

\definecolor{red}{rgb}{1,0,0}
\def\lesssim{\ \hbox{\raise 2pt \hbox{$<$} \kern -13pt
                     \lower 3pt \hbox{$\sim$}}\ }
\def\greatersim{\ \hbox{\raise 2pt \hbox{$>$} \kern -13pt
                     \lower 3pt \hbox{$\sim$}}\ }

\input epsf.tex
\def\desepsf(#1 width #2){\epsfxsize=#2 \epsfbox{#1}}

\DeclarePairedDelimiterX\braket[2]{\langle}{\rangle}{#1 \delimsize\vert #2}
\newcolumntype{C}{>{\centering\arraybackslash}p{1.5cm}}
\newif\ifcomment

\newcommand{\pPb}        {p--Pb}
\newcommand{\pp}         {pp}

\newcommand{\snn}        {\ensuremath{\sqrt{s_{\rm NN}}}}
\newcommand{\RpPb}       {\ensuremath{R_\mathrm{pPb}}}

\newcommand{\pT}         {\ensuremath{p_{\rm T}}}

\newcommand{\GeVc}       {\ensuremath{\mathrm{GeV}/c}}

\newcommand{\Fig}[1]     {Fig.~\ref{#1}}

\renewcommand{\Ref}[1]   {Ref.~\cite{#1}}

\newcommand{\h}[1]       {\relax}
\begin{document}

\title{{\LARGE White Paper on Forward Physics, BFKL,  Saturation Physics and Diffraction }  }

\author{Martin Hentschinski (editor) }
\affiliation{Departamento de Actuaria, F\'isica y Matem\'aticas, Universidad de las Am\'ericas Puebla,
Ex-Hacienda Santa Catarina Martir S/N, 
San Andr\'es Cholula
72820 Puebla, Mexico}
\author{Christophe Royon (editor)}
\affiliation{The University of Kansas, Department of Physics and Astronomy, 1251 Wescoe Hall Dr.
Lawrence, KS 66045}

\author{Marco Alcazar Peredo}
\affiliation{Departamento de Actuaria, F\'isica y Matem\'aticas, Universidad de las Am\'ericas Puebla
Ex-Hacienda Santa Catarina Martir S/N, 
San Andr\'es Cholula
72820 Puebla, Mexico}
\author{Cristian Baldenegro}
\affiliation{LLR, CNRS, Ecole Polytechnique, Institut Polytechnique de Paris, 91128 Palaiseau, France}
\author{Andrea Bellora}
\affiliation{INFN Sezione di Torino, 10125 Torino, Italy}
\author{Renaud Boussarie}
\affiliation{CPHT, CNRS, Ecole Polytechnique, Institut Polytechnique de Paris, 91128 Palaiseau, France}
\author{Francesco Giovanni Celiberto}
\affiliation{European Centre for Theoretical Studies in Nuclear Physics and Related Areas (ECT*),
I-38123 Villazzano, Trento, Italy}
\affiliation{Fondazione Bruno Kessler (FBK), I-38123 Povo, Trento, Italy}
\affiliation{INFN-TIFPA Trento Institute of Fundamental Physics and Applications, I-38123 Povo, Trento, Italy}
\author{Salim Cerci}
\affiliation{Adiyaman University, Faculty of Arts and Sciences, Department of Physics, 02040-Adiyaman, Turkey}
\author{Grigorios Chachamis}
\affiliation{Laborat{\' o}rio de Instrumenta\c{c}{\~ a}o e F{\' \i}sica Experimental de Part{\' \i}culas (LIP),\\
Av. Prof. Gama Pinto, 2, P-1649-003 Lisboa, Portugal}
\author{J.~G.~Contreras}
\affiliation{Faculty of Nuclear Sciences and Physical Engineering,\\
Czech Technical University in Prague, Prague, Czech Republic}
\author{Sylvain Fichet}
\affiliation{CTP South American Institute for Fundamental Research \& IFT-UNESP,
R. Dr. Bento Teobaldo Ferraz 271, S\~ao Paulo, Brazil}
\author{Michael Fucilla}
\affiliation{Dipartimento di Fisica, Universit\`a della Calabria, I-87036 Arcavacata di Rende, Cosenza, Italy}
\affiliation{Istituto Nazionale di Fisica Nucleare, Gruppo collegato di Cosenza, I-87036 Arcavacata di Rende, Cosenza, Italy}
\affiliation{Universit\'{e} Paris-Saclay, CNRS, IJCLab,  91405 Orsay, France}
\author{Gero von Gersdorff}
\affiliation{Department of Physics, PUC-Rio, 22451-900 Rio de Janeiro, Brazil}
\author{Pablo González}
\affiliation{University of M\"unster
  Institute for Theoretical Physics
  Wilhelm-Klemm-Str. 9
  D-48149 Muenster, Germany}
\author{Andreas van Hameren}
\affiliation{Institute of Nuclear Physics, Polish Academy of Sciences, Radzikowskiego 152, 31-342 Krakow, Poland}
\author{Jamal Jalilian-Marian}
\affiliation{Department of Natural Sciences, Baruch College, 17 Lexington Avenue, New York, NY 10010, USA}
\affiliation{City University of New York Graduate Center, 365 Fifth Avenue, New York, NY 10016, USA}
\author{Mats Kampshoff}
\affiliation{University of M\"unster
  Institute for Theoretical Physics
  Wilhelm-Klemm-Str. 9
  D-48149 Muenster, Germany}
\author{Valery Khoze}
\affiliation{IPPP, Durham University, Durham DH1 3LF, UK}
\author{Michael Klasen}
\affiliation{University of M\"unster
  Institute for Theoretical Physics
  Wilhelm-Klemm-Str. 9
  D-48149 Muenster, Germany}
\author{Spencer Robert Klein}
\affiliation{Lawrence Berkeley National Laboratory, Berkeley CA 94720 USA}
\author{Georgios Krintiras}
\affiliation{The University of Kansas, Department of Physics and Astronomy, 1251 Wescoe Hall Dr.
Lawrence, KS 66045}
\author{Piotr Kotko}
\affiliation{AGH University Of Science and Technology, Physics Faculty, Mickiewicza 30, 30-059 Krak\'ow, Poland}
\author{Krzysztof Kutak}
\affiliation{Institute of Nuclear Physics, Polish Academy of Sciences, Radzikowskiego 152, 31-342 Krakow, Poland}
\author{Jean-Philippe Lansberg}
\affiliation{Universit\'{e} Paris-Saclay, CNRS, IJCLab,  91405 Orsay, France}
\author{Emilie Li}
\affiliation{Universit\'{e} Paris-Saclay, CNRS, IJCLab,  91405 Orsay, France}
\author{Constanin Loizides}
\affiliation{ORNL, Physics Division, Oak Ridge, TN, USA}
\author{Mohammed M.A. Mohammed}
\affiliation{Dipartimento di Fisica, Universit\`a della Calabria, I-87036 Arcavacata di Rende, Cosenza, Italy}
\affiliation{Istituto Nazionale di Fisica Nucleare, Gruppo collegato di Cosenza, I-87036 Arcavacata di Rende, Cosenza, Italy}
\author{Maxim Nefedov}
\affiliation{National Centre for Nuclear Research (NCBJ), Pasteura 7, 02-093 Warsaw, Poland}
\author{Melih~A.~Ozcelik}
\affiliation{Institute for Theoretical Particle Physics, KIT, 76128 Karlsruhe, Germany}
\author{Alessandro Papa}
\affiliation{Dipartimento di Fisica, Universit\`a della Calabria, I-87036 Arcavacata di Rende, Cosenza, Italy}
\affiliation{Istituto Nazionale di Fisica Nucleare, Gruppo collegato di Cosenza, I-87036 Arcavacata di Rende, Cosenza, Italy}
\author{Michael Pitt}
\affiliation{CERN, CH\- 1211, Geneva 23, Switzerland}
\author{Agustin Sabio Vera}
\affiliation{Instituto de F\'isica Te\'orica UAM/CSIC, Nicol\'as Cabrera 15, E-28049 Madrid, Spain}
\affiliation{Theoretical Physics Department, Universidad Aut\'onoma de Madrid, E-28049 Madrid, Spain}
\author{Jens Salomon}
\affiliation{Instituto de Física Teórica UAM/CSIC, Nicolás Cabrera 15, E-28049 Madrid, Spain}
\author{Sebastian Sapeta}
\affiliation{Institute of Nuclear Physics, Polish Academy of Sciences, Radzikowskiego 152, 31-342 Krakow, Poland}
\author{Gustavo Gil da Silveira}
\affiliation{Universidade Federal do Rio Grande do Sul, Porto Alegre-RS, 91501-970, Brazil}
\author{Victor Paulo Gonçalves}
\affiliation{Universidade Federal de Pelotas, Pelotas-RS, 96010-610, Brazil}
\author{Mark Strikman}
\affiliation{Pennsylvania State University, University Park, PA, 16802, USA}
\author{Deniz Sunar Cerci}
\affiliation{Adiyaman University, Faculty of Arts and Sciences, Department of Physics, 02040-Adiyaman, Turkey}

\author{Lech Szymanowski }
\affiliation{National Centre for Nuclear Research (NCBJ), Pasteura 7, 02-093 Warsaw, Poland}
\author{Daniel Tapia Takaki}
\affiliation{The University of Kansas, Department of Physics and Astronomy, 1251 Wescoe Hall Dr.
Lawrence, KS 66045}
\author{Marek Ta\v{s}evsk\'{y}}
\affiliation{Institute of Physics of the Czech Academy of Sciences, Na Slovance 2, 18221 Prague, Czech Republic}
\author{Samuel Wallon}
\affiliation{Universit\'{e} Paris-Saclay, CNRS, IJCLab,  91405 Orsay, France}

\begin{abstract}
  The goal of this white paper is to give a comprehensive overview of the rich field
of forward physics.  We discuss the occurrences of BFKL resummation effects in special final states, such as Mueller-Navelet jets, jet gap jets, and heavy quarkonium production.It further addresses TMD factorization at low x and the manifestation of a semi-hard saturation scale in  (generalized) TMD PDFs. More theoretical aspects of low x physics, probes of the quark gluon plasma, as well as the possibility to use photon-hadron collisions at the LHC to constrain hadronic structure at low x, and the resulting complementarity between  LHC and the EIC are also presented.  We also briefly discuss diffraction at colliders as well as the possibility to explore further the electroweak theory in central exclusive events using the LHC as a photon-photon collider.
\end{abstract}

\maketitle 

\tableofcontents

\section{Introduction}
For successful runs at any colliders, such as the LHC at CERN or the incoming EIC at BNL, and future projects such as FCC at CERN, it is fundamental to understand fully the complete final states. This obviously includes the central part of the detector that is used in searches for beyond standard model physics but also the forward part of the detector, the kinematic region close to the outgoing particles after collision. The detailed understanding of final states with high forward multiplicities, as well as those with the absence of energy in the forward region (the so-called rapidity gap), in elastic, diffractive, and central exclusive processes is of greatest importance.
Some of these configurations originate from purely non-perturbative reactions, while others can
be explained in terms of multi-parton chains or other extensions of the perturbative QCD parton
picture such as the Balitsky-Fadin-Kuraev-Lipatov (BFKL) formalism. Future progress in this fundamental area in high energy physics requires the combination of  experimental
measurements and theoretical work. \\

Forward Physics addresses physics that takes place in the forward region of detectors, which at first is defined as the region complementary to the central region. The latter is the region dominantly employed  in the search for new physics at  {\it e.g.} the Large Hadron Collider. It is then also the central region where collinear factorization of hard processes in terms of an partonic cross-section, convoluted with corresponding collinear parton distribution functions is well defined. 'Hard process' refers here to a certain reaction subject to strong interactions, which is characterized by the presence of a hard scale $M$ with $M \gg \Lambda_{\text{QCD}}$ with $\Lambda_{\text{QCD}}$ the characteristic scale of Quantum Chromodynamics (QCD) of the order of a few hundred MeV.  Physics in the forward region is on the other hand at first characterized by production at large values of rapidity with respect to the central region. For a hard reactions, where the underlying partonic sub-process is resolved, one therefore deals with the interplay  of partons with a relative large proton momentum fraction $x_1$, with $x_1 \sim 0. 1 \ldots 1$, and partons with very small  proton momentum fractions $x_2$  down to $10^{-6}$ in the most extreme scenario. Such small momentum fractions lead generally to a break down of the convergence of the perturbative expansion and require resummation, which is achieved by Balitsky-Kuraev-Fadin-Lipatov (BFKL) evolution. The latter gives rise to the so-called hard or BFKL Pomeron, which predicts a strong and power-like rise of the gluon distribution in the proton in the region where $x_2 \to 0$. While such a rise is clearly seen in data,  unitarity bounds prohibit such a rise to continue forever: at a certain value of $x_2$, this rise must slow down and eventually come to hold. The latter is strongly related with the formation of an over-occupied system of gluons, known as the Color Glass Condensate, whose exploration is one of the central physics goals of a future Electron Ion Collider (EIC). While at an EIC a dense QCD state will be achieved through scattering of electrons on heavy ions, forward physics at LHC allows here for a complementary exploration, since high gluon densities are here at first produced through the low $x$ evolution of the gluon distribution in the proton. Forward physics allows therefore for the exploration of both BFKL evolution (perturbative evolution towards the low $x$ region) as well as to search and investigate effects related to the on-set of gluon saturation.
\\

Besides the direct interaction of partons at very low proton momentum fraction, forward physics also allows for the observation of a different class of events, so-called diffractive events. The latter are characterized through the presence of large rapidity gaps and therefore  probe physics beyond conventional collinear factorization. In the case of hard events, they give access to complementary information on the physics of high gluon densities as well as corrections due to soft re-scattering. At the same time such processes are themselves of direct interest for the exploration of electroweak physics and physics beyond the Standard Model: due to the presence of rapidity gaps, such events are characterized through a very few numbers of particles in the final state and allow therefore for very clean measurements with a strongly reduced background, in comparison to conventional LHC measurements. Closely related to such diffractive events are photon induced reactions which can be observed at the LHC. While such reactions can produce final states both in the central and forward region, control of the forward region is of particular importance for those reactions, since it allows us to control whether in a certain the event the scattering proton stayed indeed intact and acts in this way as the photon source. In such events, either one or both of the two scattering protons or ions at the LHC at as a photon source. The former  allows for the study of  exclusive photon-hadron interaction at highest center of mass energies and yields therefore yields another tool for the study of highest gluon densities, with high precision; as for inclusive reactions, such exclusive reactions are complementary to measurements at the future Electron Ion Collider, since at the LHC high parton densities are predominately generated due to high energy evolution, while an EIC relies due to its lower center of mass energy on the nuclear enhancement. Photon-photon interactions are on the other hand of high interest, since they provide very clear  probes of electroweak and Beyond-the-Standard-Model physics. With both scattering hadrons intact after the interaction, QCD background is suppressed to a minimum in such a reaction and complements in this way LHC searches for new physics based on inclusive events in the central region. 
\\

The outline of this white paper is as follows:  Sec.~\ref{sec:bfkl} is dedicated to attempts to pin down BFKL evolution at the LHC as well as its actual use for phenomenology. Sec.~\ref{sec:structure} deals with high gluon densities, saturation as well as their relation to TMD PDFs. Sec.~\ref{sec:imprints} deals with the investigation of high gluon densities at the LHC and their  phenomenological consequences. Sec.~\ref{sec:upc} is dedicated to ultra-peripheral collisions, and Sec.~\ref{sec:odder-disc-diffr} to the recent Odderon discovery as well as diffractive jets.  Sec.~\ref{sec:EW} deals with electroweak physics. In Sec.~\ref{sec:conclusion} we draw our conclusions.

\section{Manifestations of  BFKL evolution}
\label{sec:bfkl}
\noindent \textbf{Main Contributors:}  Cristian Baldenegro, Francesco Giovanni Celiberto, Salim Cerci, Grigorios Chachamis, Michael Fucilla, Pablo González, Mats Kampshoff, Michael Klasen, Jean-Philippe Lansberg, Mohammed M.A. Mohammed, Maxim Nefedov, Melih Ozcelik, Alessandro Papa, Christophe Royon, Deniz Sunar Cerci, Agustin Sabio Vera,  Jens Salomon  \\

In the following section we describe both attempts to pin down BFKL evolution and the underlying Multi-Regge-Kinematics in multi-jet events, Sec.~\ref{sec:mn_jets}, as well as reactions characterized by two events widely separated in rapidity, Sec.~\ref{sec:HAS_HE-QCD_precision}. While in those cases BFKL evolution takes place within a hard event which extends over several units of rapidity, Sec.~\ref{sec:HAS_HE-QCD_UGD} addresses the case where BFKL evolution is absorbed into the gluon distribution, resulting into an unintegrated gluon distribution. Sec.~\ref{sec:Quarkonia_HEF} addresses finally the case where BFKL evolution is matched with collinear factorization and used to resum large logarithms curing an instability of the latter in the limit of large partonic center of mass energies. 

\subsection{Mueller-Navelet jets}
\label{sec:mn_jets}

\noindent An active area of research in QCD phenomenology at high energies 
is to pin down novel observables where the dominant contributions 
stem from the Balitsky-Fadin-Kuraev-Lipatov (BFKL) domain~\cite{Kuraev:1977fs,Kuraev:1976ge,Fadin:1975cb,Lipatov:1976zz,Balitsky:1978ic,Lipatov:1985uk,Fadin:1998py,Ciafaloni:1998gs}. This is a challenging task since for typical observables calculations based on matrix elements computed at fixed order 
along with the Dokshitzer-Gribov-Lipatov-Altarelli-Parisi (DGLAP) evolution~\cite{Gribov:1972ri,Gribov:1972rt,Lipatov:1974qm,Altarelli:1977zs,Dokshitzer:1977sg} to account for the scale dependence of Parton Distribution Functions  (PDFs) tend to describe the bulk of the data adequately. It is then needed to move towards corners of the phase space to isolate BFKL effects. This can be done by studying the structure of final states in Mueller-Navelet (MN) jets events~\cite{Mueller:1986ey}, namely, events that have two jets with similar and large enough transverse momentum $p_{\perp}$ that can serve as a hard scale, $\Lambda^2_{QCD} \ll p_{\perp}^2 \ll s$, where $s$ is the c.o.m. energy squared. The two tagged jets should also be separated by a large rapidity interval $Y$ while there is a rich mini-jet activity in between.  Numerous studies took place on MN jets both at leading-order (LO) BFKL as well as at next-to-leading order (NLO). The main quantity of interest in most studies was the azimuthal decorrelation between the two outermost jets, for a non exhaustive list of theoretical works see Refs.~\cite{DelDuca:1993mn,Stirling:1994he,DelDuca:1994ng,Orr:1997im,Kwiecinski:2001nh,Andersen:2001kta,Vera:2006un,Vera:2007kn,Marquet:2007xx,Bartels:2001ge,Bartels:2002yj,Colferai:2010wu,Caporale:2011cc,Ducloue:2013hia,Ducloue:2013bva,Caporale:2014gpa,Celiberto:2015yba,Mueller:2015ael,LHCForwardPhysicsWorkingGroup:2016ote,Colferai:2017bog} while relevant experimental analyses by ATLAS and CMS can be found in Refs.~\cite{ATLAS:2011yyh,CMS:2012xfg,ATLAS:2014lzu,CMS:2016qng}.

\begin{figure}
\begin{subfigure}{.5\textwidth}
  \centering
  \includegraphics[width=1\linewidth]{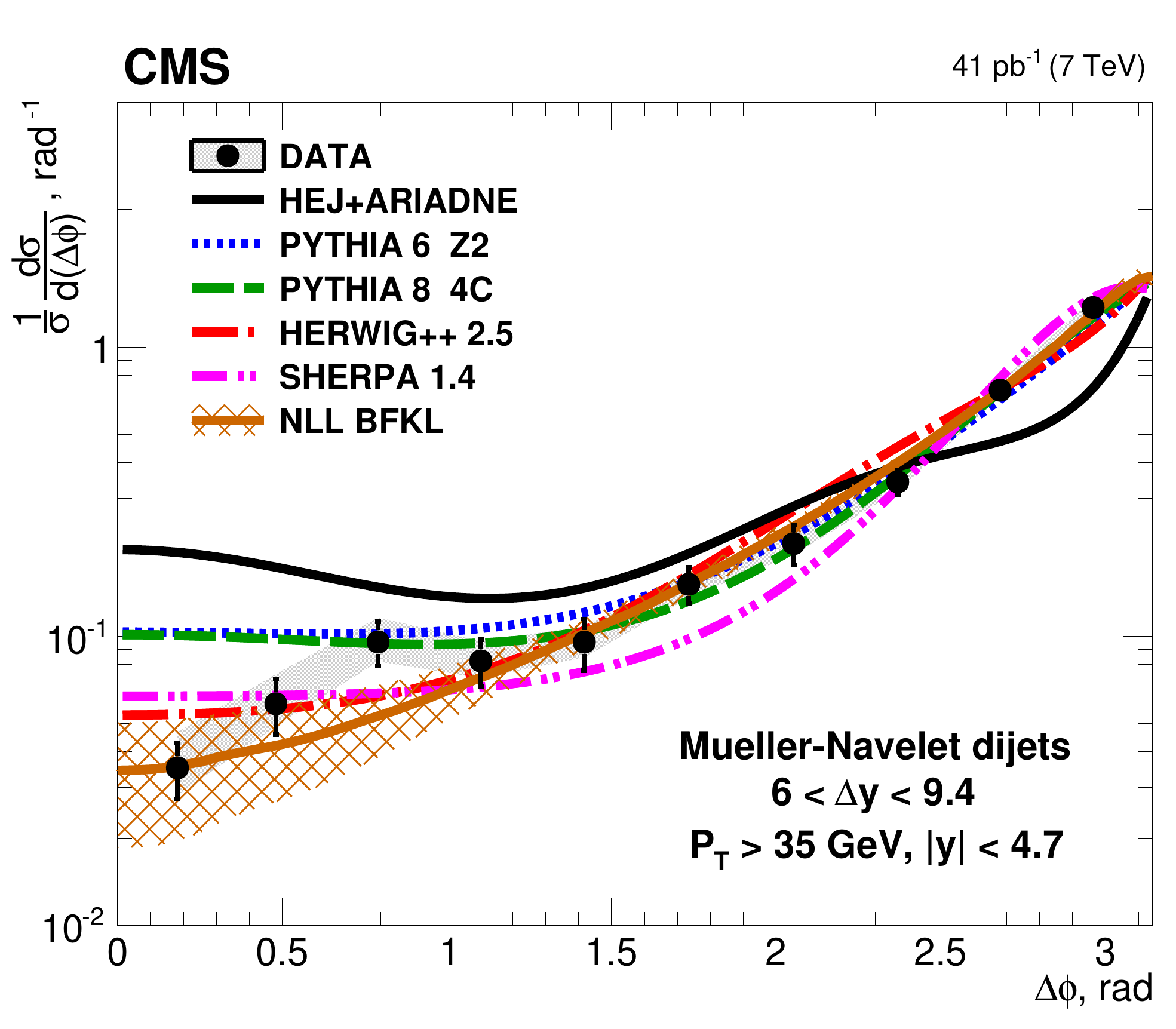}
\end{subfigure}%
\begin{subfigure}{.5\textwidth}
  \centering
  \includegraphics[width=1\linewidth]{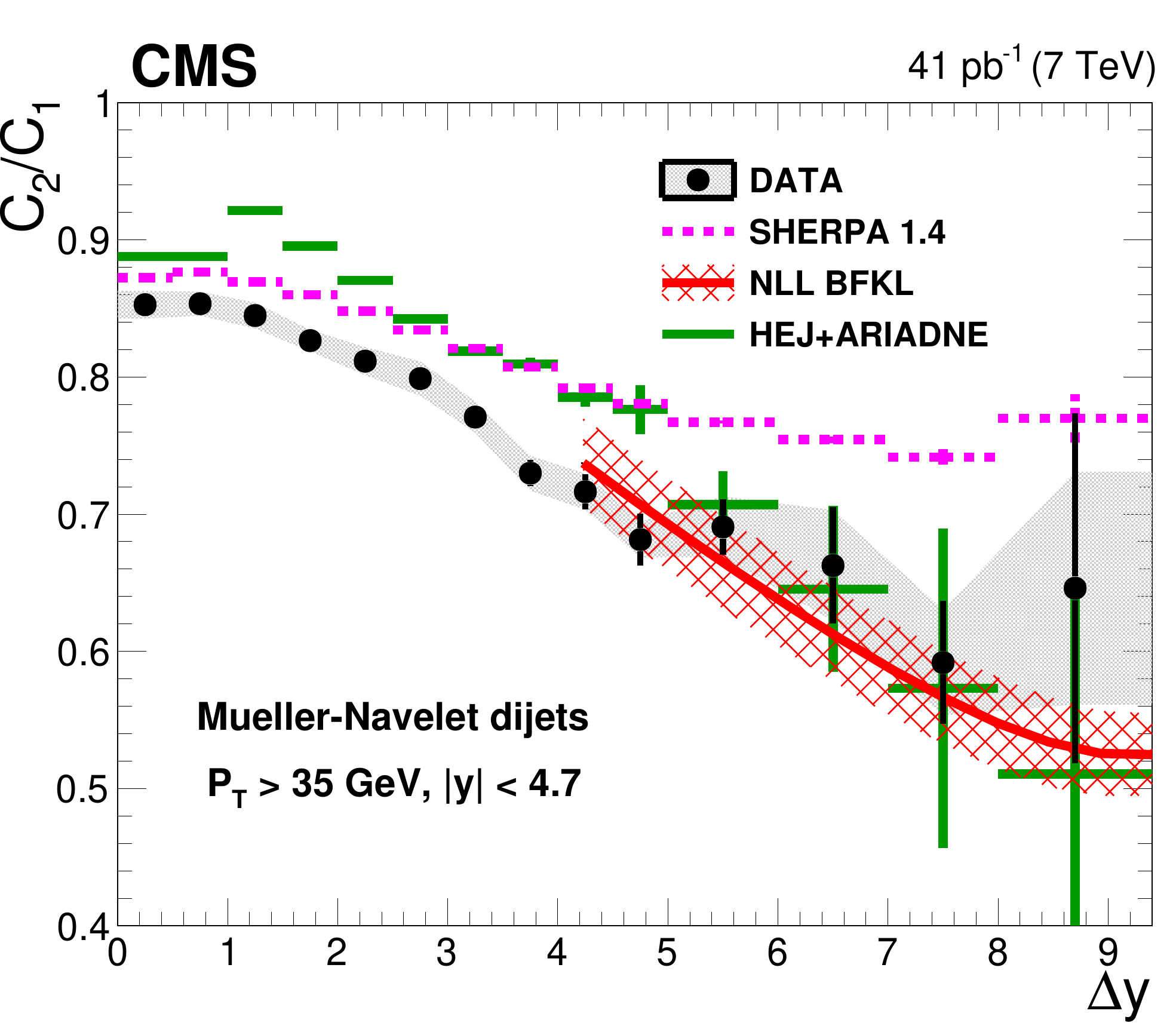}
\end{subfigure}
\caption{Left: The azimuthal-angle difference distribution measured for Mueller-Navelet jets in the rapidity interval $6.0 < \Delta y < 9.4$. Right: Comparison of the measured ratio $C_2/C_1$ as a function of rapidity difference $\Delta y$ to SHERPA, HEJ+ARIADNE and analytical NLL BFKL calculations at the parton level~\cite{CMS:2016qng}.}
\label{fig:cms}
\end{figure}

The CMS Collaboration reported a measurement of azimuthal angle decorrelation between the most forward and the most backward jets (so-called Mueller-Navelet jets) in proton-proton collisions at $\sqrt s = 7$ TeV~\cite{CMS:2016qng}. In the analysis, jets with transverse momentum, $p_T > 35$ GeV and absolute rapidity, $|y| < 4.7$ are considered. The normalised cross sections are compared with various Monte Carlo generators and analytical predictions based on the DGLAP and BFKL parton evolution equations.
In Fig.~\ref{fig:cms}, the azimuthal angle decorrelation of dijets and ratio of its average cosines ($C_{n} = <\mathrm{cos}(n(\pi - \phi_{dijet}))>$) are shown as a function of rapidity separation between the jets, $\Delta y$, reaching up to $\Delta y = 9.4$ for the first time. At higher centre-of-mass energies, new measurements of azimuthal angular decorrelations in MN jet events need to be defined and performed. It will be fundamental to test the dependence of the $\Delta\phi$ correlations as a function of the outermost jets $p_T$, in addition to the $\Delta y$ scan. The region of applicability of the BFKL formalism is expected to occur in cases where the outermost jets have similar $p_T$. At the same time, we are interested in studying the radiation pattern between the jets presented in the form of ``mini-jets'' between the outermost jets. Indeed, as the rapidity interval increases there is more phase space available for extra radiation to be emitted, so it is natural for the average jet multiplicity to increase. The number of mini-jets as well as the emission pattern in $y$--$\phi$ space could potentially be used in addition to the azimuthal angular decorrelation to further characterize MN dijet events. The main focus will be given in the definition of more ``exclusive'' observables that exploit the two-jet angular correlations between the mini-jets and the outermost jets in $y$--$\phi$ space, together with a measurement of $\langle \cos(\Delta \phi)\rangle$ between the outermost jets.
 
From the phenomenological studies so far, it became apparent that more precise theoretical work is needed (e.g. see~\cite{CMS:2016qng}).  In Refs.~\cite{Chachamis:2015ico,deLeon:2021ecb} new observables were proposed aiming at probing novel multi-Regge kinematics signatures. \\

In order to define properly the proposed observables, we assume that a MN event is characterized by:
\begin{eqnarray}
 k_a, \,k_b&:  & \text{transverse momenta of the MN jets}   \nonumber
 \\
  y_0 = y_a = Y,\, y_{N+1} = y_b = 0&:  & \text{rapidities of the MN jets}  \nonumber
 \\
 k_1, k_2, ... , k_N&:  &  \text{transverse momenta of the minijets}   \nonumber
 \\
 y_1, y_2, ... , y_N&: &   \text{rapidities of the minijets with } y_{i-1} > y_i\,.
\label{eq:MN}
\end{eqnarray}
Then the observables are
\begin{eqnarray}
\langle p_{\perp} \rangle &=& \frac{1}{N}  \sum_{i=1}^{N} p_{\perp, i}\,,
\label{eq:observablept}
\end{eqnarray}
\begin{eqnarray}
\langle {\mathcal R}_y \rangle &=& \frac{1}{N-1}  \sum_{i=1}^{N-1} \frac{y_i}{y_{i-1}}\,,
\label{eq:observableRy}
\end{eqnarray}
and 
\begin{eqnarray}
\langle {\mathcal R}_{k y} \rangle &=& \frac{1}{N-1}  \sum_{i=1}^{N-1} \frac{k_i e^{y_i}}{k_{i-1} e^{y_{i-1}}} \,.
\label{eq:observableRky}
\end{eqnarray}
Eq.~(\ref{eq:observableRy}) differs from the original definition in~\cite{Chachamis:2015ico} since now $i$ runs over the minijets and excludes the leading MN jets.  $\langle {\mathcal R}_{k y} \rangle$ incorporates a $p_{\perp}$ dependence which  carries information related to the decoupling between transverse and longitudinal components of the emitted gluons.
For the proposed observables, one can see  that events where the minijets have relatively low $p_{\perp}$,  
contrary to what one would naively expect, give a very significant contribution to the gluon Green's function (Fig. 1 in~\cite{Chachamis:2015ico}) and consequently to the cross-section. The experimental analyses however  (mainly to deal with jet energy reconstruction uncertainties) impose a veto on the $p_{\perp}$ of any resolved minijet. Usually the $p_{\perp}$ veto value for ATLAS and CMS is $Q_0 = 20$ GeV which is rather large if we compare it to the 
$Q_0 = 1$ GeV value which was the jet $p_{\perp}$ infrared cutoff for the plots in~\cite{Chachamis:2015ico}.

Here we are performing a first comparison between the predictions from a fixed order calculation and a BFKL based computation
for the observables described in Eqs.~\ref{eq:observablept},~\ref{eq:observableRy} and ~\ref{eq:observableRky}.
We focus on events where two jets with rapidities $y_a$ in the forward direction and $y_b$  in the backward direction can be clearly identified. 
In order for the BFKL dynamics to be relevant, 
the difference $Y=y_a-y_b$ needs to be large enough so that terms of the form $\alpha_s^n Y^n$ be important 
order-by-order to get a good description of the partonic cross-section which can be written in the factorized form 
\begin{eqnarray}
\hat{\sigma} (Q_1,Q_2,Y) = \int d^2 \vec{k}_A d^2 \vec{k}_B \, {\phi_A(Q_1,\vec{k}_a) \, 
\phi_B(Q_2,\vec{k}_b)} \, {f (\vec{k}_a,\vec{k}_b,Y)}.
\end{eqnarray}
In this expression $\phi_{A,B}$ are impact factors depending on the external scales, $Q_{1,2}$, and the off-shell reggeized gluon momenta, $\vec{k}_{a,b}$.  The gluon Green function $f$ depends on $\vec{k}_{a,b}$ and the center-of-mass energy in the scattering $\sim e^{Y/2}$.

Here, we will work at leading order (LO) with respect to an expansion in the strong coupling constant $\alpha_s$, however, for BFKL phenomenology at the LHC it is mandatory to work within the next-to-leading order (NLO) approximation for both the 
impact factors and the gluon Green's function which introduces the dependence on physical scales such as the one associated to the running of the coupling and the one related to the choice of energy scale in the resummed logarithms~\cite{Forshaw:2000hv,Chachamis:2004ab,Forshaw:1999xm,Schmidt:1999mz}. It is possible to write the gluon Green function in an iterative way in transverse momentum and rapidity space at LO~\cite{Schmidt:1996fg} and NLO~ \cite{Andersen:2003an,Andersen:2003wy}.  
The iterative solution at LO has the form (for the NLO expressions see Refs.~\cite{Andersen:2003an,Andersen:2003wy})
\begin{eqnarray}
f &=& e^{\omega \left(\vec{k}_A\right) Y}  \Bigg\{\delta^{(2)} \left(\vec{k}_A-\vec{k}_B\right) + \sum_{N=1}^\infty \prod_{i=1}^N \frac{\alpha_s N_c}{\pi}  \int d^2 \vec{k}_i  
\frac{\theta\left(k_i^2-\lambda^2\right)}{\pi k_i^2} \nonumber\\
&&\hspace{-.6cm}\times \int_0^{y_{i-1}} \hspace{-.3cm}d y_i e^{\left(\omega \left(\vec{k}_A+\sum_{l=1}^i \vec{k}_l\right) -\omega \left(\vec{k}_A+\sum_{l=1}^{i-1} \vec{k}_l\right)\right) y_i} \delta^{(2)} \hspace{-.16cm}
\left(\vec{k}_A+ \sum_{l=1}^n \vec{k}_l - \vec{k}_B\right) \hspace{-.2cm}\Bigg\} \, , 
\label{BFKL_iter}
 \end{eqnarray}
where 
\begin{eqnarray}
\omega \left(\vec{q}\right) &=& - \frac{\alpha_s N_c}{\pi} \ln{\frac{q^2}{\lambda^2}} 
\end{eqnarray}
corresponds to the gluon Regge trajectory which carries a regulator, $\lambda$, of infrared divergences. All these expressions have been implemented  in the Monte Carlo code {\tt BFKLex} which has already been used for different applications ranging from collider phenomenology to more formal studies in the calculation of scattering amplitudes in supersymmetric theories~\cite{Chachamis:2011rw,Chachamis:2011nz,Chachamis:2012fk,Chachamis:2012qw,Caporale:2013bva,Chachamis:2015zzp}. 

\begin{figure}
\begin{subfigure}{.5\textwidth}
  \centering
  \includegraphics[width=1\linewidth]{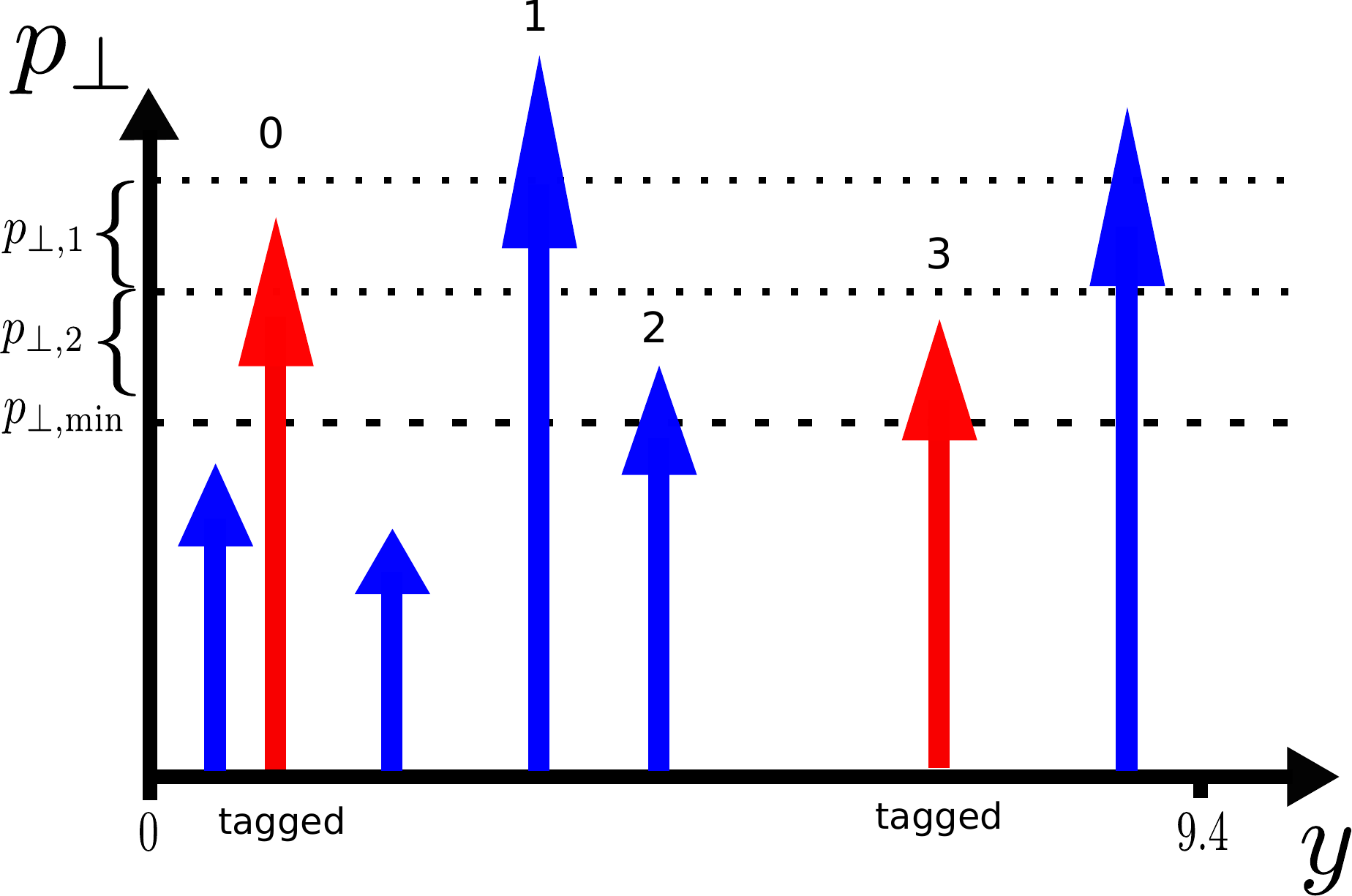}
\end{subfigure}%
\begin{subfigure}{.5\textwidth}
  \centering
  \includegraphics[width=1\linewidth]{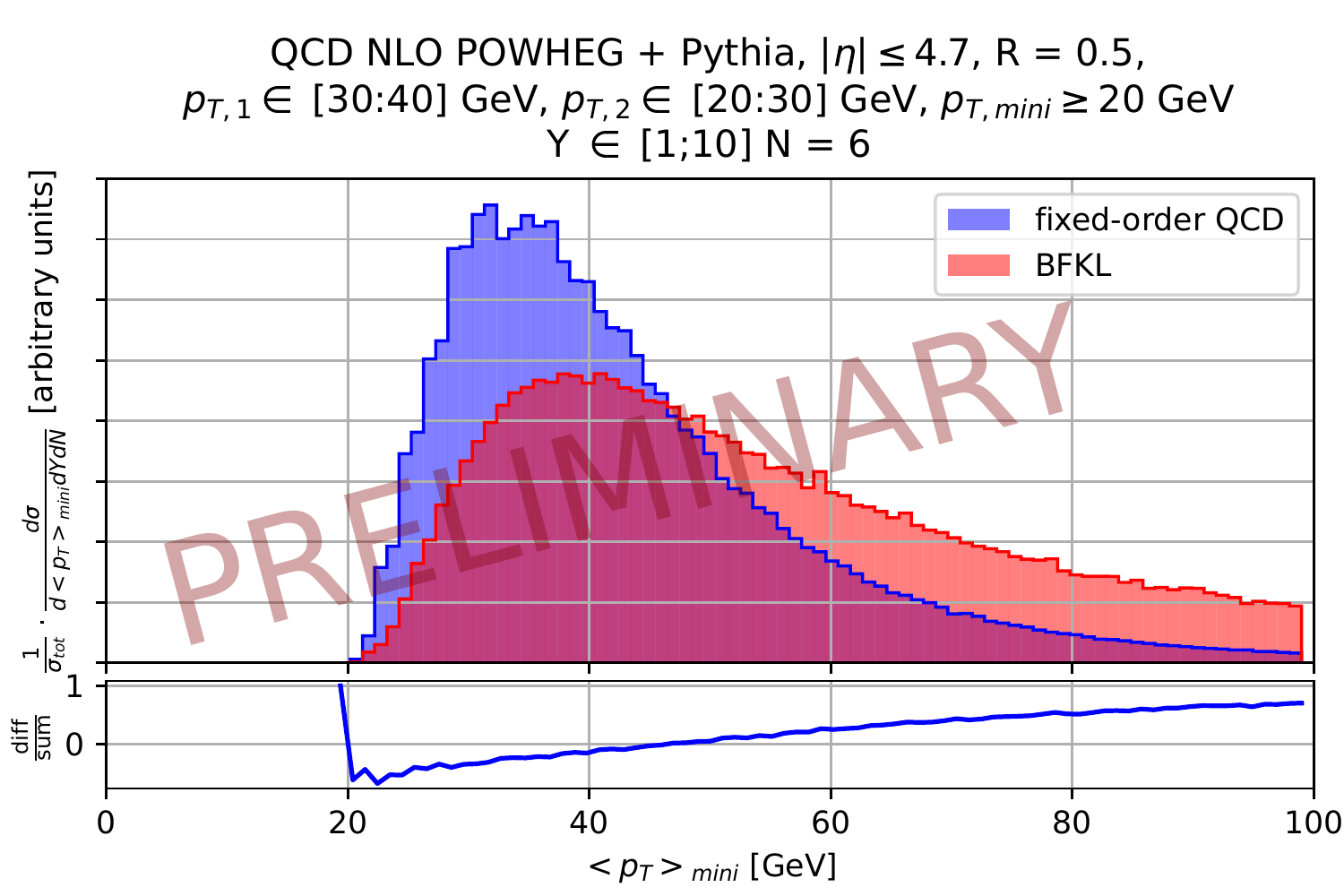}
\end{subfigure}
\caption{Left: Minijets below $p_{\perp, mim} = 20$ GeV are ignored for the calculation of the observables. Right: The observable
from Eqs.~\ref{eq:observablept}.}
\label{fig:cuts-pt}
\end{figure}

\begin{figure}
\begin{subfigure}{.5\textwidth}
  \centering
  \includegraphics[width=1\linewidth]{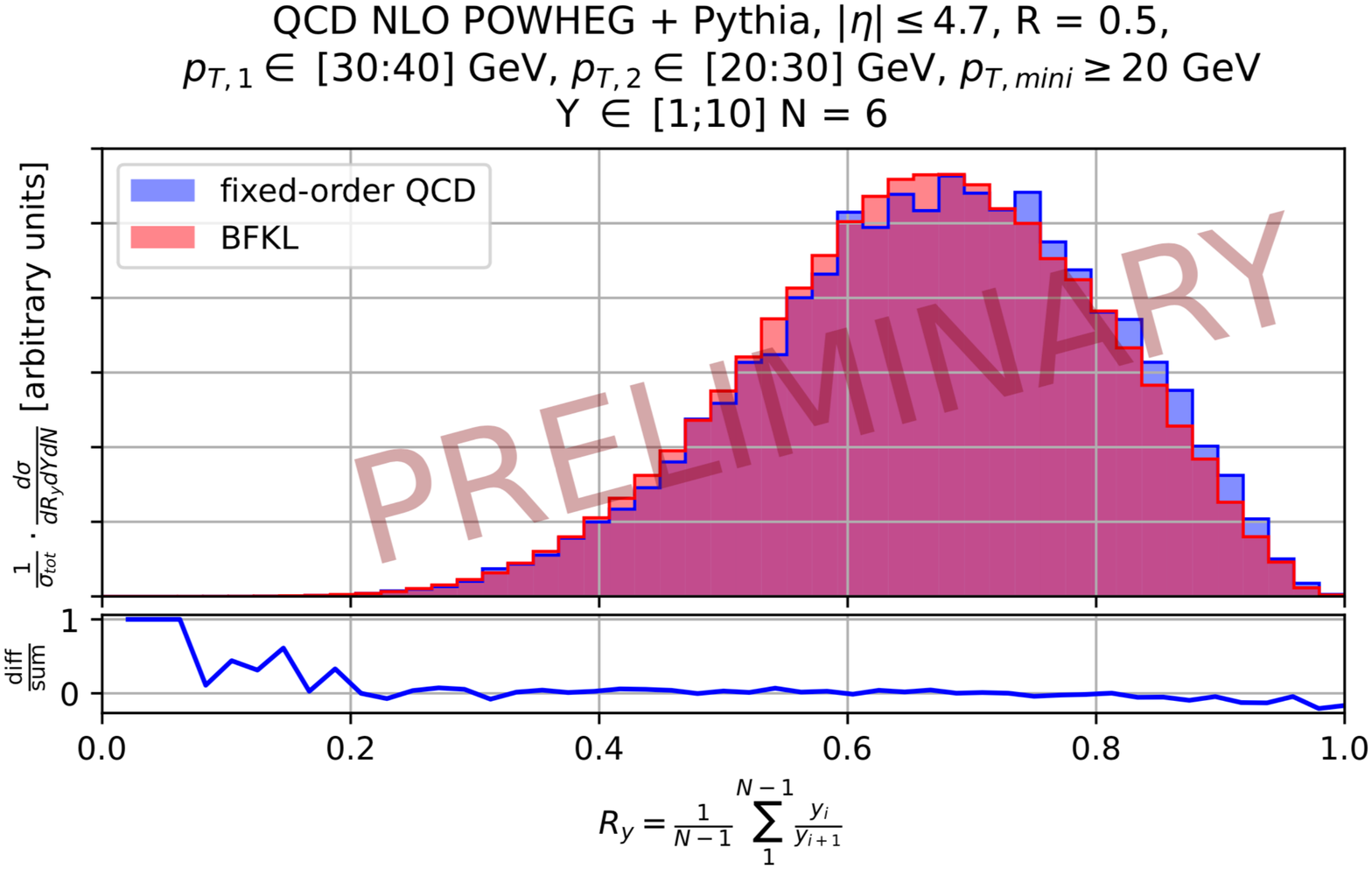}
\end{subfigure}%
\begin{subfigure}{.5\textwidth}
  \centering
  \includegraphics[width=1\linewidth]{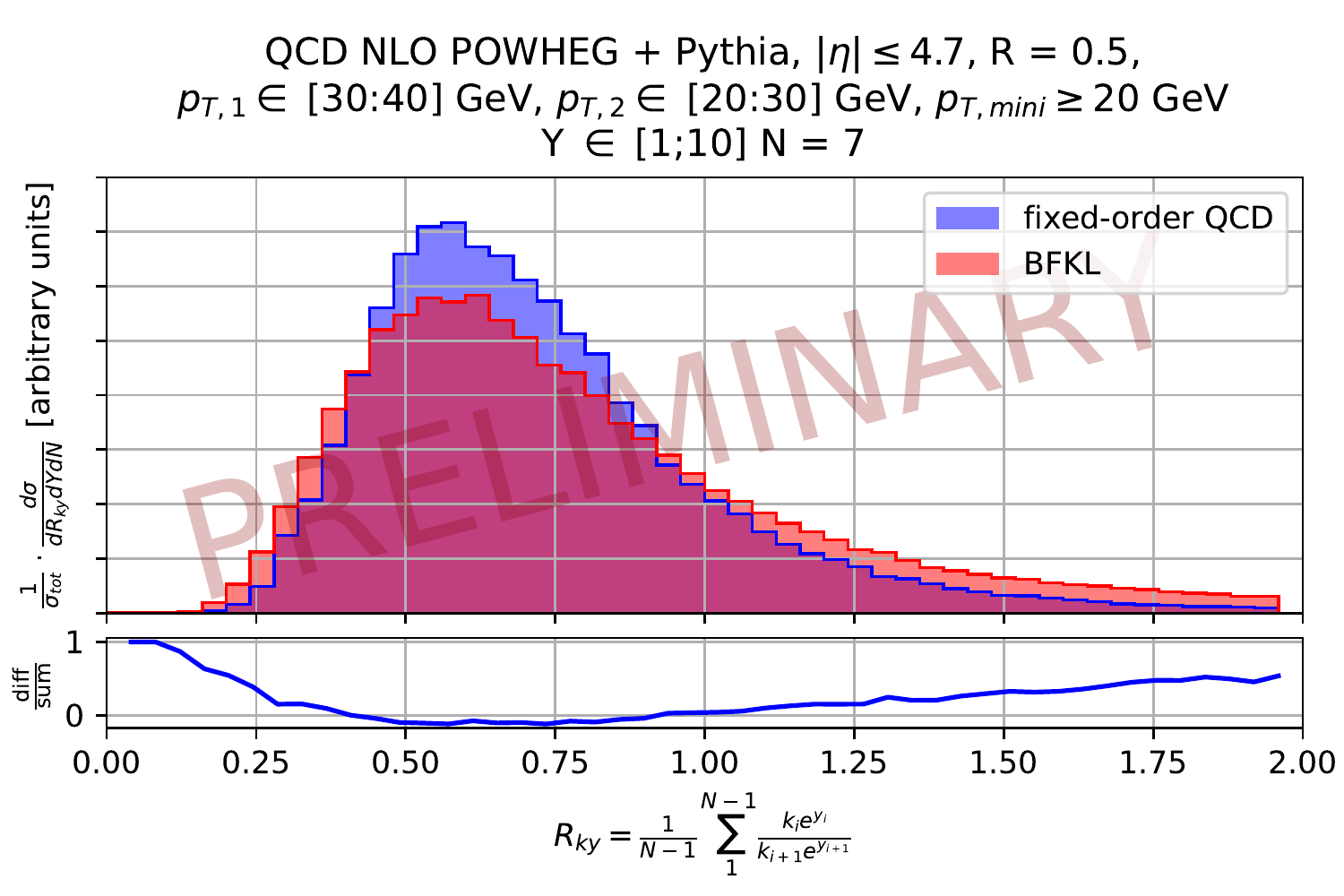}
\end{subfigure}
\caption{Left: The observable from Eq.~\ref{eq:observableRy}. Right: The observable from Eq.~\ref{eq:observableRky}.}
\label{fig:ry-rky}
\end{figure}

For the fixed order QCD computation of the observables in Eqs.~\ref{eq:observablept},~\ref{eq:observableRy} and ~\ref{eq:observableRky}
we use POWHEG~\cite{Nason:2004rx,Frixione:2007vw,Alioli:2010xd} and Pythia 8~\cite{Sjostrand:2014zea}. 
In both the BFKL based computation and the fixed order one, the anti-$kt$ jet clustering algorithm has been used as implemented in
{\tt fastjet}~\cite{Cacciari:2011ma,Cacciari:2005hq}.
We use the following kinematic 
cuts:
\begin{equation}
\begin{aligned}
&p_{\perp 0} \in[30 ; 40] \,\mathrm{GeV} \\
&p_{\perp n-1} \in[20 ; 30]\, \mathrm{GeV} \\
&p_{\perp \min } \geq 20\, \mathrm{GeV} \\
&Y \in[-4.7 ; 4.7]\,,
\end{aligned}
\end{equation}
whereas the jet radius was taken to be $R=0.5$ and the NNPDF31~\cite{NNPDF:2017mvq} PDF sets were used.

In Figs.~\ref{fig:cuts-pt} and~\ref{fig:ry-rky}
we present some preliminary plots of the observables defined in Eqs.~\ref{eq:observablept},~\ref{eq:observableRy} and ~\ref{eq:observableRky}. At the moment, there are no clear conclusions
to draw here, this is still work in progress and
the final results will be reported elsewhere.

\subsection{Toward precision studies of BFKL dynamics}
\label{sec:HAS_HE-QCD_precision}

\begin{figure}[b]
\centering

   \includegraphics[scale=0.50,clip]{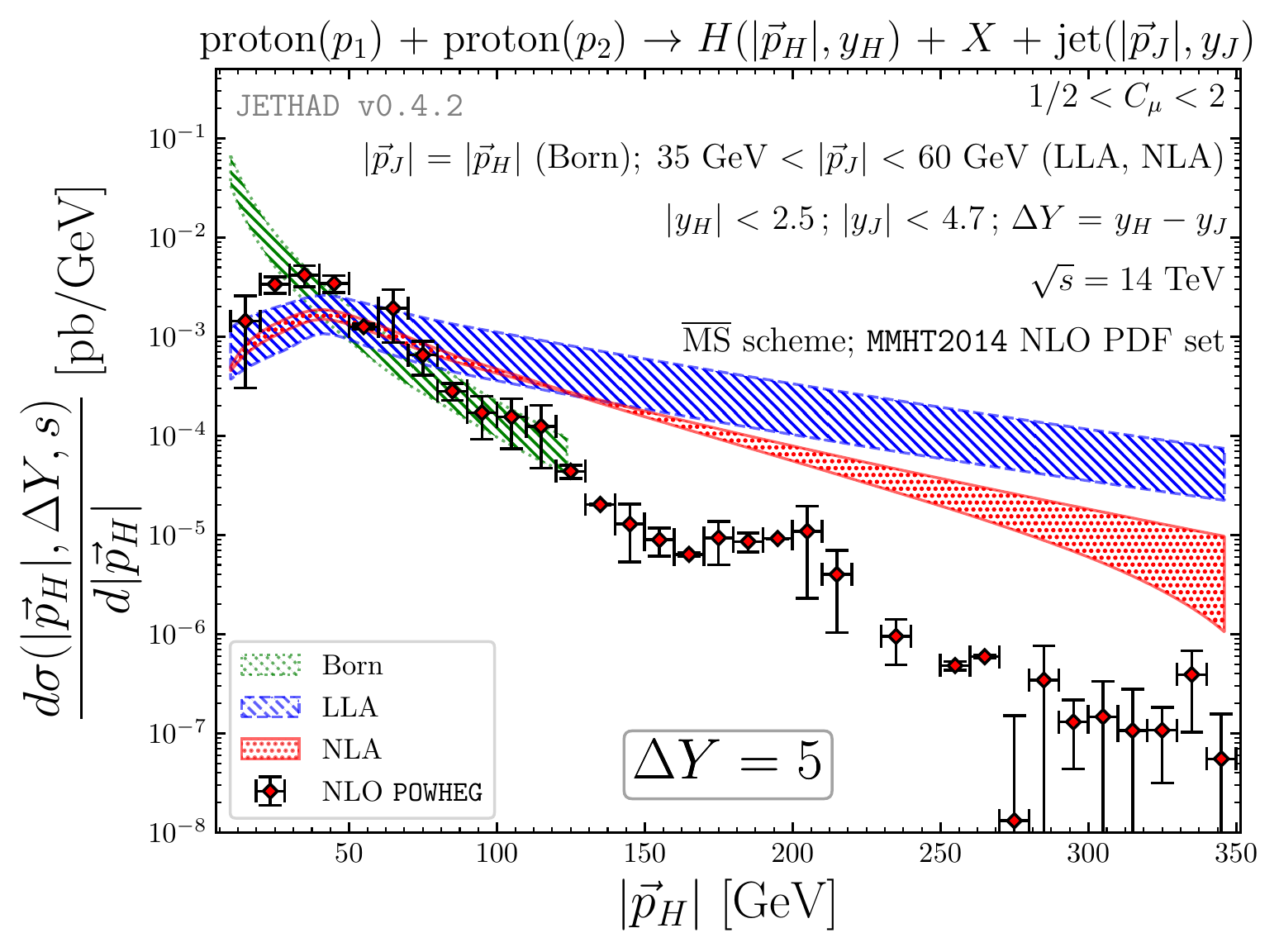}
   \hspace{0.25cm}
   \includegraphics[scale=0.40,clip]{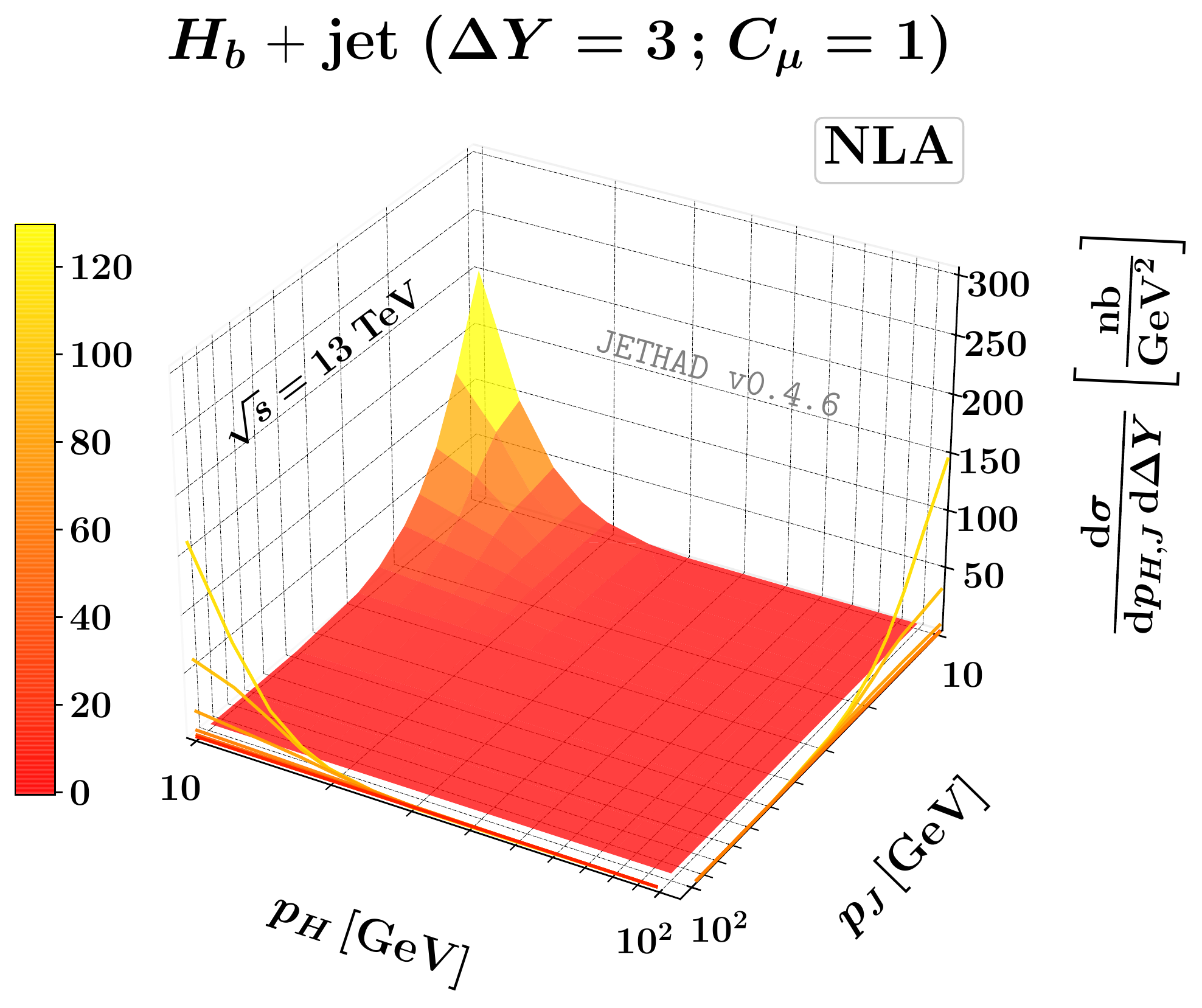}

\caption{
Left panel: 
$p_T$-dependence of the NLA cross section for the inclusive hadroproduction of a Higgs~$+$~jet system at $\Delta Y=5$ and $\sqrt{s} = 14$ TeV. Shaded bands give the uncertainty effect of coming from $\mu_{R,F}$ scale variation.
Right panel:
NLA doubly differential $p_T$-distribution for the inclusive emission of a $H_b$~$+$~jet system at $\Delta Y=3$ and $\sqrt{s} = 13$ TeV. Calculations are done at natural scales.
Figures from Refs.~\cite{Celiberto:2020tmb,Celiberto:2021fdp}.
}
\label{fig:HAS_HE-QCD_Higgs_b-flavor}
\end{figure}




Over the last decade, predictions for a large number of semi-hard observables in unpolarized hadronic collisions have been obtained.
Among them, azimuthal correlations between two jets emitted with high transverse momenta and large separation in rapidity (Mueller--Navelet dijet channel~\cite{Mueller:1986ey}) have been identified as promising observables whereby discriminating between BFKL-resummed and fixed-order-inspired calculations~\cite{Celiberto:2015yba,Celiberto:2015mpa}.
Several phenomenological studies have been conducted so far~\cite{Marquet:2007xx,Colferai:2010wu,Caporale:2012ih,Ducloue:2013hia,Ducloue:2013bva,Caporale:2013uva,Caporale:2014gpa,Caporale:2015uva,Mueller:2015ael,Celiberto:2016ygs,Celiberto:2016vva,Caporale:2018qnm}, which are in fair agreement with the only set of data available, \emph{i.e.} the one collected by the CMS collaboration for \emph{symmetric} ranges of the jet transverse momenta~\cite{Khachatryan:2016udy}.
In Ref.~\cite{Celiberto:2020wpk} (see also Refs.~\cite{Celiberto:2017ius,Celiberto:2017uae,Celiberto:2017ydk,Bolognino:2018oth,Bolognino:2019yqj,Bolognino:2019cac,Celiberto:2020rxb,Celiberto:2021xpm}) a clear evidence was provided that the high-energy resummed dynamics can be sharply disengaged from the fixed-order pattern at LHC energies when \emph{asymmetric} cuts for transverse momenta are imposed both in dijet and in jet plus light-hadron final states.
A wealth of inclusive hadronic semi-hard reactions have been considered as testfields for the BFKL resummation: di-hadron correlations~\cite{Celiberto:2016hae,Celiberto:2016zgb,Celiberto:2017ptm}, multi-jet emissions~\cite{Caporale:2015vya,Caporale:2015int,Caporale:2016soq,Chachamis:2016qct,Caporale:2016vxt,Chachamis:2016lyi,Caporale:2016pqe,Caporale:2016xku,Celiberto:2016vhn,Caporale:2016djm,Caporale:2016lnh,Caporale:2016zkc,Chachamis:2017vfa,Caporale:2017jqj}, $J/\psi$-plus-jet~\cite{Boussarie:2017oae,Celiberto:2022dyf}, heavy-quark pair~\cite{Celiberto:2017nyx,Bolognino:2019yls,Bolognino:2019ouc}, and forward Drell–Yan di-lepton production with backward-jet detection~\cite{Golec-Biernat:2018kem} and more.\\

One well know issue in the BFKL approach is that NLO corrections to the Green's function turn out to be large and with opposite sign with respect to the LO contribution. This is generally true also for the impact factors, depicting the transition in the fragmentation region of the colliding particles, all that resulting is a strong instability of the high-energy series. A notable example in this respect is represented by the
Mueller--Navelet reaction, where instabilities can be dumped by 
unnaturally large values of the renormalization and factorization scales~\cite{Ducloue:2013bva,Caporale:2014gpa,Celiberto:2020wpk},
chosen within suitable optimization schemes, such as the Brodsky--Lepage--Mackenzie (BLM) method~\cite{Brodsky:1996sg,Brodsky:1997sd,Brodsky:1998kn,Brodsky:2002ka}. 
This brings to a substantial lowering of cross sections and hampers any chance of making precision studies.

Recently, however, a set of semi-hard reactions was singled out exhibiting
a first, clear stability, in the typical BFKL observables, under higher-order corrections calculated at {\em natural} scales. It is the case of forward
emission of objects with a large transverse mass, such as Higgs bosons~\cite{Celiberto:2020tmb,Celiberto:2021fjf,Celiberto:2021tky,Celiberto:2021txb} and heavy-flavored jets~\cite{Bolognino:2021mrc,Bolognino:2021hxx,Bolognino:2021zco}, studied with partial NLO accuracy. Strong stabilizing effects in full NLO emerged in recent studies on inclusive emissions of $\Lambda_c$ baryons~\cite{Celiberto:2021dzy,Celiberto:2021txb} and bottom-flavored hadrons~\cite{Celiberto:2021fdp}. Here, a corroborating evidence was provided that the characteristic behavior of variable-flavor-number-scheme (VFNS) collinear fragmentation functions (FFs) describing the production of those heavy-flavored bound states at large transverse momentum~\cite{Kniehl:2020szu,Kniehl:2008zza,Kramer:2018vde} acts as a fair stabilizer of high-energy dynamics. 
We refer to this property, namely the existence of semi-hard reactions that can be studied in the BFKL approach without applying any optimization scheme nor artificial improvements of the analytic structure of cross section, as \emph{natural stability} of the high-energy resummation.
Figure~\ref{fig:HAS_HE-QCD_Higgs_b-flavor}(left) summarizes the key features
of a well behaved perturbative series in the case of the $p_T$-distribution of a forward Higgs inclusively produced together with a backward jet (rapidity difference $\Delta Y=5$) in proton-proton collisions at $\sqrt s$ = 14 TeV: Born and NLO fixed order predictions are clearly separated from LO and NLO BFKL, and the latter show a very moderate dependence on scale variation. Similar features are seen if a bottom-flavored hadron is detected
instead of a Higgs boson in the forward region -- see Fig.~\ref{fig:HAS_HE-QCD_Higgs_b-flavor}(right). This supports the statement that high-energy emissions in forward regions of rapidities bring along a high discovery potential and a concrete opportunity to widen our understanding of hadronic structure and, more in general, of strong interactions at new-generation colliders, such as the EIC~\cite{Accardi:2012qut,AbdulKhalek:2021gbh,Khalek:2022bzd}, HL-LHC~\cite{Chapon:2020heu}, the International Linear Collider (ILC)~\cite{AlexanderAryshev:2022pkx}, the Forward Physics Facility~(FPF)~\cite{Anchordoqui:2021ghd,Feng:2022inv}, and NICA-SPD~\cite{Arbuzov:2020cqg,Abazov:2021hku}.
\subsection{Unintegrated gluon distribution (UGD)}
\label{sec:HAS_HE-QCD_UGD}

\begin{figure}
\centering

\includegraphics[scale=0.52,clip]{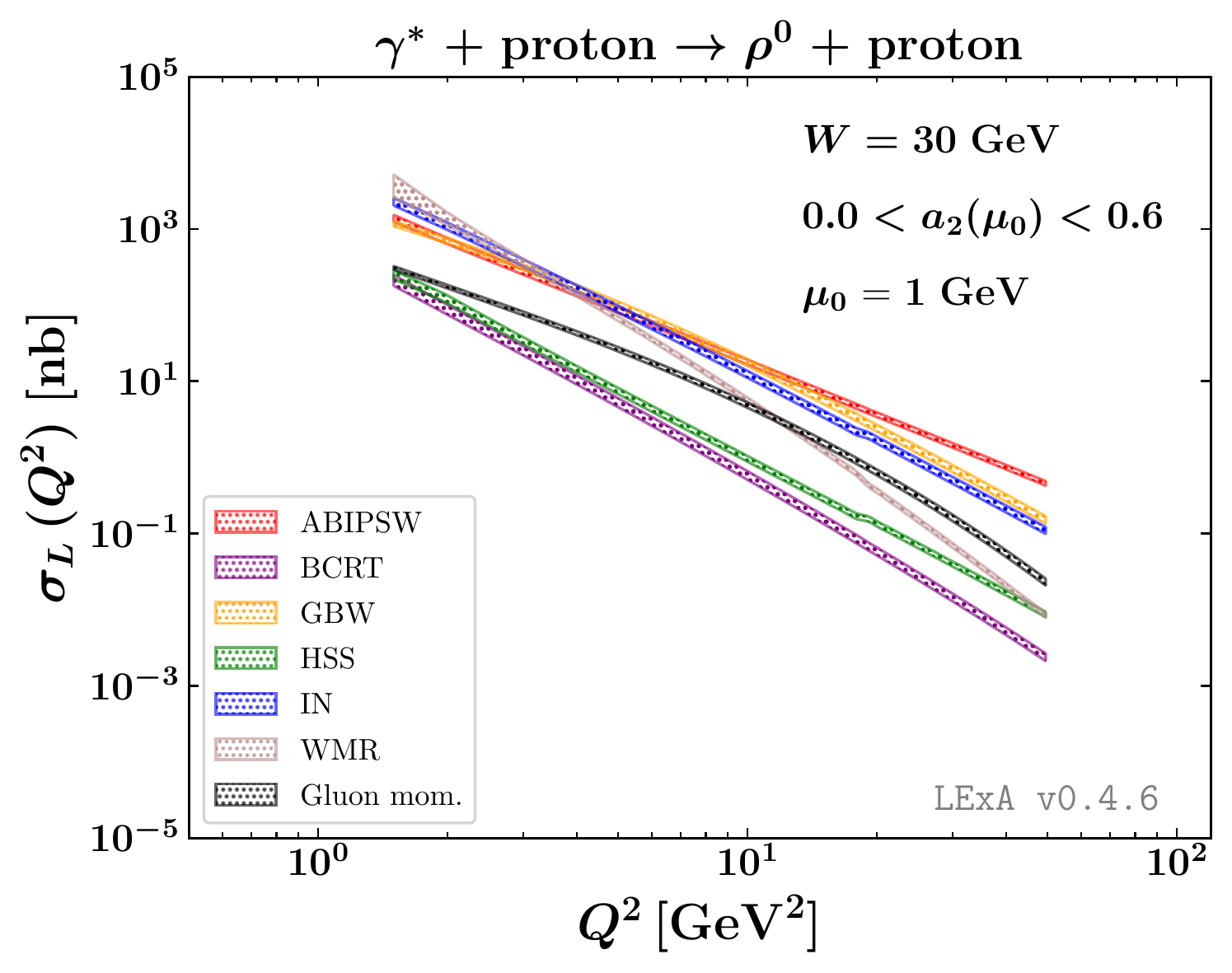}
\hspace{0.25cm}
\includegraphics[scale=0.52,clip]{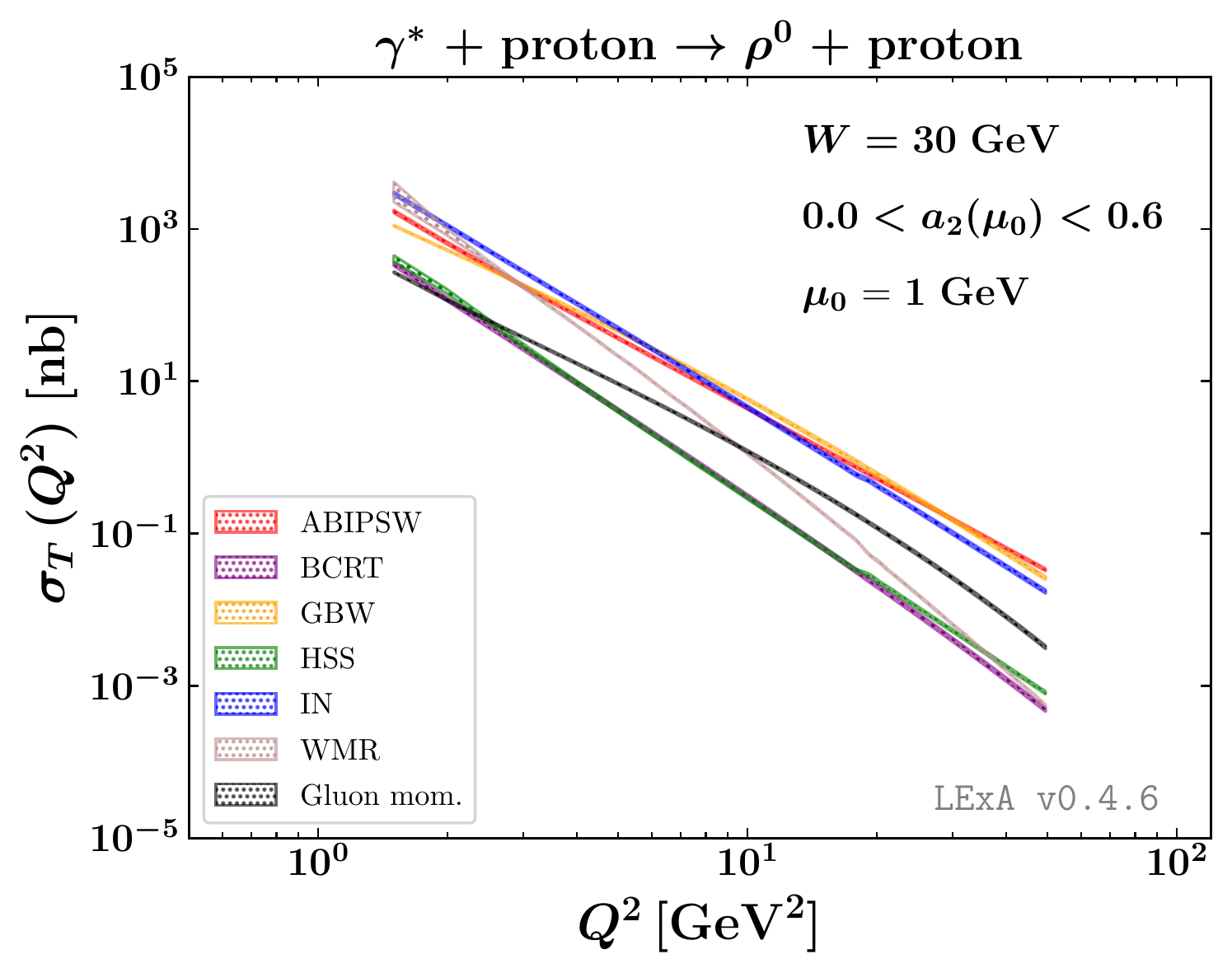}

\caption{
$Q^2$-dependence of the longitudinally (left) and transversely (right) polarized cross section, $\sigma_{L,T}$, for all the considered UGD models, at the EIC reference energy of $W = 30$ GeV. Uncertainty bands describe the effect of varying non-perturbative parameters inside distribution amplitudes depicting the exclusive emission of a $\rho$ meson.
Figures from Ref.~\cite{Bolognino:2021niq}.
}
\label{fig:HAS_HE-QCD_rho}
\end{figure}

Inclusive emissions of single forward particles represent a golden channel to access the proton content at low-$x$ via an unintegrated gluon distribution.
The original definition of the unintegrated gluon distribution relies on high energy factorization and  Balitsky--Fadin--Kuraev--Lipatov (BFKL) evolution~\cite{Fadin:1975cb,Kuraev:1976ge,Kuraev:1977fs,Balitsky:1978ic}. It takes the form of a convolution in transverse momentum space between the BFKL Green's function and the proton impact factor.
The Green's function is process-independent and accounts for the resummation of small-$x$ logarithms, while the proton impact factor represents the non-evolved part of the density and is of non-perturbative nature.
Our knowledge of the proton impact factor is very limited and different models for it and for the unintegrated gluon distribution itself have been proposed so far.
First analyses of unintegrated gluon distributions were performed in the context of deep-inelastic-scattering (DIS) structure functions~\cite{Hentschinski:2012kr,Hentschinski:2013id}. Subsequently, the unintegrated gluon distribution was probed via the exclusive electro- or photo-production of vector mesons at HERA~\cite{Anikin:2009bf,Anikin:2011sa,Besse:2013muy,Bolognino:2018rhb,Bolognino:2018mlw,Bolognino:2019bko,Bolognino:2019pba,Celiberto:2019slj,Bautista:2016xnp,Garcia:2019tne,Hentschinski:2020yfm} and the EIC~\cite{Bolognino:2021niq,Bolognino:2021gjm,Bolognino:2022uty}, the single inclusive heavy-quark emission at the LHC~\cite{Chachamis:2015ona}, and the forward Drell--Yan production at LHCb~\cite{Motyka:2014lya,Brzeminski:2016lwh,Motyka:2016lta,Celiberto:2018muu}.

The connection between the uintegrated gluon distribution and the collinear gluon PDF was investigated through a high-energy factorization framework set up in Refs.~\cite{Catani:1990xk,Catani:1990eg,Collins:1991ty}, and via the Catani--Ciafaloni--Fiorani--Marchesini (CCFM) \emph{branching} scheme~\cite{Ciafaloni:1987ur,Catani:1989sg,Catani:1989yc,Marchesini:1994wr,Kwiecinski:2002bx}. Then, first determinations of small-$x$ improved PDFs \emph{\`a la} Altarelli--Ball--Forte (ABF)~\cite{Ball:1995vc,Ball:1997vf,Altarelli:2001ji,Altarelli:2003hk,Altarelli:2005ni,Altarelli:2008aj,White:2006yh} were recently achieved~\cite{Ball:2017otu,Abdolmaleki:2018jln,Bonvini:2019wxf}.
A first connection between the unintegrated gluon distribution and the unpolarized and the linearly polarized gluon TMDs, $f^g_1$ and $h_1^{\perp g}$,
was investigated in Refs.~\cite{Dominguez:2011wm,Hentschinski:2021lsh,Nefedov:2021vvy}. Recent studies~\cite{Altinoluk:2019fui,Fujii:2020bkl,Boussarie:2021ybe} on the hadronic structure in the \emph{saturation} regime have highlighted the significance of the interplay between the Color Glass Condensate (CGC), the low-$x$ improved TMD (iTMD) framework~\cite{Kotko:2015ura,vanHameren:2016ftb} and the BFKL dynamics, see also the discussion in Sec.~\ref{sec:structure}. Here, both the genuine and the kinematic twists play a key role in shedding light on the transition regions among these approaches. \\

In Fig.~\ref{fig:HAS_HE-QCD_rho} we show the dependence on the hard scale $Q^2$  seven different unintegrated gluon distributions, presented in Section~3 of Ref.~\cite{Bolognino:2021niq}. To be specific, we  single exclusive production of a $\rho$-meson in lepton-proton collisions via the sub-process
\begin{equation}
\label{eq:HAS_HE-QCD_subprocess}
 \gamma^*_{\lambda_i} (Q^2) \, p \; \to \; \rho_{\lambda_f} p \;,
\end{equation}
where a photon with virtuality $Q^2$ and polarization $\lambda_i$ is absorbed by a proton and a $\rho$-meson with polarization $\lambda_f$ is detected in the final state. The two spin states $\lambda_{i,f}$ can be longitudinal $(0)$ or transverse $(1)$. The $(00)$ combination gives rise to the longitudinal cross section, $\sigma_L(Q^2)$, while the $(11)$ one to the transverse cross section, $\sigma_T(Q^2)$.
Here the semi-hard scale ordering, $W^2 \gg Q^2 \gg \Lambda^2_{\rm QCD}$ (with $W$ the hard-scattering center-of-mass energy), is stringently preserved, and the small-$x$ regime, $x = Q^2/W^2$, is accessed. We further present new results for the EIC~\cite{Accardi:2012qut,AbdulKhalek:2021gbh,Khalek:2022bzd} at the reference energy of $W = 30$ GeV (right panel). We make use of the twist-2 (twist-3) distribution amplitudes s for the longitudinal (transverse) configuration, and we gauge the impact of the collinear evolution of the distribution amplitudes describing the exclusive emission of the $\rho$  via a variation of the non-perturbative parameter $a_2(\mu_0 = 1\,$\rm GeV$)$ in the range 0.0 to 0.6 (see Section~2 of Ref.~\cite{Bolognino:2021niq} for further details).

We point out that our predictions are spread over a large range. This provides us with a clear evidence that polarized cross sections for the exclusive production of light-vector mesons (such as the $\rho$-particle) in lepton-proton collisions act as a discriminator for the unintegrated gluon distribution. We expect that future studies at the EIC will substantially extend our knowledge of the gluon content of the proton at small-$x$.


\subsection{BFKL resummation of NLO collinear factorization:  Heavy quarkonium production}\label{sec:Quarkonia_HEF}


Another way BFKL dynamics manifests itself is within a  direct
resummation of contributions enhanced by logarithms of partonic center of mass energy in collinear factorization. Such an approach allows to combine high energy resummation with
theoretical fixed order predictions which are in general available at a
higher perturbative order than their counterparts obtained within high
energy factorization. In the following we focus on the production of
heavy quarkonia -- bound states of $c\bar{c}$ or $b\bar{b}$ heavy
quark pairs, see~\cite{Brambilla:2010cs, Brambilla:2014jmp, Lansberg:2019adr} for a recent review.  New quarkonium-related measurements had been proposed for the
experimental programs of the High-Luminosity LHC~\cite{Chapon:2020heu}
and Spin Physics Detector at NICA~\cite{Arbuzov:2020cqg}, as well as
for fixed-target program at the LHC~\cite{Hadjidakis:2018ifr,Brodsky:2012vg}. It is believed, that the non-relativistic nature of
these bound states  should allow for the description of hadronization of the
heavy quark-antiquark pair into an observed quarkonium state with a
modest number of free parameters. Despite the availability of several factorisation approaches such as the 
Color-Singlet Model~\cite{Gastmans:1986qv}, Non-Relativistic QCD
Factorisation approach~\cite{Bodwin:1994jh} and more recent
potential-NRQCD~\cite{Brambilla:2021abf} and Soft-Gluon
factorisation~\cite{Ma:2017xno,Li:2019ncs}, 
none of them is yet capable to fully
describe the rich phenomenology of inclusive heavy quarkonium
production observables, which includes differential cross sections and
polarisation observables in proton-proton, lepton-proton collisions
and $e^+e^-$ annihilation, as described in more deltail in the
reviews~\cite{Brambilla:2010cs, Brambilla:2014jmp,Lansberg:2019adr,Arbuzov:2020cqg,Chapon:2020heu} cited above.\\

\begin{figure}
    \centering
    \parbox{0.45\textwidth}{\includegraphics[width=0.45\textwidth]{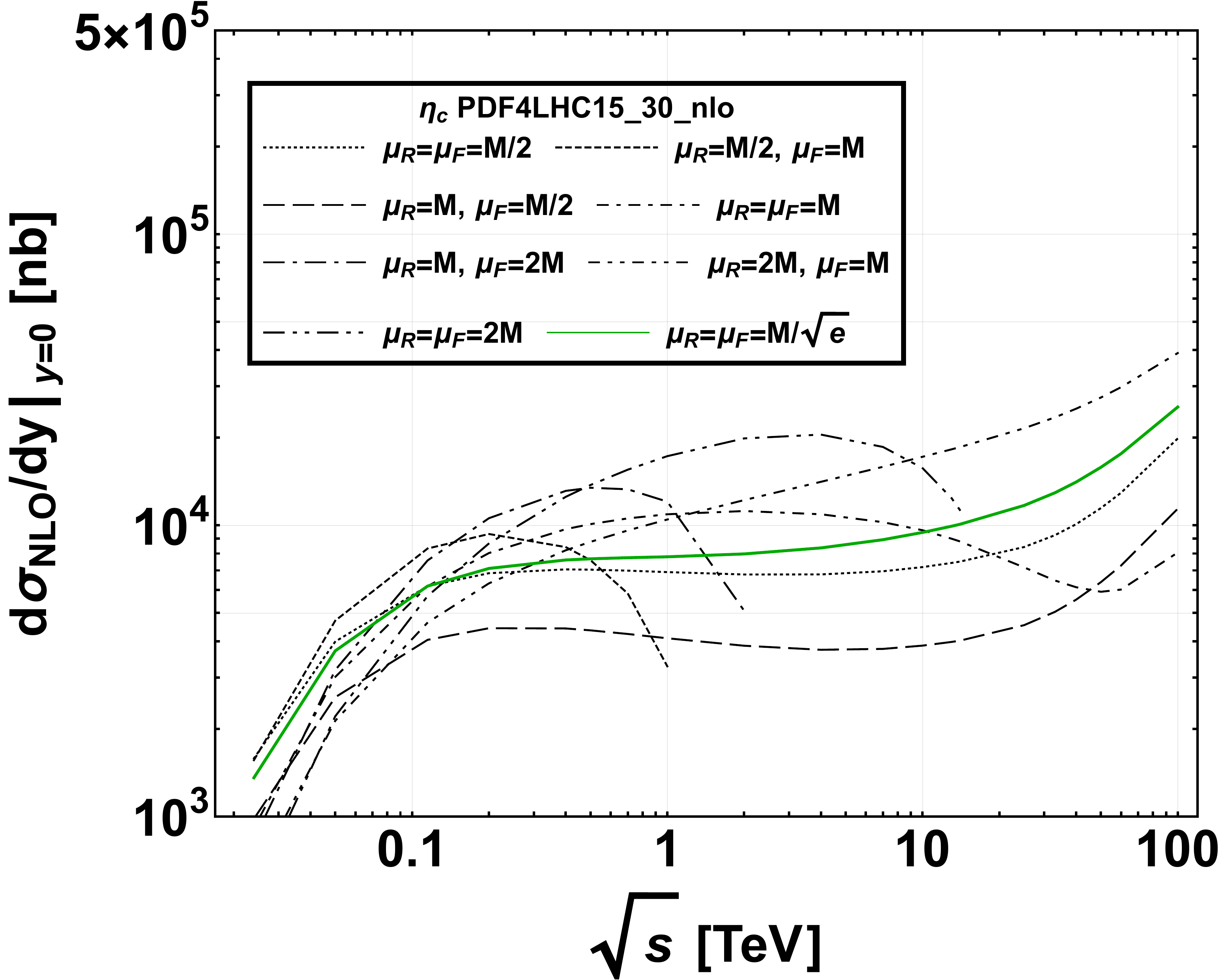}}\hspace{0.05\textwidth}
    \parbox{0.45\textwidth}{\includegraphics[width=0.45\textwidth]{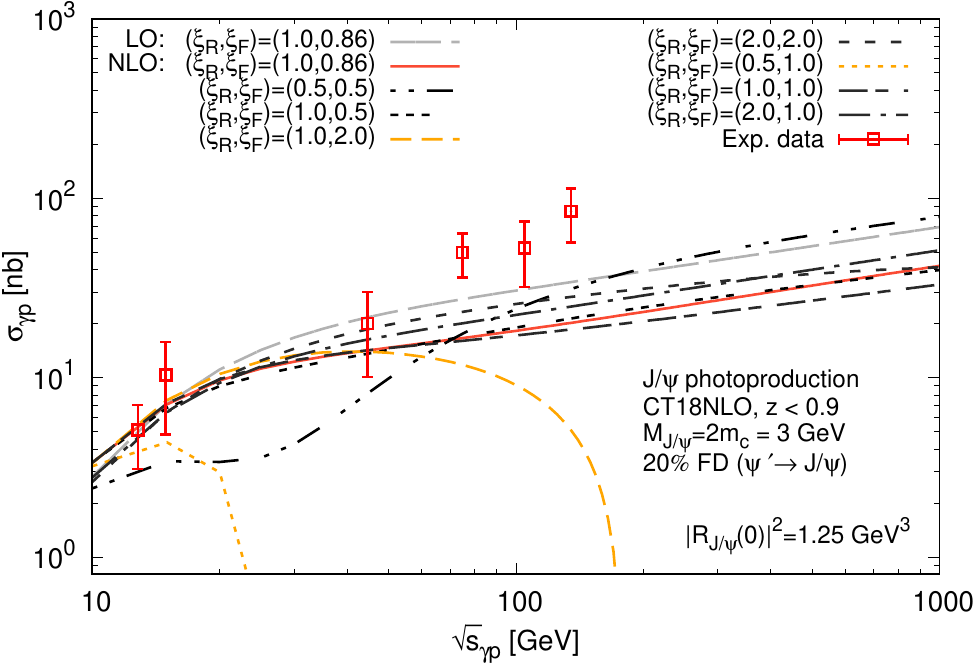}}
    \caption{Collision energy dependence of $p_T$-integrated prompt inclusive $\eta_c$ hadro-production cross section (left panel, adopted form Ref.~\cite{Lansberg:2020ejc}) and prompt inclusive $J/\psi$ photoproduction cross section (right panel, adopted from Ref.~\cite{ColpaniSerri:2021bla}) at NLO in $\alpha_s$ in the Color-Singlet Model for various choices of factorization ($\mu_F=\xi_F M$) and renormalization ($\mu_R=\xi_R M$) scales.}
    \label{fig:etac_jpsi_s-dep}
\end{figure}
 As pointed out in Ref.~\cite{Lansberg:2020ejc} for the case of
$p_T$-integrated prompt $\eta_c$ hadro-production cross section and
Ref.~\cite{ColpaniSerri:2021bla} for the total inclusive

photo-produciton cross section of prompt $J/\psi$, the collinear NLO
calculations of these quantities, based on the Color-Singlet(CS)
Model, become unreliable if the collision energy $\sqrt{s}$
significantly exceeds the heavy quarkonium mass $M$, see
Fig.~\ref{fig:etac_jpsi_s-dep} leading to negative cross-sections for reasonable values of factorisation scale such as $\mu_F=2M$. A
careful analysis of the perturbative partonic cross-section and the
convolution integral of the former with parton distribution functions,
allows to trace this instability back to the behaviour of the NLO partonic coefficient function at large partonic center of mass energy $\hat{s}\gg M^2$. Beyond NLO, the high energy
logarithmic corrections $\sim \alpha_s^n \ln^{n-1}(\hat{s}/M^2)$ arise in this limit. The necessary resummation can be addressed using the BFKL
resummation in a form of High-Energy Factorisation (HEF), provided by Refs.~\cite{Catani:1990eg,Catani:1994sq,Catani:1990xk,Collins:1991ty}. The
formalism described in these papers allows one to perform a resummation
of Leading Logarithmic (LL) corrections
$\sim \alpha_s^n \ln^{n-1}(\hat{s}/M^2)$ to the partonic cross-section in
all orders in $\alpha_s$. As shown in \cite{Lansberg:2021vie}, in order for this resummation 
to be consistent with the factorzation scale dependence of  collinear PDFs, subject to standard NLO DGLAP evolution, it is needed to truncate the  full LL($\ln \hat{s}/M^2$) resummation
for the partonic cross-section, by taking into account only
Doubly-Logarithmic (DL) terms $\sim \alpha_s^n \ln^{n-1}(\hat{s}/M^2) \ln^n ({\bf q}_T^2/\mu_F^2)$ in the resummation functions of high energy logarithms. This allows to obtain the double logarithmic resummed
expression -- $\hat{\sigma}_{ij}^{\rm (HEF)}(\hat{s},\mu_F,\mu_R)$ ($i,j=q,\bar{q},g$),
which is guaranteed to reproduce the leading logarithmic terms in the $\hat{s}\gg M^2$
asymptotics of the exact partonic cross-section up to NNLO in $\alpha_s$; it serves therefore for  an approximation for the latter one in
the Regge limit. In Ref.~\cite{Lansberg:2021vie}  the resummed expression was then  combined with the exact collinear NLO result, through introducing a  smooth weight functions
$0<w_{ij}^{\rm (CF)}(\hat{s})<1$:
\begin{equation}
\hat{\sigma}_{ij}(\hat{s})= \sigma_{ij}^{\text{(CF, LO)}}(\hat{s}) + \alpha_s w_{ij}^{\rm (CF)}(\hat{s}) \hat{\sigma}^{\text{(CF, NLO)}}_{ij}(\hat{s}) + (1-w_{ij}^{\rm (CF)}(\hat{s})) \hat{\sigma}^{\rm (HEF)}_{ij}(\hat{s}), \label{eq:NLO+HEF-matching}    
\end{equation}
which we construct by suitably adapting the Inverse Squared Errors Weighting (InEW) matching method of the Ref.~\cite{Echevarria:2018qyi}, in such a way, that the NLO CF term is suppressed when $\hat{s}\gg M^2$ and the resummation term is suppressed outside of the Regge limit. 

\begin{figure}
    \centering
    \parbox{0.45\textwidth}{\includegraphics[width=0.45\textwidth]{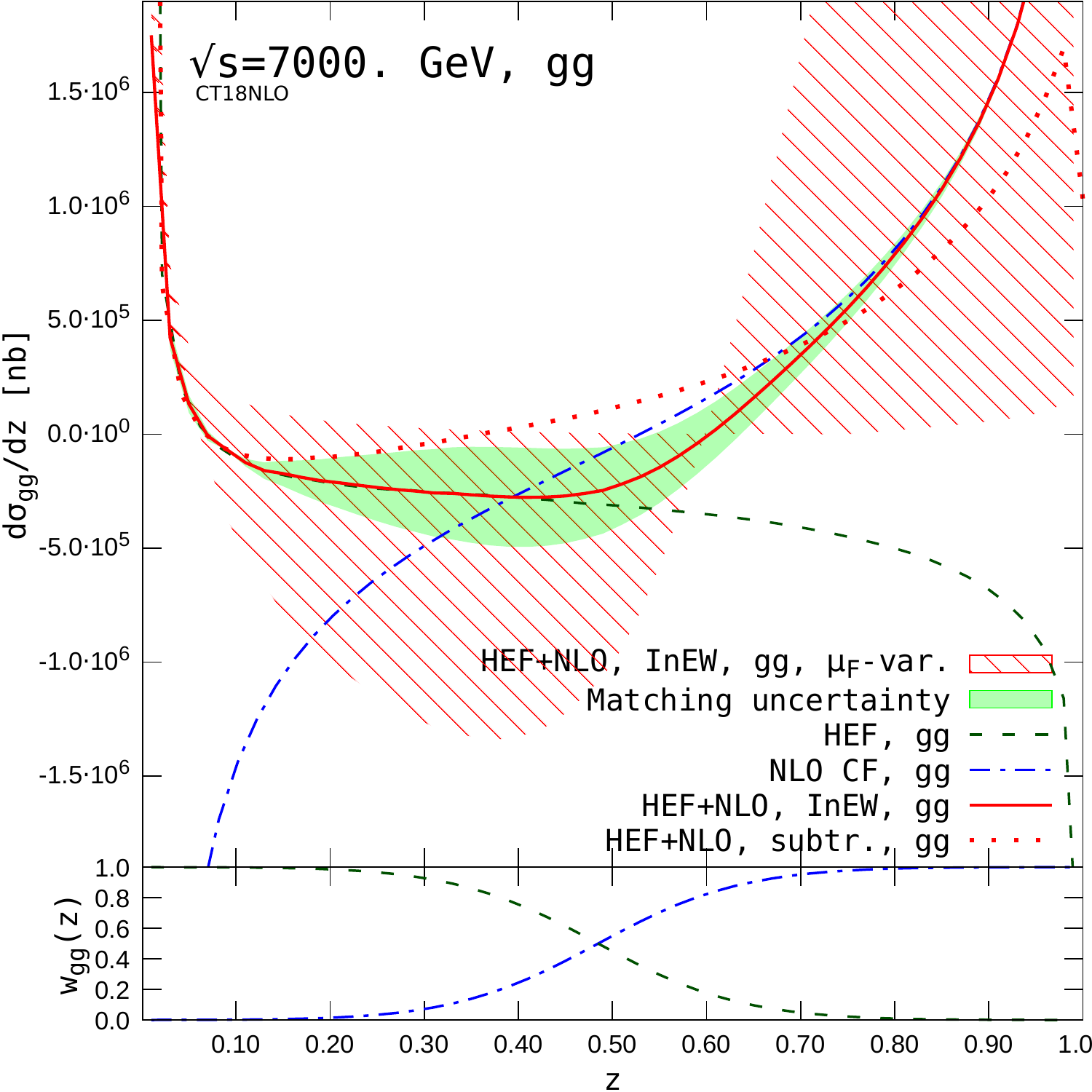}}\hspace{0.05\textwidth}
    \parbox{0.48\textwidth}{\includegraphics[width=0.45\textwidth]{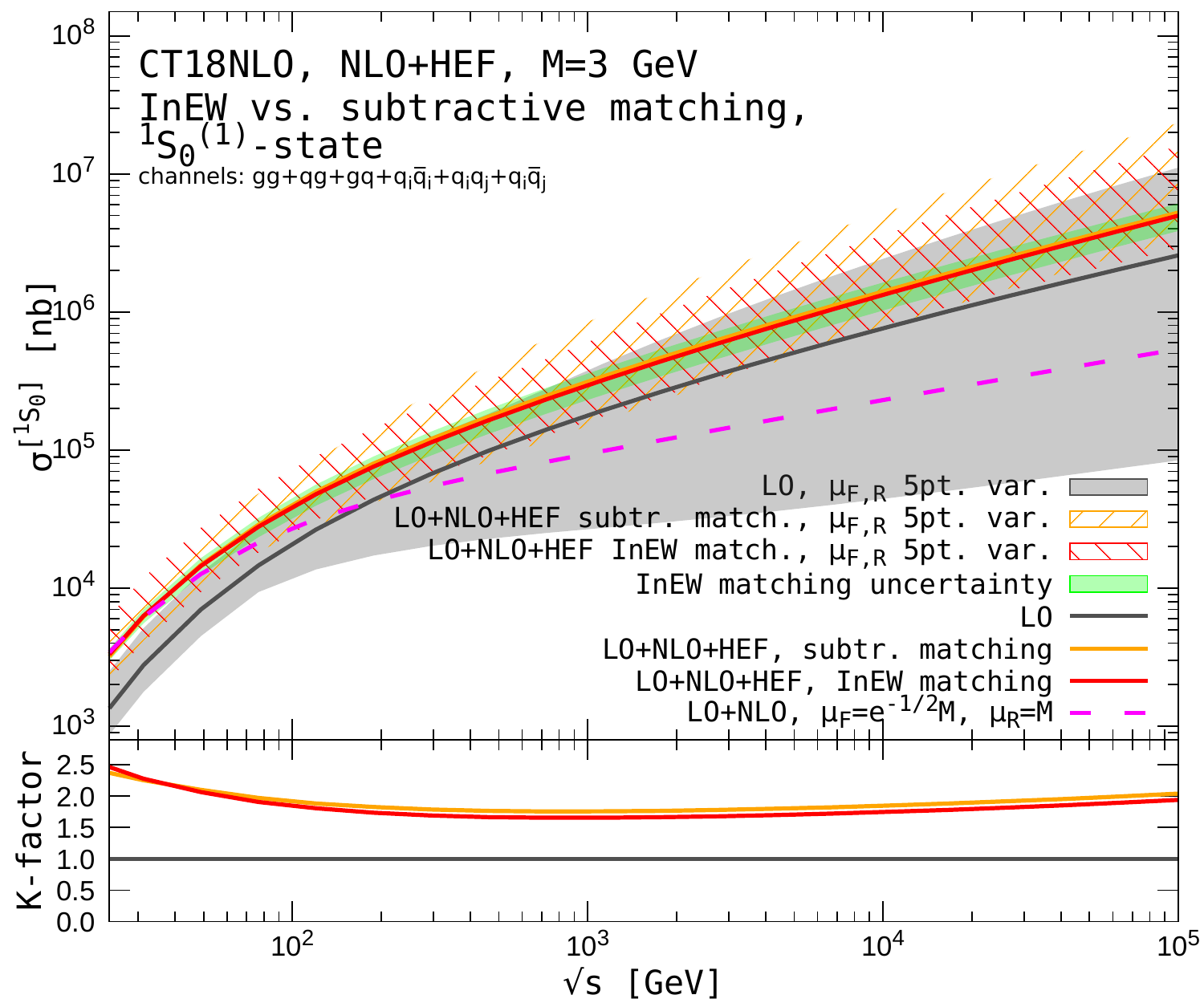}}
    \caption{Left panel: the integrand of the total $\eta_c$ hadro-production cross section as function of $z=M^2/\hat{s}$ in the $gg$-channel. Dashed line -- DL HEF contribution, dash-dotted line -- NLO CF contribution, solid line -- their matching, according to the Eq.~(\ref{eq:NLO+HEF-matching}), the weight function is shown in the inset. Right panel: The total cross section of hadro-production production of the $c\bar{c}\left[{}^1S_0^{(1)}\right]$-state as function of $\sqrt{s_{pp}}$. Plots adopted from the Ref.~\cite{Lansberg:2021vie}. }
    \label{fig:NLO+HEF-matching}
\end{figure}

  The numerical results of such matching calculation are illustrated by plots in the Fig.~\ref{fig:NLO+HEF-matching}. The left panel shows how the InEW matching of NLO collinear factorization and the resummed contributions works. In the right panel, the plot of the InEW-matched total cross section (red line) is shown together with its $\mu_F$ and $\mu_R$ scale-variation uncertainty (shaded band). One can see that the scale uncertainty of the matched prediction does not show any pathological behaviour at high energy, unlike the scale-variation plots in the Fig.~\ref{fig:etac_jpsi_s-dep}, and it is reduced compared to the scale uncertainty of the LO cross section. \\
  
  Thus we conclude, that problems with high-energy behaviour of $p_T$-integrated cross sections of quarkonium production described in Refs.~\cite{Lansberg:2020ejc, ColpaniSerri:2021bla} are manifestations of the {\it necessity} to perform the BFKL-type resummation of high-energy logarithms in the partonic coefficient function of these processes. Nevertheless  such resummed partonic cross section is only part of the full answer, and at realistic energies $\sqrt{s}$, the region of $M^2/\hat{s}\sim 1$ gives comparably large contribution, as it is clear from the left plot in the Fig.~\ref{fig:NLO+HEF-matching}, so both contributions should be matched. Interestingly, at high energies the $\hat{\mu}_F$-prescription of the Refs.~\cite{Lansberg:2020ejc, ColpaniSerri:2021bla} predicts much lower cross section\footnote{This happens for pseudo-scalar quarkonia because of the low value of $\hat{\mu}_F<M$ which leads to less evolved gluon PDFs and therefore smaller cross sections. In contrast to this, for other processes such as Higgs production, one may encounter larger $\hat{\mu}_F>M$ values and consequently larger cross sections due to PDF evolution \cite{Lansberg:2020ejc}.} (dashed line in the right panel of the Fig.~\ref{fig:NLO+HEF-matching}) than the matched NLO+ resummed calculation, which shows the importance of systematic resummation formalism, see the Sec. 2.5 of the Ref.~\cite{Lansberg:2021vie} for more detailed discussion.\\
  
In the future these  calculations based on double logarithmic resummation of high energy logarithms matched with NLO collinear factorization  in the scheme described above need to be extended to the $J/\psi$ photoproduction case, studied in the Ref.~\cite{ColpaniSerri:2021bla}, as well as to the case of rapidity and $p_T$-differential hadroproduction cross sections of prompt $\eta_c$ and $\chi_{c0,1,2}$ mesons, including the Color-Octet contributions for the latter ones.  Another important intermediate-term goal is to study how the DL HEF+NLO CF calculation changes the prompt $J/\psi$ $p_T$-spectrum in the Color-Singlet Model. And crucially, we must find the way to extend out formalism beyond the double logarithmic approximation to be able to reduce scale-variation uncertainties of our results to make them useful e.g. for determination of the gluon PDFs at low scales and small $x$. Experimentally, the inclusive $J/\psi$ photoproduction in wider range of $\sqrt{s_{\gamma p}}$ than was available at HERA could be accessed using ultra-peripheral collisions at the LHC, as well as future EIC data which will lie at lower energies but will be more precise due to the increased luminosity. The $\chi_{c0,1,2}$ production cross section in wide range of energies could be studied by the fixed-target experiments using LHC beams~\cite{Brodsky:2012vg, Hadjidakis:2018ifr, Barschel:2020drr}.  

%
%
%

\subsection{Diffraction: Gaps between jets}
\label{sec:jetgapjet}

Another jet probe of BFKL dynamics at the LHC is the production of two high-$p_\text{T}$ jets separated by a large (pseudo)rapidity interval void of particle activity, as proposed by Mueller and Tang nearly 30 years ago~\cite{Mueller:1992pe}. The rapidity gap signature between the jets is indicative of an underlying $t$-channel color-singlet exchange mechanism. The hard scale of the process, justified by the high jet $p_\text{T}$, allows for a treatment of this exchange in terms of perturbation theory. A natural mechanism in QCD to explain this process is BFKL Pomeron exchange between partons. This description is expected to be more justified as the jets become more separated in rapidity. Contributions based on DGLAP evolution are expected to be strongly suppressed in dijet events with a central rapidity gap by virtue of a Sudakov form factor that needs to be supplemented to the calculation. Thus, the jet-gap-jet process may allow us to directly access the small-$x$ dynamics of interest, complementary to other standard probes of this regime of QCD interactions.

Measurements of jet-gap-jet events have been presented by the CDF, D0, and CMS at $\sqrt{s} =$ 0.63, 1.8, 7, and 13 TeV~\cite{Abachi:1994hb, Abe:1998ip, Sirunyan:2017rdp, Sirunyan:2021oxl}. At the Tevatron and at the LHC, the pseudorapidity gap between the jets is defined as the absence of particles in $|\eta|<1$ with $p_\text{T} > 200$ MeV (or $300$ MeV in some cases) between the highest $p_\text{T}$ jets. The threshold is constrained by the capability of the detectors to reconstruct charged-particle tracks and by the calorimeter noise energy threshold. Experimentally, these events are very clean and can be separated from the overwhelming color-octet exchange dijet background using data-driven methods or with Monte Carlo generators. The observable that is extracted in these measurements is the fraction of color-singlet exchange dijet events in the inclusive dijet sample,

\begin{eqnarray}
f_\text{CSE} \equiv \frac{\text{d}\sigma_\text{CSE}}{\text{d}\sigma_\text{inclusive}}
\end{eqnarray}

The $f_\text{CSE}$ fraction is measured as a function of the second-leading jet $p_\text{T}^\text{jet2}$, the pseudorapidity separation between the jets $\Delta\eta_\text{jj} \equiv |\eta_\text{jet1}-\eta_\text{jet2}|$, and in some cases a measure of momentum imbalance between the jets, such as $\Delta\phi_\text{jj} \equiv |\phi_\text{jet1}-\phi_\text{jet2}|$. Theoretical uncertainties related to the choice of PDF and the variation of renormalization and factorization scales partially cancel in $f_\text{CSE}$. Correlated experimental uncertainties related to jet energy corrections, luminosity, acceptance and efficiency effects, cancel in the ratio. The $f_\text{CSE}$ fractions are of the order of 0.5--1\%, depending on the collision energy and dijet kinematics. This means that about 0.5--1\% of the inclusive dijet cross section is due to hard color-singlet exchange.

Previous phenomenological studies of the jet-gap-jet process were based on PYTHIA6 and HERWIG6 Monte Carlo generator~\cite{Cox:1999dw, Motyka:2001zh, csp, Kepka:2010hu, cspLHC}. This is a good motivation to re-visit the phenomenological predictions in light of recent developments on the event generator tuning at the LHC and with the advent of NLO + PS generators for the calculation of the cross section for inclusive dijet production. This helps us assess the possible theoretical shortcomings and ideas for future experimental measurements.

\begin{figure}
\includegraphics[width = 0.45\textwidth, page = 1]{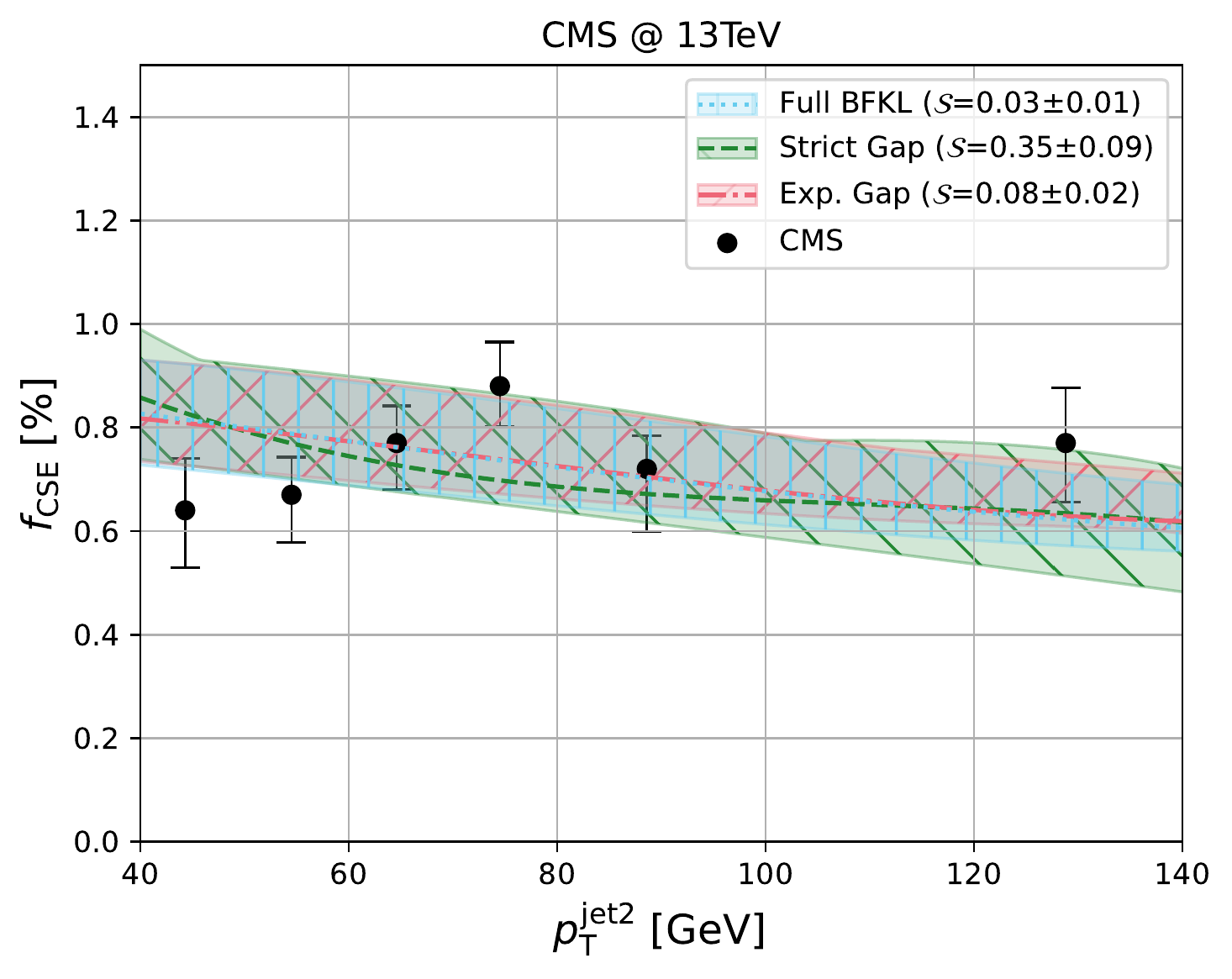}
\includegraphics[width = 0.45\textwidth, page = 2]{figures/LHC13_plots.pdf}
\includegraphics[width = 0.45\textwidth, page = 3]{figures/LHC13_plots.pdf}

\caption{Measurement of the color-singlet exchange fraction $f_\text{CSE}$ at 13 TeV by the CMS Collaboration~\cite{Sirunyan:2021oxl}. Theoretical predictions based on BFKL calculations at next-to-leading logarithmic accuracy assuming experimental gap (``exp gap''), a theoretical-like gap (``strict gap''), and no gap requirement (``full BFKL'') are presented. The bands represent uncertainties related to factorization and renormalization scale variations. The gap survival probabilities, indicated in the legend, were fit with a $\chi^2$ scan.  \label{fig:jetgapjet_predictions}}
\end{figure}

To understand these measurements in the context of BFKL dynamics, we have embedded the BFKL pomeron exchange amplitudes at NLL with LO impact factors in the PYTHIA8 event generator. We use a recent CP1 tune of PYTHIA8, which has an improved phenomenology of initial- and final-state radiation, multiple parton interactions, and hadronization for a wide range of energies and collision systems, including 13 TeV pp collisions~\cite{CMS:2019csb}. We use POWHEG+PYTHIA8 for the NLO+PS calculation of the inclusive dijet cross section using the CP5 tune of PYTHIA8. We compared our calculations to the measurements by the Tevatron and LHC experiments using the same rapidity gap selection as the experiments ($p_T > 200$ MeV in $|\eta|<1$) and with a rapidity gap definition that is closer to the theoretical expectation ($|\eta|<1$, no $p_\text{T}$ requirement). In Fig.~\ref{fig:jetgapjet_predictions}, we show a few predictions for 13 TeV together on top of the measurement by CMS. In doing these studies, we discovered that there is an important role of initial-state radiation effects in the destruction of central gaps as one goes to larger $\sqrt{s}$. We find that the description using a theoretical-like gap and the experimental gap agree with each other, modulo a global normalization factor, at 1.8 TeV and 7 TeV, but a clear disagreement is observed at 13 TeV. The theoretical gap prediction gives a better description of the data at 13 TeV.

In investigating the source behind the phenomenological differences between the gap definitions, we find that there is sensitivity in the modeling of fragmentation with the additional production of color charges with ISR in PYTHIA8. The Run-2 tunes of CMS were fit to reproduce charged particle spectra measurements in minimum-bias events split into single-diffractive (forward rapidity gap), non-diffractive, or inelastic topologies~\cite{Sirunyan:2019nog}. For a better phenomenological interpretation of the jet-gap-jet process, additional experimental input on measurements of minimum-bias events with central rapidity gap topologies, similar to the ones used for the jet-gap-jet process but without the requirement of high-$p_\text{T}$ jets, will be necessary to further tune ISR and fragmentation modelling effects. To get more insight on this aspect, future phenomenological calculations should be done by embedding the BFKL calculations in the HERWIG7 generator, which has a different evolution variable in the parton shower and a different fragmentation model.

The present BFKL calculations for the jet-gap-jet process account for the resummation of logarithms of energy at NLL accuracy using LO impact factors. The NLO impact factors were calculated in recent years~\cite{hentschinski1, hentschinski2}, but they have not been incorporated into the phenomenological analysis yet. In addition, to improve the phenomenological description, it will be important to take into account the effect of wide angle, soft gluon emissions into the gap region. These lead to so-call non-global logarithms, which are not resummed in the BFKL framework. The resummation is known exactly in the large-$N_C$ limit, and is described by the Banfi--Marchesini--Smye equation\cite{Banfi:2002hw}. The effect of these non-global logarithms for the jet-gap-jet topology is a suppression of the gluon-gluon processes relative to quark-gluon and quark-quark processes~\cite{uedahatta}.

Experimentally, future measurements could benefit from exploring different definitions of the pseudo-rapidity gap between the jets. For example, by scanning the $p_\text{T}$ threshold used to define the pseudo rapidity gap in order to better control the aforementioned ISR effects, or by defining a ``sliding'' pseudo rapidity gap interval event-by-event. To suppress the underlying event activity, one could target hadron-hadron collisions where at least one of the colliding hadrons remains intact due to Pomeron exchange. Such a measurement has been presented by CMS and TOTEM ~\cite{Sirunyan:2021oxl}, demonstrating the feasibility for such studies. However, the sample size was rather limited and did not allow for a differential measurement of the $f_\text{CSE}$. Special runs at the LHC at $\sqrt{s} = 14$ TeV with single proton-proton collisions with an integrated luminosity of $\mathcal{L} = 10$ pb$^{-1}$ would allow for a highly differential measurement in a controlled hadronic environment.

To this date, we still do not have a global and satisfactory description of jet-gap-jet events from the point of view of QCD. In principle, given the hard scale of the process, the jet-gap-jet process should be describable in terms of perturbation theory, and potentially be a venue for understanding BFKL dynamics. It is counter-intuitive that such a simple signature is more complicated to describe than the significantly ``busier'' inclusive dijet events.

According to the Tevatron and LHC measurements, about 0.5--1\% of the inclusive dijet cross section is due to $t$-channel hard color-singlet exchange. The subprocess for QCD hard color-singlet exchange is not currently implemented as a standard subprocess in modern Monte Carlo event generators. As the experimental precision increases for inclusive jet cross section measurements, the absence of $t$-channel color-singlet exchange subprocesses becomes more important, for example for PDF or $\alpha_s$ extractions. Thus, for the next years in high-energy physics, it will be important to have a proper understanding of this process, both experimentally and theoretically.

\section{Hadronic Structure at low \texorpdfstring{$x$}{x} and gluon saturation}
\label{sec:structure}

\noindent \textbf{Main Contributors:} Renaud Boussarie,   Francesco Giovanni Celiberto,  Michael Fucilla, Andreas~van Hameren, Jamal Jalilian-Marian,  Piotr Kotko, Krzysztof Kutak, Mohammed M.A. Mohammed, Alessandro Papa,   Sebastian Sapeta, Lech Szymanowski, Samuel Wallon \\

\subsection{Color Glass Condensate and high gluon densities}

Even though BFKL resummation can stabilize  collinear factorization, this approach is expected to eventually breakdown due to the resulting high gluon occupation numbers. While the gluon distribution is rapidly growing at small $x$ due to radiation of more and more small $x$ gluons due to the availability of large longitudinal phase space at high energy, at some point the number of partons (gluons) occupying the same transverse area in the target hadron or nucleus will be large and the QCD-improved parton model where nearby partons are treated as not interacting with each other will cease to be applicable to high energy collisions. This is the phenomenon of the so called gluon saturation. In the Color Glass Condensate formalism one treats
this state with a large gluon occupation number as a classical color field generated by the large $x$ color degrees of freedom generically called sources of color charge $\rho$.   In this formalism a high energy hadronic or nuclear collision is then treated as a collision of two highly contracted classical color fields, i.e. shock waves. This a highly non-trivial and so far not amenable to analytic solutions in general. A somewhat easier problem is to consider scattering of a dilute system of partons on a dense system of gluons described as a saturated state of gluons. In this so-called dilute-dense collision the relevant degrees of freedom are Wilson lines in fundamental or adjoint representation,
\begin{equation}
    \label{eq:WLCGC}
    V (x_t) \equiv \hat{P}\, \exp \, \bigg\{i g \int_{-\infty}^{+\infty} d x^+ \, A^-_a\, t_a\bigg\}
\end{equation}

re-summing multiple scatterings of a quark or gluon parton projectile on the classical color field $A^-$ describing the target dynamics in light cone gauge $A^+ = 0$. 
Production cross sections in this approach involve two and four point correlation functions of Wilson lines known as dipoles and quadrupoles (these are the only two correlation functions in leading $N_c$ approximation). Quantum loop effects are then incorporated into this formalism via a functional renormalization group equation known as the JIMWLK equation \cite{Jalilian-Marian:1996mkd,Jalilian-Marian:1997qno,Jalilian-Marian:1997jhx,Jalilian-Marian:1997ubg,Jalilian-Marian:1998tzv,
Kovner:2000pt, Weigert:2000gi, Kovner:1999bj, Iancu:2000hn,Iancu:2001ad,Ferreiro:2001qy}  which in the Gaussian and large $N_c$ approximation reduces a close equation known as the BK equation \cite{Balitsky:1995ub,Kovchegov:1999yj}. While applications of the Color Glass Condensate to high energy collisions at HERA, RHIC and the LHC have yielded tantalizing hints of gluon saturation effects there is still no firm evidence.

\subsection{Transverse Momentum Dependent Parton Distribution Functions}

Unlike collinear factorization, such an approach is based on high energy factorization, which yields cross-sections as convolutions in transverse momenta or coordinates, in contrast to convolutions in hadron momentum fraction, encountered for collinear factorization. 
In~\cite{Dominguez:2011wm} it has been shown that neglecting any higher twist correction one finds in the small $x$ description of semi-inclusive observables  the sought after distributions which contains the information on transverse momentum in hadron: the Transverse Momentum Dependent (TMD) gluon distributions. These distributions allow for 3 dimensional imaging of hadrons: one accesses one longitudinal direction and two transverse dimensions of momentum inside them. Such TMD distributions are not only of interest to characterize effects related to the presence of high gluon densities in hadron. More generally they allow to obtain a 3D imaging of the proton content and to answer  fundamental questions of the dynamics of strong interactions, such as the origin of proton mass and spin calls, see Refs.~\cite{Collins:2011zzd,Collins:1981uk} and references therein. The complete list of unpolarized and polarized gluon TMDs at leading twist (twist-2) was afforded for the first time in Ref.~\cite{Mulders:2000sh}. Note that this is list is generic and is at first not restricted to high energy factorization and/or the presence of high parton densities. They however provide useful tools to map the information contained in correlators of multiple Wilson lines Eq.~(\ref{eq:WLCGC}). Tab.~\ref{tab:gluon_TMDs} contains the eight twist-2 gluon TMDs for a spin-1/2 target, using the nomenclature defined in Refs.~\cite{Meissner:2007rx,Lorce:2013pza}.
The two functions on the diagonal in Tab.~\ref{tab:gluon_TMDs} respectively stand for the density of unpolarized gluons inside an unpolarized nucleon, $f_1^g$, and of circularly polarized gluons inside a longitudinally polarized nucleon, $g_1^g$.
They are the counterparts to the well-known unpolarized and helicity gluon PDFs obtained within collinear factorization.
According to TMD factorization, {\it i.e.} factorization in the limit where a certain transverse momentum $k_\perp$ is significantly smaller than a certain hard scale $M $, $M \gg k_\perp$, all these densities embody the resummation of transverse-momentum logarithms, which constitute their perturbative input. Much is know about this resummation~\cite{Bozzi:2003jy,Catani:2010pd,Echevarria:2015uaa}, but very little is known about the non-perturbative content of these TMD distribution. It is then this non-perturbative content (from the point of view of TMD factorization) which promises to give information on the Color Glass Condensate. \\

The distribution of linearly polarized gluons in an unpolarized hadron, $h_1^{\perp g}$, is particularly relevant at low-$x$, since it leads to spin effects in collisions of unpolarized hadrons~\cite{Boer:2010zf,Sun:2011iw,Boer:2011kf,Pisano:2013cya,Dunnen:2014eta,Lansberg:2017tlc}, whose size is expected to become more and more relevant when $x$ diminishes. The Sivers function, $f_{1T}^{\perp g}$, gives on the other hand  information about unpolarized gluons in a transversely polarized nucleon, and is relevant to study transverse-spin asymmetries emerging in collisions with polarized-proton beams; within the context of low $x$ physics it is of particular interest due to its connection  with the QCD Odderon~\cite{Boussarie:2019vmk}.

{
\renewcommand{\arraystretch}{1.7}

 \begin{table}
\centering
 \hspace{1cm} gluon pol. \\ \vspace{0.1cm}
 \rotatebox{90}{\hspace{-1cm} nucleon pol.} \hspace{0.1cm}
 \begin{tabular}[c]{|m{0.5cm}|c|c|c|}
 \hline
 & $U$ & circular & linear \\
 \hline
 $U$ & $f_{1}^{g}$ & & \textcolor{blue}{$h_{1}^{\perp g}$} \\
 \hline	
 $L$ & & $g_{1}^{g}$ & \textcolor{red}{$h_{1L}^{\perp g}$} \\
 \hline	
 $T$ & \textcolor{red}{$f_{1T}^{\perp g}$} & \textcolor{blue}{$g_{1T}^{g}$} & \textcolor{red}{$h_{1}^{g}$}, \textcolor{red}{$h_{1T}^{\perp g}$} \\
 \hline
  \end{tabular}
 \caption{A table of leading-twist gluon TMDs for spin-$1/2$ targets. 
 $U$, $L$, $T$ stand for unpolarized, longitudinally polarized and transversely polarized hadrons, whereas
 $U$, `circular', `linear' depict unpolarized, circularly polarized and linearly polarized gluons, respectively. 
 $T$-even (odd) functions are given in blue (red). 
 Black functions are $T$-even and survive the integration over the gluon transverse momentum.}
 \label{tab:gluon_TMDs}
 \end{table}

}



First attempts at phenomenological analyses of $f_1^g$ were done in Refs.~\cite{Lansberg:2017dzg,Gutierrez-Reyes:2019rug,Scarpa:2019fol}. The phenomenology of $f_{1T}^{\perp g}$ was discussed in Refs.~\cite{Adolph:2017pgv, DAlesio:2017rzj,DAlesio:2018rnv,DAlesio:2019qpk}.
Due to the shortage of experimental data on the gluon-TMD sector, exploratory analyses of gluon TMDs via simple and flexible models are required. Pioneering analyses along this direction were conducted by the hands of the so-called \emph{spectator framework}~\cite{Lu:2016vqu,Mulders:2000sh,Pereira-Resina-Rodrigues:2001eda}.
Originally employed to model quark TMD distributions~\cite{Bacchetta:2008af,Bacchetta:2010si,Gamberg:2005ip,Gamberg:2007wm,Jakob:1997wg,Meissner:2007rx}, it relies on the assumption that from the struck hadron a gluon is extracted, and what remains is considered as an effective on-shell spin-1/2 object.
Spectator-model $T$-even gluon TMDs at twist-2 were recently obtained in Ref.~\cite{Bacchetta:2020vty} (see also Refs.~\cite{Celiberto:2021zww,Bacchetta:2021oht,Celiberto:2022fam}), while a preliminary calculation of the $T$-odd ones can be found in Refs.~\cite{Bacchetta:2021lvw,Bacchetta:2021twk,Bacchetta:2022esb}. The $T$-even gluon correlator is taken at tree level and does not account for the gauge-line dependence, which appears in our model in the $T$-odd case.


For an unpolarized proton, we identify the unpolarized distribution 
\begin{equation}
 x \rho (x, p_x, p_y) = x f_1^g (x, \boldsymbol{p}_T^2) 
\label{eq:HAS_HE-QCD_TMD_unpol}
\end{equation}
as the probability of extracting unpolarized gluons at given $x$ and $\boldsymbol{p}_T$, while the Boer--Mulder density 
\begin{equation}
 x \rho^{\leftrightarrow} (x, p_x, p_y) = \frac{1}{2} \bigg[ x f_1^g (x, \boldsymbol{p}_T^2) + \frac{p_x^2 - p_y^2}{2 M^2} \, x h_1^{\perp g} (x, \boldsymbol{p}_T^2) \bigg]
\label{eq:HAS_HE-QCD_TMD_BM}
\end{equation}
represents the probability of extracting linearly-polarized gluons in the transverse plane at $x$ and $\boldsymbol{p}_T$.

\begin{figure}[tb]
 \centering
 \includegraphics[scale=0.24,clip]{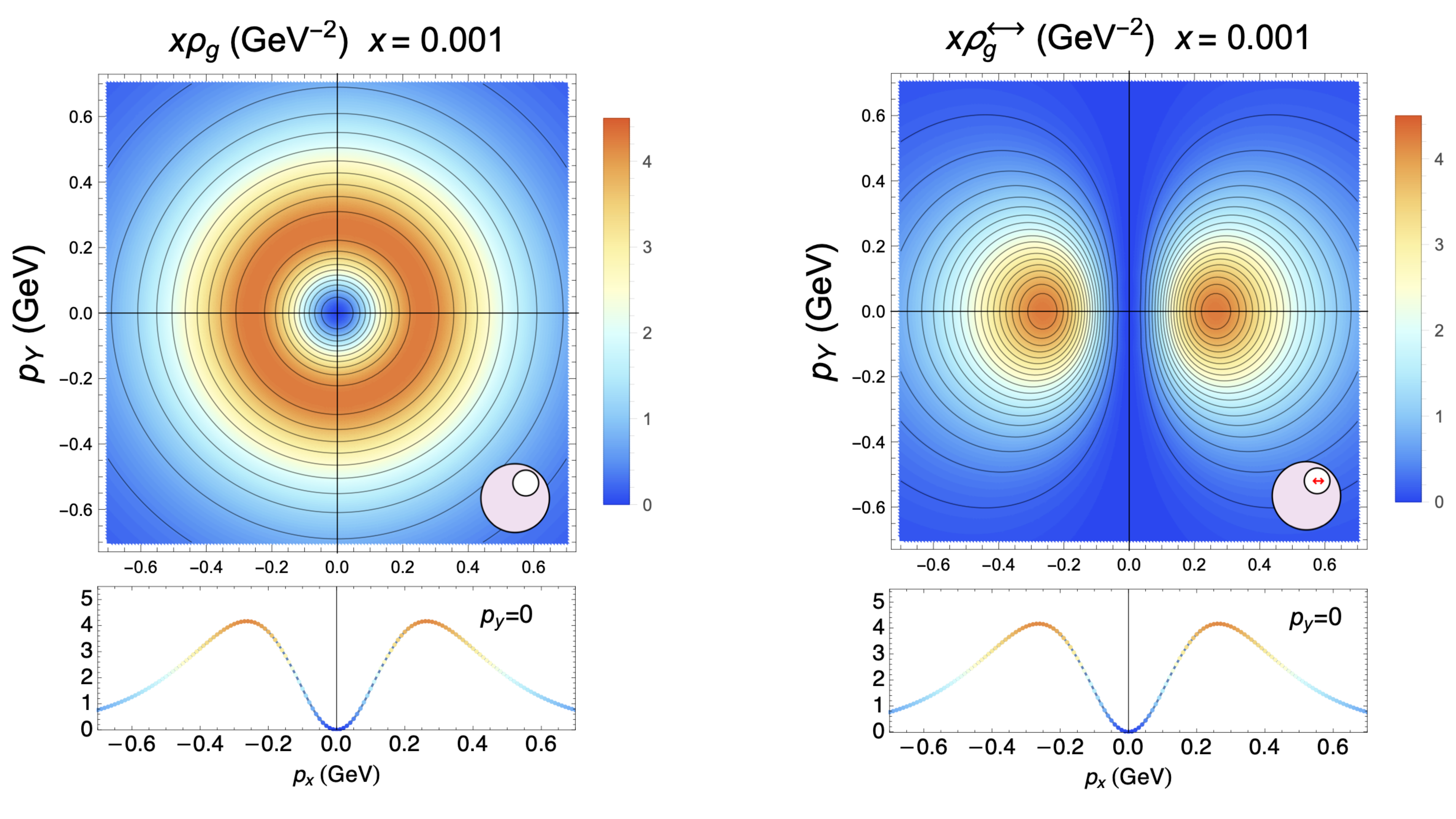}

 \caption{
 3D tomographic imaging of the proton  unpolarized (left) and Boer--Mulders (right) gluon TMD densities as functions of the gluon transverse momentum, for $x = 10^{-3}$ and at the initial energy scale, $Q_0 = 1.64$ GeV. 1D ancillary panels below main contour plots show the density at $p_y = 0$.
 Figures from Ref.~\cite{Bacchetta:2020vty}.
 }
 \label{fig:HAS_HE-QCD_gluon_TMDs}
\end{figure}

Contour plots in Fig.~\ref{fig:HAS_HE-QCD_gluon_TMDs} refer to the behavior in $\boldsymbol{p}_T$ of the $\rho$-densities in Eqs.~(\ref{eq:HAS_HE-QCD_TMD_unpol}) and~(\ref{eq:HAS_HE-QCD_TMD_BM}), respectively, obtained at $Q_0 = 1.64$ GeV and $x=10^{-3}$ for an unpolarized proton virtually moving towards the reader. The color code is related to the size of the oscillation of each density along the $p_x$ and $p_y$ directions. To better catch these oscillations, ancillary 1D plots representing the corresponding density at $p_y = 0$ are shown below each contour plot. As expected, the density of Eq.~(\ref{eq:HAS_HE-QCD_TMD_unpol}) has a cylindrical pattern around the direction of motion of the proton. Conversely, the Boer--Mulders $\rho$-density in Eq.~(\ref{eq:HAS_HE-QCD_TMD_BM}) presents a dipolar structure which reflects the fact that gluons are linearly polarized. The running away from the cylindrical symmetry is emphasized at small $x$, because the Boer--Mulders function is particularly large. 
From the analytic point of view, the ratio between $f_1^g$ and $h_1^{\perp g}$ TMDs turns to a constant in the asymptotic limit $x \to 0^+$.
This is in line with the prediction coming from the linear BFKL evolution, \emph{i.e.} that at low-$x$ the ``number" of unpolarized gluons equals the number of linearly-polarized ones, up to higher-twist effects (see, \emph{e.g.}, Refs.~\cite{Dominguez:2011br,Marquet:2016cgx,Taels:2017shj,Marquet:2017xwy,Petreska:2018cbf}). 
Therefore, a connection point between our gluon TMDs and the high-energy QCD dynamics has been established.
\\

effects~\cite{Fujii:2020bkl}.

To access  more dimensions of partonic content, exclusive processes are necessary.  The richest distributions one encounters in perturbative QCD processes are the so-called Generalized TMD distributions (GTMD). They parameterize master correlators of 5 parameters $(x,\xi,\vec{k}_\perp^2,\vec{\Delta}_\perp^2,\vec{k}_\perp\cdot\vec{\Delta}_\perp)$ where $x,\vec{k}_\perp$ are respectively the fraction of the longitudinal target momentum and the transverse momentum carried by a parton, and $\xi$ and $\vec{\Delta}_\perp$ are the longitudinal and transverse momentum transfer to the hadron. For $(\xi=0,\vec{\Delta}_\perp=\vec{0}_\perp)$ the GTMD becomes a TMD. In a given process, the set of GTMD distributions which are accessed depends on final state kinematics, but also on the parton and hadron polarizations with repercussions on the measured final state. Multidimensional tomography of hadrons as a probe for Wigner distributions and TMD distributions has been a major focus of theoretical and experimental efforts, and it will be an important share of the physics goals of future colliders such as the Electron Ion Collider~\cite{Hatta:2016dxp}.
 Examples of relevant processes -- which also allow at least in principle for an analysis at next-to-leading order in the perturbative expansion --  are 
 Meson production in $\gamma^{(\ast)}p$ collisions~\cite{Boussarie:2016bkq}, inclusive and diffractive  dijets ~\cite{Boussarie:2021ybe, Caucal:2021ent, Hatta:2016dxp,Boussarie:2014lxa,Boussarie:2016ogo,Hagiwara:2017fye}, see also~\cite{Boussarie:2019ero} for a first phenomenological application to HERA data, as well as  exclusive pion production in {\it unpolarized} electron-proton scattering in the forward region, which is a direct probe of both the gluon Sivers function and the QCD Odderon~\cite{Boussarie:2019vmk}. \\

\subsection{Complementarity between EIC and LHC}

There exists a very nice complementarity between EIC and LHC experiments. Indeed it is possible  to gather important input on such (G)TMD distributions from photon-hadron reactions, which at the LHC is accessible through  ultraperipheral proton-nucleus and nucleus-nucleus collisions, see  Sec.~\ref{sec:upc} for a detailed discussion. A photo production processes which provides the necessary hard scale for an analysis based on perturbative QCD expansion is  the production of quarkonia, in particular diffractive exclusive photoproduction of $J/\psi$ charmonia with negative charge parity which allows to study the distribution of gluons in the target. On the other hand, the same type of production of charmonia with positive charge  parity as $\eta_c$ or $\chi_c$ permits studying gluonic exchanges with negative charge parity corresponding to Odderon exchanges.   A related process is 
exclusive diffractive meson production at large momentum transfer $|t|$, which could provide a new set of observables to reveal saturation effects. While  HERA measurements are limited by  statistics,  UPC reactions at LHC might allow for the observation of a different scaling in $t$, when passing from a high energy kinematics governed by BFKL-like descriptions to extreme rapidities in which saturation effects are expected. 
Instead of using $t$ as a hard scale, it is also possible to study diffractive states which contain at least one meson carrying a large $p_\perp,$ originating from collinear fragmentation of the virtual $q \bar{q}$ pair produced by the photon. The large value of $p_\perp$ ensures that each of the $q$ and $\bar{q}$ carries an (opposite) large $q_\perp$ through the usual ordering of collinear fragmentation.  Relying on the recent results obtained for impact factors $\gamma \to q \bar{q} g$ at leading order and $\gamma \to q \bar{q}$ at next-to-leading order, which were used for the computation of the $\gamma \to dijet$ impact factor at NLO,  a complete NLO study could be done in the future. Most probably, the pion channel which is the best known for fragmentation functions, would be the most promising. \\

Last but not least it is worth to mention  large invariant mass systems: In the spirit of time-like Compton scattering, in which the hard scale is provided by the virtuality of the emitted virtual photon, exclusive production  at JLab of a $\gamma$-meson pair of large invariant mass is a very promising process to access to Generalized   Parton Distributions (GPD)~\cite{Boussarie:2016qop,Duplancic:2018bum},  another particular facet of the master correlators obtained by integrating out their $\vec{k}_\perp$ dependency. It turns out that the same process in the large $s$ limit belongs to the class of diffractive processes, therefore furnishing another probe for studying gluonic saturation. Thus, by covering a very wide kinematical range, the same process studied at JLab, EIC (in photoproduction), and LHC (in UPC), would allow to pass from a description based on collinear factorization involving a nucleon GPD to a high energy description in which linear and non-linear resummation effects are expected.

\subsection{Forward dijets: from LHC to EIC}

Dijets produced in the forward direction of the detectors are characterized by  final  states at large rapidities and hence they
trigger events in which the partons from the nucleus carry small
longitudinal momentum fraction $x$.  This kinematic setup is well suited to
investigate the properties of dense partonic system and the phenomenon
of saturation. The following study is based on the so-called small-$x$  Improved Transverse Momentum Dependent 
factorization framework~\cite{Kotko:2015ura,Altinoluk:2019fui,Bury:2018kvg,Bury:2020ndc, Boussarie:2020vzf,Altinoluk:2021ygv} which accounts for  
 exact kinematics of the scattering process with off-shell initial-state
  gluons, 
gauge invariant formulation of the TMD gluon densities as well as  off-shell partonic amplitudes, and complete  kinematic twists, while neglecting genuine twists. The framework therefore covers both 
  $k_T$-factorization~\cite{Catani:1990eg,Deak:2009xt} in the limit of large
  off-shellness of the initial-state gluon from the nucleus
  small-$x$ TMD factorization \cite{Dominguez:2011wm} in the limit where momenta of the final-state jets are much larger then the momentum of
  the incoming off-shell gluon. Furthermore~\cite{Fujii:2020bkl,Boussarie:2021ybe} demonstrate a very good agreement of this approach 
with the full CGC result  in the region dominated by hard jets i.e.\ $k_T,p_T\,>\,Q_s$.

While the original Improved TMD framework includes gluon saturation
effects, it does not account for the complete set of  contributions proportional to the logarithms of
the hard scale set by the  large transverse momenta of jets -- the so-called
Sudakov logarithms. As shown in Refs.~\cite{vanHameren:2014ala,vanHameren:2015uia}, inclusion of the
Sudakov logarithms is necessary in order to describe the LHC jet data in the
region of small $x$. In the low-$x$
domain, the resummation leading to the Sudakov logarithms has been developed in
Refs.~\cite{Mueller:2013wwa,Mueller:2012uf}, see also \cite{Kutak:2014wga}. In
Ref.~\cite{vanHameren:2019ysa}, it was shown for the first time that
the interplay between the saturation effects and the resummation of the Sudakov
logarithms is essential to describe the small-$x$ forward-forward dijet data.

In this section we present two results that demonstrate relevance 
of both effects, i.e.\ nonlinearity, accounting for saturation, and
the Sudakov effects, accounting for emissions of soft gluons. 
We shall consider two processes:
\begin{itemize}
\item  the inclusive dijet production
\begin{equation}
  \mathrm{p} \left(P_{\mathrm{p}}\right) + \mathrm{A} \left(P_{\mathrm{A}}\right) \to j_1 (p_1) + j_2 (p_2)+ X\ ,
\end{equation}
\item the dijet production in deep inelastic scattering
\begin{equation}
  \mathrm{e} \left(P_{e}\right) + \mathrm{A} \left(P_{\mathrm{A}}\right) \to \mathrm{e} \left(p_{e}\right) + j_1 (p_1) + j_2 (p_2)+ X\ ,
\end{equation}
\end{itemize}
where $A$ can be either the lead nucleus or a proton. 

To describe the former process, we use a hybrid approach where one
assumes that the proton $p$ is a dilute projectile, whose partons are collinear
to the beam and carry momenta $p=x_{\mathrm{p}} P_{\mathrm{p}}$.  The hadron $A$
is probed at a dense state.  The jets $j_1$ and $j_2$ originate from the hard
partons produced in a collision of the probe $a$ with a gluon belonging to the
dense system $A$. This gluon is off-shell, with momentum $k=x_{\mathrm{A}}
P_{\mathrm{A}} + k_T$ and $k^2=-|\vec{k}_T|^2$.  The ITMD factorization formula for the above process can be found in Ref.~\cite{Kotko:2015ura},
while the formula for the $e\,\textrm{-}A$ collision can be found in Ref.~\cite{vanHameren:2021sqc} 

As it has been argued above, in order to provide realistic cross section
predictions one needs to include also the Sudakov effects. For dijet production
at the LHC, we used a DGLAP based Sudakov form factor~\cite{vanHameren:2019ysa}.

\begin{figure}[t]
  \begin{center}
    \includegraphics[width=0.99\textwidth]{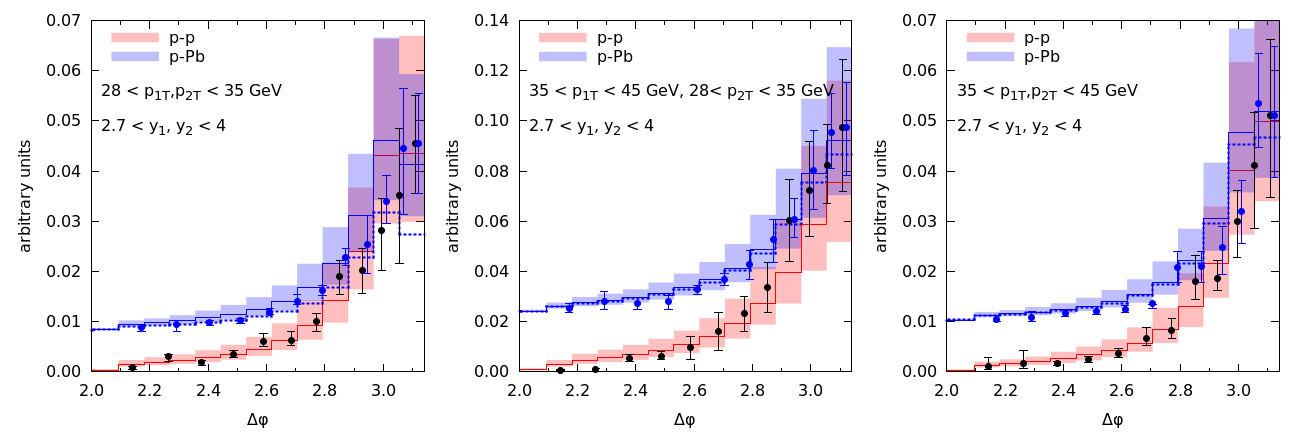}
  \end{center}
  \caption{
    Broadening of  azimuthal decorrelations in $p\,\textrm{-}Pb$ collisions vs $p\,\textrm{-}p$
    collisions for different sets of cuts imposed on the jets' transverse
    momenta.  The plots show  normalized cross sections as functions of the
    azimuthal distance between the two leading jets, $\Delta\phi$.  The points
    show the experimental data \cite{ATLAS:2019jgo} for $p\,\textrm{-}p$ and $p\,\textrm{-}Pb$, where
    the $p\,\textrm{-}Pb$ data were shifted by a pedestal, so that the values in the bin
    $\Delta\phi\sim \pi$ are the same.  Theoretical calculations are represented
    by the histograms with uncertainty bands coming from varying the scale by
    factors 1/2 and 2.
  }
  \label{fig:broadening}
\end{figure}

\begin{figure}[t]
  \begin{center}
   \includegraphics[width=0.4\textwidth]{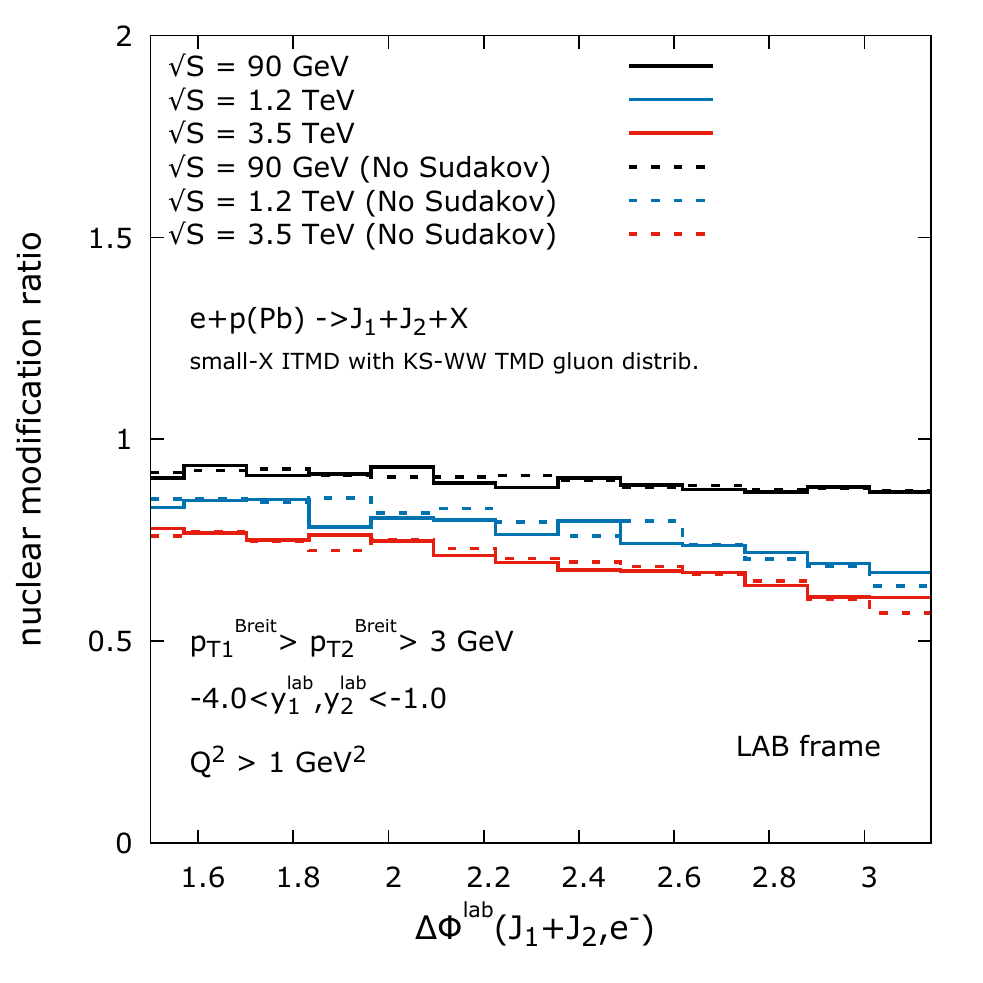}
   \includegraphics[width=0.4\textwidth]{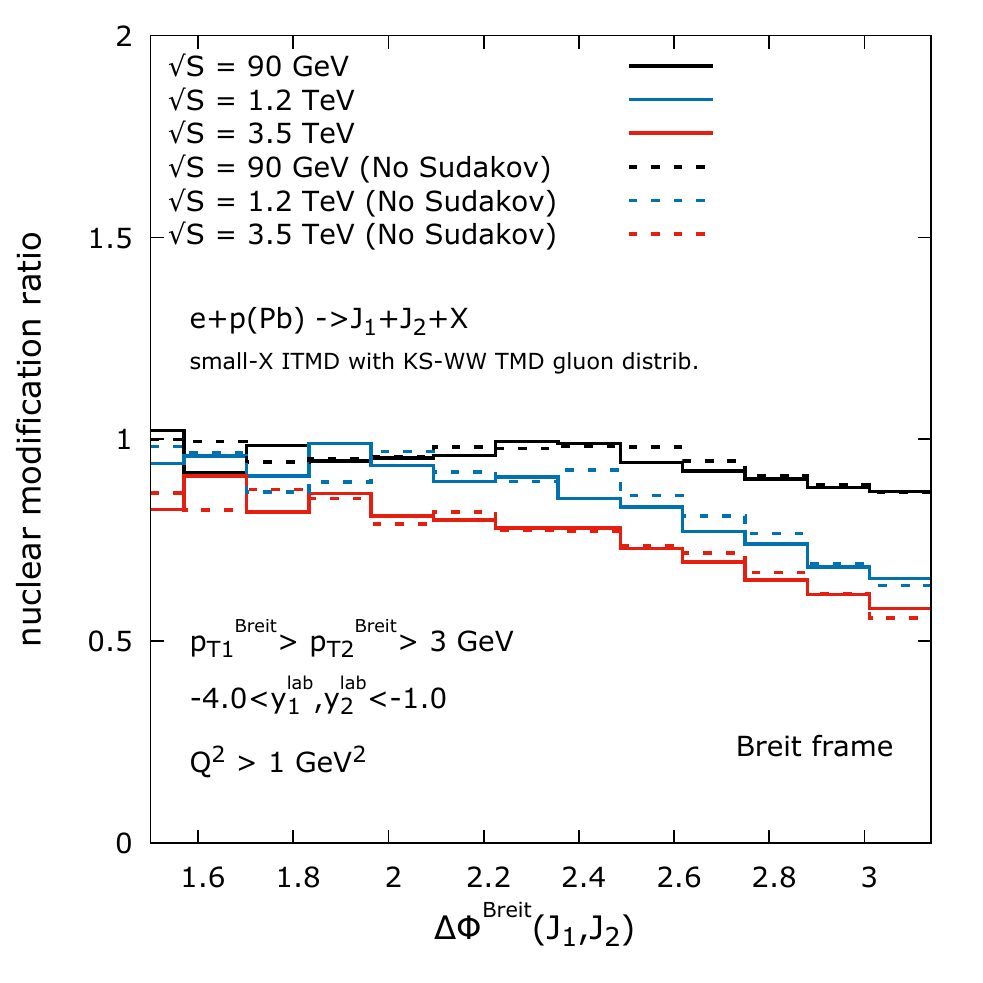}
   \hspace{30pt}
  \end{center}
  \caption{
  Nuclear modification factor $R_{pA}$ as a function of the azimuthal angle between the jet system and the scattered electron in LAB frame~(left) and as a function of the azimuthal angle between the jets in the Breit frame~(right). The calculations are done for three different CM energies per nucleon.
  Solid lines correspond to calculations using Sudakov form factor and KS-based
  WW gluon density.
  }
  \label{fig:RpA}
\end{figure}

In Fig.~\ref{fig:broadening}  we show normalized cross sections as functions of
$\Delta\phi$ in $p\,\textrm{-}p$  and $p\textrm{-}Pb$ collisions.  The three panels correspond to
three different cuts on the transverse momenta of the two leading jets.  The
points with error bars represent experimental data from
Ref.~\cite{ATLAS:2019jgo}.  The main results for the $p\,\textrm{-}Pb$ collisions  are
represented by the blue solid lines.
The broadening of the distributions as we go from $p\,\textrm{-}p$ to $p\,\textrm{-}Pb$ comes from the
interplay of the non-linear evolution of the initial state and the Sudakov
resummation.  

In the Fig.~\ref{fig:RpA} we show 
predictions for nuclear modification in DIS process as one increases the energy of the collision.
In particular, we present the results for three values of the energy that
are relevant for the EIC, LHeC and FCCeh. 
We see that for larger energies, the suppression due to saturation is larger, and that the Sudakov form factor cancels too a large degree.
This results demonstrates that having results for the absolute
cross sections \cite{vanHameren:2021sqc} and nuclear modification ratio, one can in principle isolate
effects due to saturation from the effects coming from to Sudakov resummation.\\

\subsection{Color Glass Condensate beyond high energy factorization}

\begin{figure}
    \centering
    \parbox{0.8\textwidth}{\includegraphics[width=0.8\textwidth]{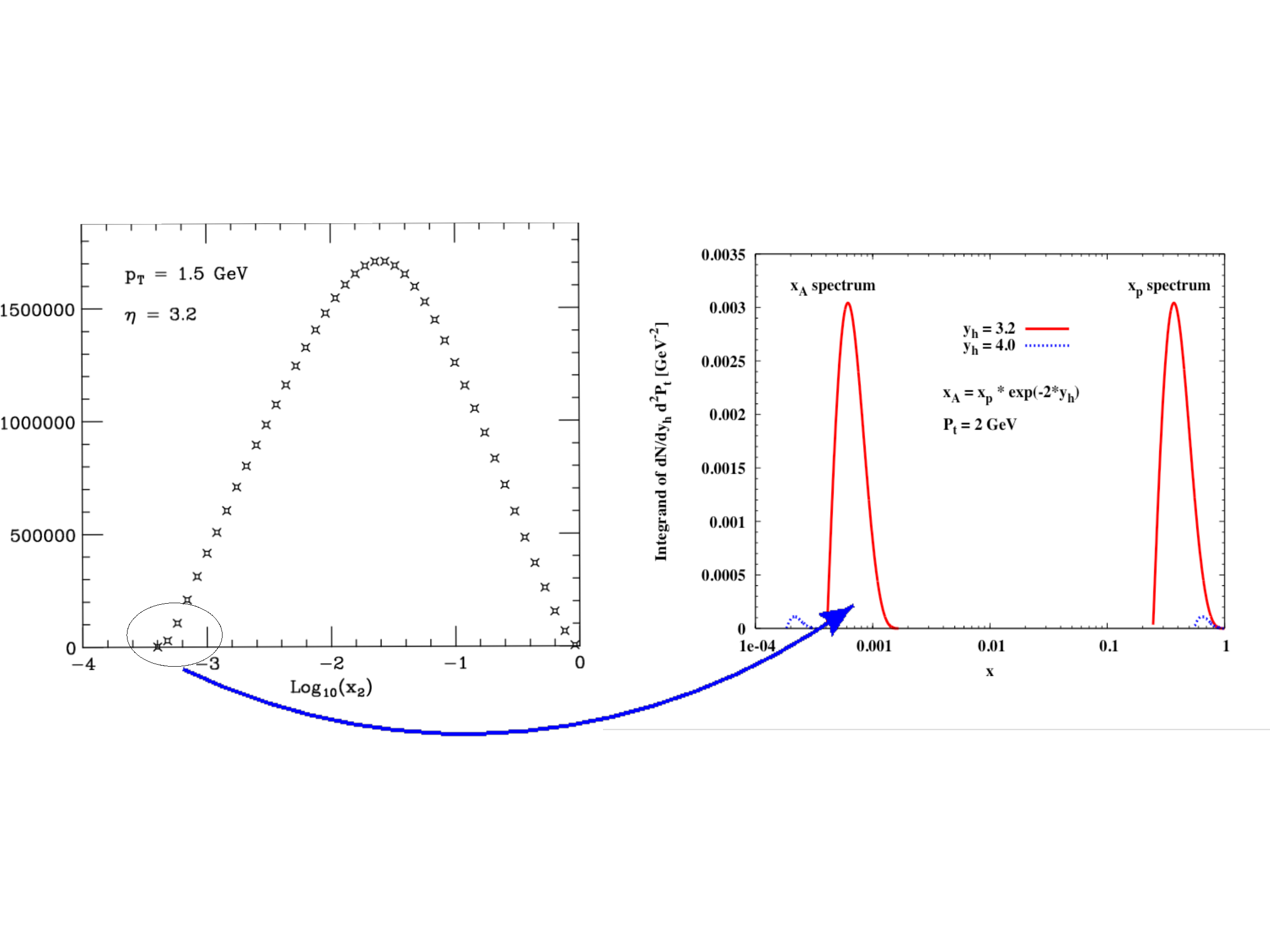}}
    \caption{A comparison of of target $x$ range contributing to single inclusive pion production in proton-proton collisions at RHIC. Left panel  is from Ref.~\cite{Guzey:2004zp}) showing the target $x$-range as predicted by the collinear factorization formalism while the right panel is the $x$-range obtained from the Color Glass Condensate formalism, taken from  Ref.~\cite{Dumitru:2005gt}}
    \label{fig:kinematics-comparison}
\end{figure}

Theory predictions for Color Glass Condensate correlators are genuinely based on high energy factorization which can be recast as the first order of an expansion in parton momentum fraction $x$. However, in particular for EIC kinematics but also if descriptions are to be extended to central rapidities,  there is the danger that one is leaving the regime of applicability of this expansion. This is most easily seen in the kinematics relation
\be
x_{1,2} =  \frac{p_\perp}{ \sqrt{s}}\, e^{\pm y}
\ee
where $x_{1,2}$ are the projectile and target momentum fraction probed in the collision while $p_\perp$ and $y$ are the transverse momentum and rapidity of the produced particle. As is seen as one looks at higher and higher transverse momenta of produced particles one is probing larger and larger values of the target $x$ which is bound to eventually become too large for CGC to be a valid description of the target.  
Furthermore it is essential to realize that the multiple scatterings of the projectile on the target which is needed due to high gluon density of the target are treated in the eikonal, i.e. recoil-less approximation which can accommodate only a small angle deflection of the projectile. This eikonal approximation to multiple scatterings is the reason one can elegantly re-sum them into a Wilson line. As one considers  particle production at higher $p_\perp$ this recoil-less approximation breaks down and partons are scattered at a large angle. This is not included in the CGC formalism.  In Fig. (\ref{fig:kinematics-comparison}) we show the difference in the target $x$ involved in pion production in proton-proton collisions in the forward rapidity region at RHIC. 
Both collinear factorization and CGC approaches fit the data well and yet the physical cross section is dominated by very different target $x$'s in the two approaches so clearly there is a discrepancy. Both approaches can not be correct at the same time so at least one or both of the two must be missing some crucial physics. This demonstrates the need for a more general formalism that include collinear factorization at high $p_\perp$ and the Color Glass Condensate formalism at small $x$. \\

In~\cite{Jalilian-Marian:2019kaf,Jalilian-Marian:2018iui,Jalilian-Marian:2017ttv} a new approach to particle production at high energies was proposed which aims to accomplish this, include large $x$ (high $p_\perp$) physics as encoded in collinear factorization and gluon saturation effects at small $x$ as described by JIMWLK equation. A related, but different approach \cite{Hentschinski:2017ayz, Gituliar:2015agu} aims at first at a unification of collinear factorization and low density high energy factorization. All these approaches imply the necessity to go beyond the eikonal approximation and allow a large angle deflection of the projectile parton. This can not happen if one considers scattering from small $x$ modes of the target as is the case in CGC. Therefore one must include scattering from the large $x$ modes of the target.  This is currently work in progress and will be reported elsewhere. 


\section{Imprints of high gluon densities at low x at the LHC.}
\label{sec:imprints}

\noindent \textbf{Main Contributors:} Georgios K. Krintiras, Constantin Loizides, Mark Strikman

\subsection{Open QCD questions at a hadron-hadron collider: Parton fragmentation, mini-jets and their interplay with  high parton densities}
\label{sec:minijets}



Apart from the study of specific observables which allow to obtain information on different TMD distribution function, there exists also more global features of QCD phenomenology at a hadron-hadron collider, which allow to study and quantify the manifestation of high and potentially saturated parton distributions. 

\subsubsection{Evidence for nonlinear QCD dynamics in the fragmentation region in pA scattering}

A first region where such evidence for QCD non-linear dynamics could be found is the 
the fragmentation region of proton nucleus scattering. In a  collision of two nucleons or nuclei, a parton with given hadron momentum fraction $x_1$ resolves partons in
another nucleon down to momentum fractions of the order
\begin{equation} 
x_2=4 {p_\perp}^2/(x_1 s),
\end{equation}
with ${p_\perp}$ the transverse momentum of the parton and $\sqrt{s}$ the center of mass energy  of the collider. At the LHC, for $x_1=0.3$ and $|{ p_\perp}|$=2 GeV/c, and $\sqrt{s} = 14$~TeV in proton-proton collisions, one is therefore sensitive down to values of approximate $x_2\sim  3\cdot 10^{-7} $. Even smaller values of $x$, down to $x\sim 10^{-9}$,  are resolved in the collisions with cosmic rays. At such low values of $x$, the secondary hadron is in generally characterized by the presence of high parton -- in particular gluon -- densities. A parton  propagating through  such a dense medium of small $x \ge x_2$  partons  acquires a significant transverse momentum,  due to interaction with the dense gluonic field and looses a finite fraction of its momentum \cite{Frankfurt:2007rn}; this is in particular  the case for central proton nucleus collisions, where parton densities are further subject to nuclear enhancement\footnote{Note, that there is a moderate  gain in the  gluon density per unit area in such proton nucleus collisions and hence in the average value of gained transverse momentum, in comparison to scattering on a proton. Nevertheless, fluctuations of the gluon density are much smaller in case of nuclei.}\\

One consequence of this is a strong suppression of the leading particle spectrum as  compared to minimal bias events: Each parton fragments independently  and splits into a couple of partons with comparable energies. This suppression  is especially pronounced  for the production of  nucleons:  for  values of Feynman $x_F \sim x_1$ above $0.1$  the differential multiplicity of pions should exceed that of nucleons, see \cite{Dumitru:2002wd} for a study which however does not include  an additional suppression due to finite fractional energy losses.   Suppression of  forward pion production was observed at RHIC in  deuteron - gold collisions  at $x_F \ge 0.3$,\cite{BRAHMS:2004xry,STAR:2006dgg}, see also the discussion in \cite{Guzey:2004zp}.  It is important to note that in this kinematic regime, perturbative QCD works pretty well. Indeed  essential values of  $x_2$ are in these reactions of the order of
$\sim 0.01$  \cite{Guzey:2004zp}, see Fig.~\ref{fig:kinematics-comparison} left-hand side, which is still far from the phase space region  where gluon densities are so high such that nonlinear effects may become important in the evolution of  nuclear parton distribution functions. Note that in the CGC scenario central collisions dominate and one has to assume the existence of a mechanism for  a very strong suppression of the scattering in the DGLAP $x_2 \ge -.01$ kinematics.
Nevertheless, the observed suppression  of the inclusive per nucleon  $d Au \to \pi^0 +X$ cross section as compared  to the   inclusive $pp \to \pi^0 +X$  cross section, $R_{dAu}$,
\begin{equation} R_{dAu}(x_F=0.5,p_\perp=2 \mbox{GeV/c})\sim 0.5,
\end{equation} 
is in a gross contradiction with the naive perturbative QCD prediction of $R_{dAu}(pQCD)=1.0$.
Moreover the analysis of the STAR data on the multiplicity of hadrons produced in the events with a forward $\pi^0$ trigger indicates that the dominant contribution to the pion yield originates from peripheral collisions. This  suggests that for central collisions  the actual suppression is much larger - about a factor of the order five.\\

The observed pattern is consistent with the scenario of an  effective fractional energy loss, which leads to a large suppression:  The $\pi$ inclusive cross section strongly drops with increasing  $x_F$.   Leaving  details aside,  the observed effect is  strong evidence for a  break down of the perturbative QCD approximation.  It is natural to suspect that  this is due to effects of strong small $x$ gluon fields in nuclei, as  the forward kinematics is sensitive to small $x$ effects.   
Overall the generic features expected in all  models in which interaction strength is 
comparable with a black disk limit are:
\begin{itemize}
    \item[i)] Strong suppression of the  large $x_F$ spectra at moderate values of $p_\perp$
    \item[ii)] Broadening of the  transverse momentum distributions of leading hadrons  at large $x_F$. 
\end{itemize}
Both effects should become more and more pronounced with increasing  collision energy and centrality of collision  and/or and increase of the number of nucleons  $A$. They  should be studied as a function of $A$ and centrality. Hence one may expect much stronger suppression  of pion spectrum at $x_F \gtrsim 0.4$ and a stronger  $p_t$ broadening  at the LHC as compared to RHIC. Note  that these effects should be much more mild for  central proton-nucleus collisions, due to weak dependence of the cross section on $x_F$. This is in line with  the observation of the ALICE experiment \cite{ALICE:2014xsp}, which measured the pion yield as a function of $p_\perp$ at central rapidities at the LHC  and found only a small enhancement of the yield  for the central collisions. Note that the rapidity interval between the pion and the  initial nucleon in this case similar to that of the RHIC experiments. This is consistent with the scenario that  at central rapidities the  only significant effect is $p_\perp$ broadening due to elastic re-scatterings, which leads to  an enhancement of the cross section rather than to its  suppression. It would be informative to measure a recoil gluon mini-jet from the underlying quark - gluon collision  to study the effect of suppression as a function of $x_g$ for fixed pion $x_F$. 
\\

Further exploration of these effects in $pp$  scattering  would be possible  by studying production of the leading mesons with a centrality trigger - like dijet production at  at central rapidities. Also 
it would be instructive to study these effects in the ultra-peripheral collisions at the LHC since in this case one can  reach  center of mass energies for the photon-nucleus reaction, which are  comparable to center of mass energies for nucleus-nucleus scattering at RHIC. \\

\subsubsection{Mini-jet dynamics at collider energies}

The leading order cross section for scattering of two partons grows within perturbative QCD rapidly 
with decreasing momentum transfer between the scattering partons.  Combining this effect with the growth of gluon densities at small $x$, one obtains an inclusive mini jet cross section which exceeds even the inelastic proton-proton cross section. This suggests that  the average mini jet multiplicity may exceed one. To tame this mini-jet cross section, Monte Carlo models either introduce a hard  (no collisions  with $p_\perp < p_0(s)$ or a soft cutoff (smoothly switching off interactions with $p_\perp$'s below a few GeV/c. Interestingly,  models seem to suggest that the suppression factor grows rather rapidly  with collision energy. Clarifying the precise  mechanism for the energy dependence of the suppression of mini-jets in both in proton-proton and proton-nucleus scattering is one of the current challenges of  high energy QCD. Note that while gluon saturation  generates such an effect for the small $x$ region, it does not explain a similar suppression for peripheral proton-nucleus and proton-proton collisions. It does also not explain suppression for $x\sim 10^{-2} \div 10^{-3}$ far from the black disk  limit~\cite{Rogers:2008ua}. A comparison of transverse momentum  distributions of hadrons produced in the  very forward region and  central rapidities would certainly help for a better understanding of this suppression mechanism.
\\

 A direct observation of mini-jets is pretty difficult, since transverse momenta of hadrons generated in the fragmentation of mini-jets   may be rather close  to the soft scale. The hadron density close to the fragmentation region of protons is however much smaller than for central rapidities, which should make extraction of the mini-jet signal easier.  One possible strategy is to select a hadron (a mini-jet) at $y \sim 2\div 4 $ or even higher with  fixed $p_\perp$ and measure  the average  transverse momentum of  hadrons produced at negative rapidities. 
A distinctive feature of this mechanism is the presence of transverse correlations between hadrons only at small rapidity intervals, $\Delta y \le 2$, which follow a Gaussian distribution in $\Delta y$ in contrast to the power law suppression of correlations  in the hard mechanism \cite{Azarkin:2018cmr}.  Another suggestion is  to use proton - nucleus scattering to distinguish the production of two pairs of mini-jets  in a two parton collisions ($2\to 4$ mechanism)  from production of  four mini-jets in two binary collisions  $4\to 4$ mechanism)\cite{Alvioli:2019kcy}. The procedure is based on the centrality dependence of two mechanisms -- the production rate in the  $2\to 4$ mechanism grows linearly with the nuclear thickness, while in the $4\to 4$ mechanism it is quadratic in the thickness \cite{Alvioli:2019kcy}. 
\\


\subsection{Forward direct photon measurements} 
\label{sec:focal}

Prompt photons provide a direct access to the parton kinematics, since they couple to quarks, and unlike hadrons are not affected by final state effects.
At leading order~(LO), the photon is produced directly at the parton interaction vertex without fragmentation.
At LHC collision energies, the quark-gluon Compton cross section is significantly larger than quark-anti-quark annihilation.
At next-to-leading order~(NLO) or higher order, photons may also be produced by bremsstrahlung or fragmentation of one of the outgoing partons. 
However, fragmentation photons are accompanied by hadronic fragmentation products and the contribution of this process can be largely suppressed by application of isolation cuts.
Isolation cuts ensure that the remaining particle production process is dominantly from Compton scattering, where the measured photon is directly sensitive to the gluon PDF~\cite{dEnterria:2012kvo}.

In this paragraph, we will focus on direct photon measurements at forward rapidities, as enabled by the LHCb experiment~\cite{Alves:2008zz,Aaij:2014jba} and the planned forward calorimeter~(FoCal) upgrade of ALICE~\cite{ALICE:2020mso}.
LHCb is a single-arm spectrometer equipped with tracking and particle-identification detectors as well as calorimeters with a forward angular coverage of about $2<\eta<5$.
The FoCal is a calorimeter at $3.4<\eta<5.8$ consisting out of a high-granularity, compact silicon-tungsten (Si+W) sampling electromagnetic part with longitudinal segmentation and a conventional high granularity metal/scintillating hadronic part providing good hadronic resolution and compensation.
\\

In fact, measurements of isolated photon spectra at forward rapidities at the LHC, are sensitive to the gluon density at small Bjorken-$x$ of up to about $x=10^{-5}$ 
over a large range of momentum transfer, $Q^2$. 
By comparing measurements in \pPb\ and \pp\ collisions one can hence extract the gluon nuclear modification at small $x$ and $Q^2$.
The parton structure of protons and nuclei is described by momentum distributions at an initial momentum scale, and the scale dependence of the structure can be calculated with linear QCD evolution equations, such as the DGLAP~\cite{Gribov:1972ri,Altarelli:1977zs,Dokshitzer:1977sg} and BFKL~\cite{Fadin:1975cb, Lipatov:1976zz, Balitsky:1978ic, Salam:1999cn} equations.
At small $x$, hadronic structure is expected to evolve non-linearly due to the presence of  high gluon densities, as predicted by the JIMWLK~\cite{Mueller:2001uk} and BK~\cite{Lappi:2016fmu} evolution equations.
These non-linear effects should affect multi-parton dynamics, resulting in phenomena beyond a reduction of inclusive yields, including for instance observable effects in coincidence measurements.
Measurements of photon--jet correlations will allow to study this effect quantitatively by constraining the parton kinematics as precisely as possible in hadron interactions. \\

\begin{figure}[t!]
\includegraphics[width=0.48\textwidth]{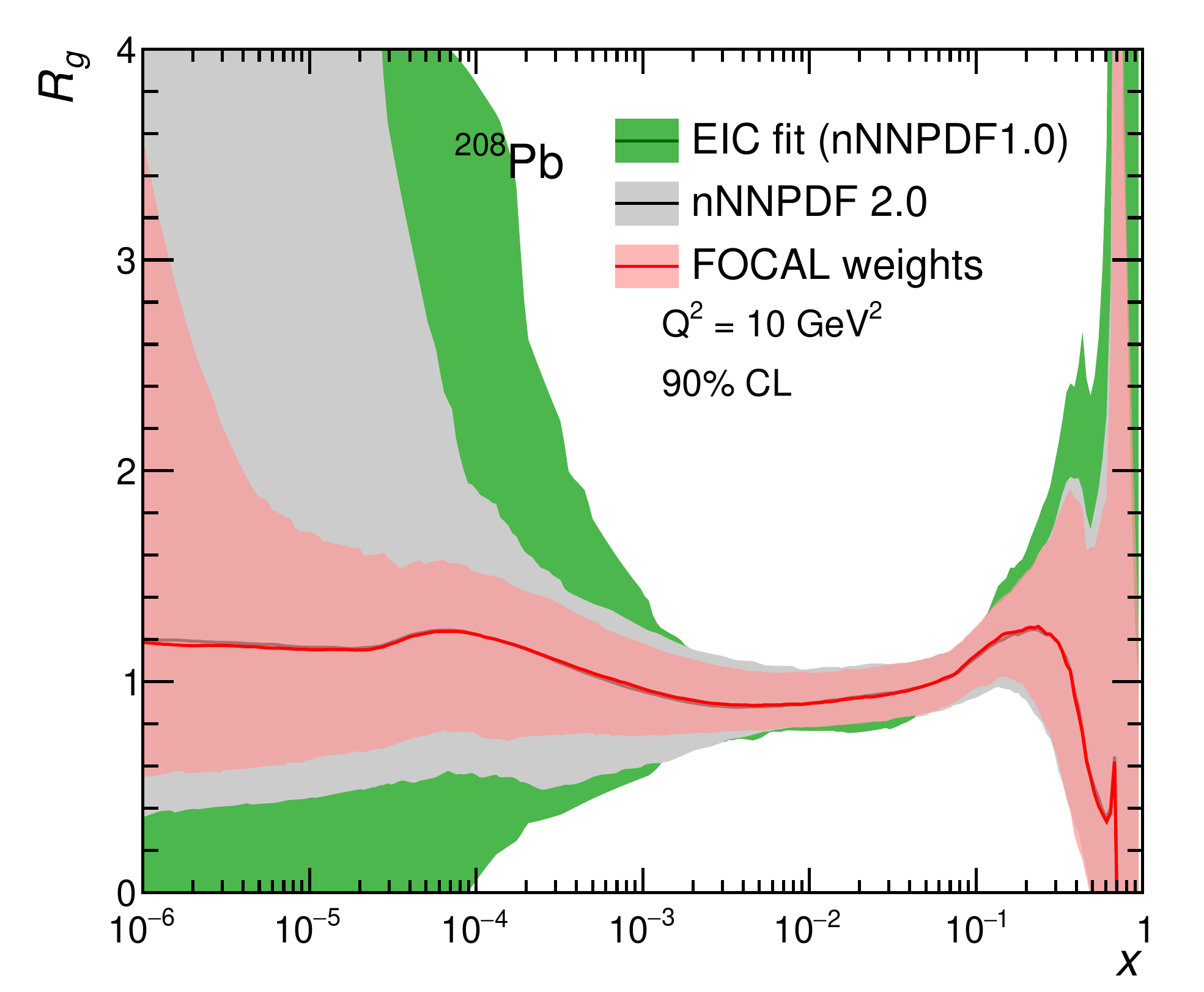}
\includegraphics[width=0.48\textwidth]{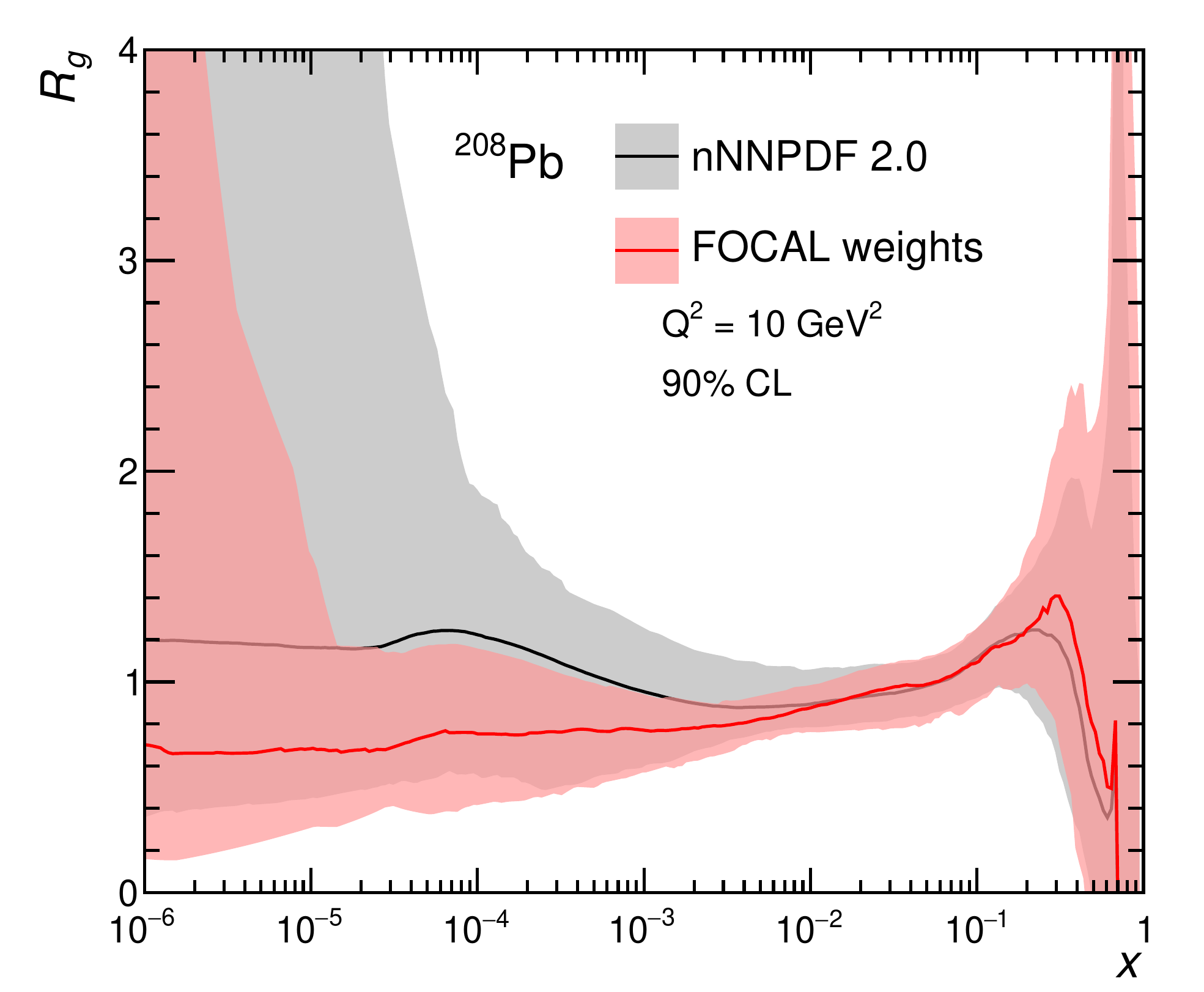}
\caption{\protect\label{fig:isophotoperf}
The nuclear modification of the gluon distribution, $R_g$, for the Pb nucleus versus $x$ at $Q^2=10$~GeV$^2/c^2$.
(Left)~Compared are the nNNPDF2.0 parameterization~\cite{AbdulKhalek:2020yuc} and fits to the FoCal pseudo-data above $4$~\GeVc\ (red band), as well as ``high energy'' EIC pseudo-data (green band; starting from the nNNDPF1.0 parametrization)~\cite{AbdulKhalek:2019mzd}.
In all cases, 90\% confidence-level uncertainty bands are drawn, and the nuclear PDFs are normalized by the proton NNPDF3.1.
(Right) Comparison of the original and the reweighted nNNPDF2.0 fits, where the FoCal pseudo-data were shifted by about 0.7~\cite{AbdulKhalek:2020yuc}.
}
\end{figure}

To illustrate the expected performance of future forward isolated photon measurements, the expected uncertainties of the gluon PDFs for the nNNPDF fit using either pseudo-data for the EIC (starting from nNNPDF1.0)~\cite{AbdulKhalek:2019mzd} or the FoCal above $4$~\GeVc\ based on nNNPDF2.0~\cite{AbdulKhalek:2020yuc}, are presented in the left panel of \Fig{fig:isophotoperf}.
As expected, the higher-energy option of the EIC~(which will be realized by eRHIC at BNL) will constrain the gluon PDF for $x$ down to about $5\cdot10^{-3}$, while isolated photon measurements provided by the FoCal would lead to significantly improved uncertainties even significantly below $10^{-4}$. 
We also note that the uncertainty on the nNNPDF2.0 parametrisation is already smaller than the EIC band.
The fit to EIC pseudo data from the nNNPDF group gives a qualitatively similar result to an earlier study based on modified EPPS16 nuclear PDFs~\cite{Aschenauer:2017oxs}, although the uncertainty estimates at small $x$ where there is no direct constraint from the pseudo data differ.
Clearly, the FoCal measurements will probe much smaller $x$ than the existing and possible future EIC measurements, and lead to high precision results due to the excellent direct photon performance. 
However, in case of the EIC, unlike at a hadron collider, the initial state is precisely known, and one can map $x$ and $Q^2$, independently, and  measure not only longitudinal but generalized parton~(GPDs) and transverse-momentum~(TMDs) distributions~\cite{AbdulKhalek:2021gbh}.
Furthermore, the EIC will allow us to scan the $A$~(nucleus) dependence using several nuclear beams. \\

So far, it was assumed in the re-weighting process that the central value of the FoCal measurement of the isolated photon $\RpPb$ would be the same as the central value of the initial baseline prediction. 
Instead in the right panel of \Fig{fig:isophotoperf}, we show the effect of the FoCal pseudo-data for a value of $\RpPb$ shifted to about $0.7$.
In this case, the FoCal data would add a significant amount of new information to the global fit, leading to a deviation of $R_g$ from the expected value in almost the entire range of $10^{-5}<x<10^{-2}$.
Therefore, this analysis indicates that FoCal measurements could be sensitive either to the gluon shadowing effects or to possible non-linear QCD dynamics. 
To disentangle one from the other, a dedicated analysis of the $\chi^2$ in global pdf fits and the nPDF behavior in the small-$x$ region would be required, following the approach of \Ref{Ball:2017otu}.\\

Significant suppression of $R_g$ arises also from forward charm measurements~\cite{LHCb:2017yua} when they are included in the determination of nuclear PDFs as recently done~\cite{Khalek:2022zqe}. 
In this case, comparing precise forward photon and charm measurements will allow us to test factorization and universality of the nuclear PDFs.


\subsection{Top quark pair production as a tool to probe  Quark-Gluon -Plasma formation}
\label{sec:ttbar_saturation}

\begin{figure}[t]
\centering
\includegraphics[width=.6\textwidth]{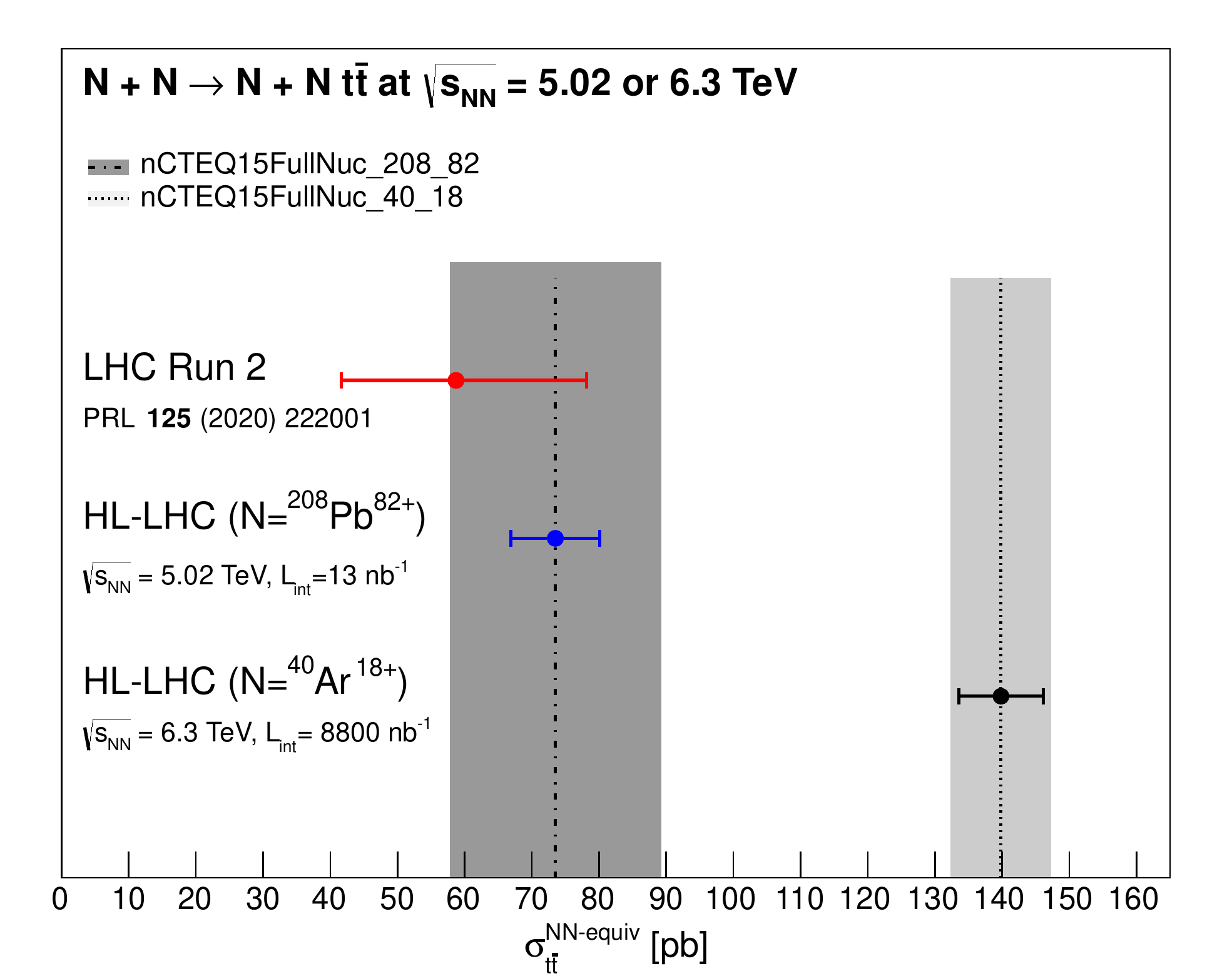}
\caption{\label{fig:ttbar_proj} Comparison of inclusive top quark pair production cross sections measured at LHC Run 2~\cite{CMS:2020aem} and projected at HL-LHC with either lead-lead or argon-argon collisions. Vertical lines and bands represent Quantum Chromodynamics predictions at next-to-next-to-leading order with soft-gluon resummation at next-to-next-to-leading logarithmic accuracy~\cite{Czakon:2011xx,Czakon:2013goa}, with nuclear modification effects and their uncertainties, respectively, as parametrized by the nCTEQ15 bound-nucleon distribution functions~\cite{Kovarik:2015cma}. The shown cross section values are their nucleon-nucleon equivalent.}
\end{figure}

Past Deep-inelastic-scattering (DIS) experiments provided accurate information  on  the  partonic  structure  of  the  free  proton. Notwithstanding   the  phenomenological success of  Quantum Chromodynamics (QCD) analyses,  a  detailed  understanding  of the partonic  structure modifications in bound  nuclei  is still lacking. Compared to the parton distribution functions (PDFs) in the proton, nuclear PDFs (nPDFs) are less constrained mainly because of the lack of data across the momentum fraction $x$--squared momentum transfer $Q^2$ plane and nuclear mass number range. The scheduled proton-lead (pPb) and lead-lead (PbPb) LHC Runs 3--4 at the LHC provide the opportunity to precisely constrain the nPDFs for the lead nucleus. This is demonstrated in Fig.~\ref{fig:ttbar_proj}, 
where a simple projected scenario of the existing CMS measurement~\cite{CMS:2020aem} is assumed, considering improvements in the statistical and currently dominant systematic uncertainties, respectively. There is even a complementarity between the physics programs at LHC and the planned Electron Ion Collider, allowing for  stringent  tests  of  the  nPDF universality too. Top quark production (inclusively or differentially) in pPb and PbPb collisions has been suggested as a valuable probe of the high-$x{\sim}10^{-2}$--$10^{-1}$ gluon distribution at very high $Q^2$ in the Pb ions~\cite{dEnterria:2015mgr}. 
\\

One powerful probe of the quark-gluon plasma (QGP) is ``jet quenching'', i.e., the study of jet modifications while passing through the QGP. Processes used so far, {\it e.g.}, dijet or Z/$\gamma$+jet production, are only sensitive to the properties of the QGP integrated over its lifetime. Hadronically decaying W bosons can provide key novel insights into the time structure of the QGP when studied in events with a top-antitop quark pair~\cite{Apolinario:2017sob} thanks to a ``time delay'' between the moment of the collision and that when the W boson decay products start interacting with the QGP. Although there seems to exist limited potential to bring the first information on the time structure of the QGP considering the baseline LHC scenario of Runs 3--4, lighter ions are potentially promising candidates despite their expected smaller quenching effects. Because of the potential for order-of-magnitudes higher effective integrated nucleon-nucleon luminosities, in this paragraph we advocate on the usage of an ``optimal'' nucleus-nucleus colliding system at HL-LHC. Such an example for the inclusive top quark pair production cross section is also shown in Fig.~\ref{fig:ttbar_proj}, considering the expected luminosity increase for the case of argon-argon collisions~\cite{Citron:2018lsq}. Substantially increased LHC partonic and photon-photon luminosities at HL-LHC (or future higher energy colliders) could be also achieved via isoscalar beams, even opening up opportunities for studies not accessible with high-pileup collisions~\cite{Krasny:2020wgx}. The high-luminosity collisions of isoscalar nuclei could provide a new environment to study the QGP and complement the QGP studies in the low-luminosity collisions of heavy nuclei.

\section{Ultra-peripheral collisions at hadronic colliders and exclusive reactions at the EIC}
\label{sec:upc}

 \textbf{Main Contributors:} Marco A. Alcazar Peredo, J. Guillermo Contreras, Martin Hentschinski, Spencer Klein, Daniel Tapia Takaki  

\subsection{Existing measurements on diffractive vector meson photoproduction in UPCs}

Photons are clean probes of the QCD structure of nuclear targets 
\cite{Klein:2020nvu,Baltz:2007kq}.  At hadron colliders, like RHIC and the LHC, photoproduction processes can be studied in ultra-peripheral collisions (UPCs), where the incoming particles pass each other at impact parameters larger than the sum of their radii, such that strong interactions are suppressed and photon-induced processes are dominant. For a recent review, see~\cite{Klein:2019qfb}.

The most measured process, but not the only one, in UPCs is the diffractive production of vector mesons.  Photons fluctuate to virtual $q\overline q$ pairs which then scatter elastically from nuclear targets, emerging as real vector mesons.  They are copiously produced.   Their decays into few charged particles provide clean experimental signatures that can be used to trigger and select the corresponding events. In the theoretical side, the different vector meson masses allow us to study QCD at different scales: the production of a $\rho^{0}$ serves to investigate the approach to the black-disc limit of QCD, while that of a $\mathrm{J/}\psi$ sheds light into aspects of perturbative QCD at high energies like saturation~\cite{Morreale:2021pnn} and shadowing~\cite{Armesto:2006ph}. Additionally, in a Good-Walker approach~\cite{Good:1960ba,Miettinen:1978jb}, coherent and incoherent processes where the photon interacts with the full target or just with a piece of it, respectively, give access to the average behaviour of the gluonic field (coherent) or to the variance of its quantum fluctuations (incoherent). 
\begin{figure}[t]
    \centering
    \includegraphics[width=.7\textwidth]{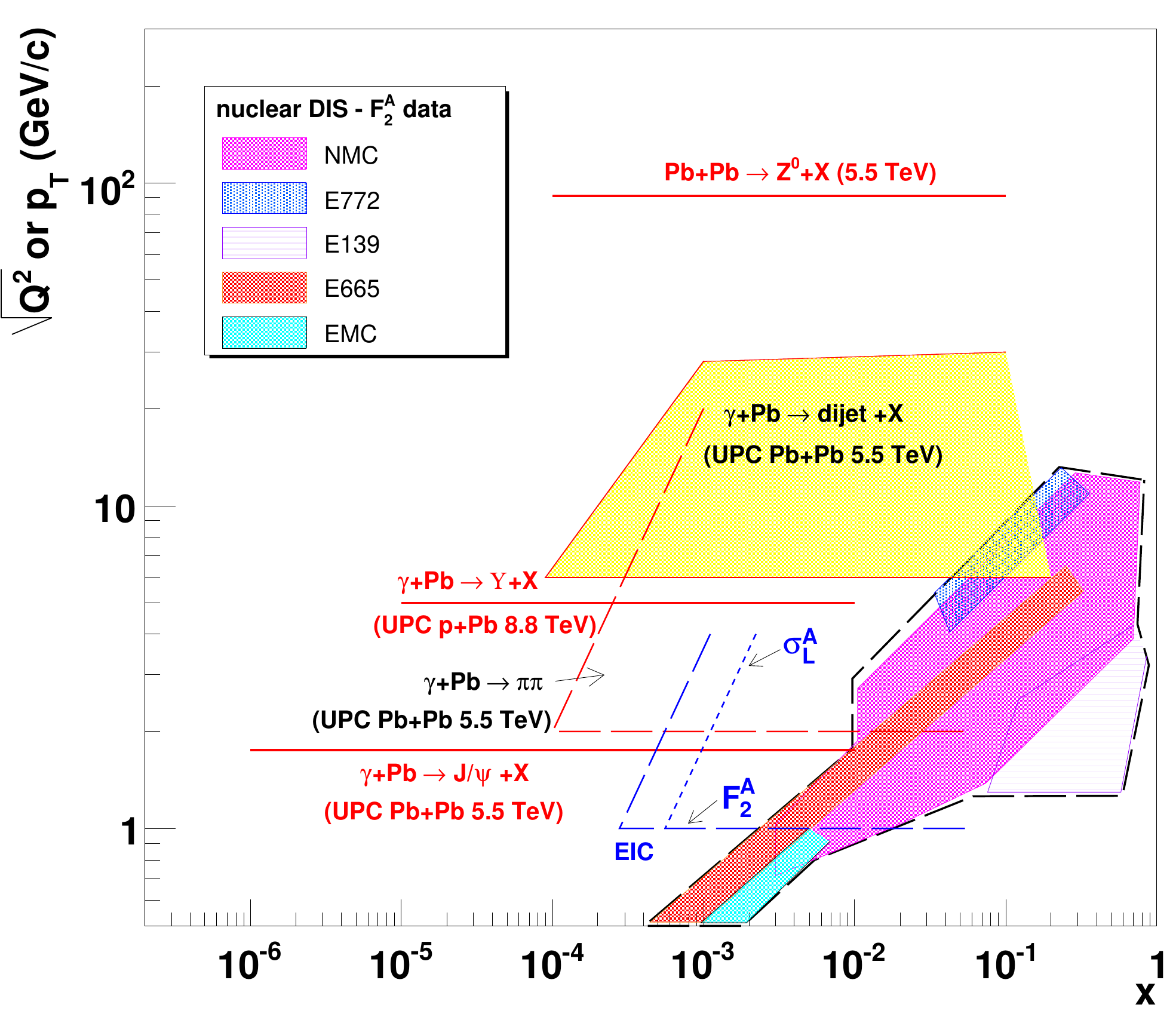}
    \caption{The kinematic range in which UPCs at the LHC can probe gluons in protons and nuclei in quarkonium production, dijet and di-hadron production. The $Q$ value for typical gluon virtuality in exclusive quarkonium photoproduction is shown
for $J/\Psi$ and $\Upsilon$. The transverse momentum of the jet or leading pion sets the scale for dijet and $\pi \pi$ production respectively. For comparison, the kinematic ranges for $J/\Psi$
at RHIC, $F_2^A$ and  $\sigma^A_L$
at eRHIC and $Z^0$ hadro-production at the LHC are also shown. Figure taken from \cite{Baltz:2007kq}}
    \label{fig:fig7Baltz2007l}
\end{figure}
Figure \ref{fig:fig7Baltz2007l} shows the $x$ and $Q^2$ ranges covered
by UPCs at the LHC. 
\\

The coherent photoproduction of $\rho^{0}$ vector mesons off nuclear targets has been extensively studied in UPCs. The STAR Collaboration at RHIC has published measurements in Au--Au UPCs  at  different centre-of-mass energies per nucleon pair, $\sqrt{s_{\rm NN}}$: 62.4 GeV~\cite{Agakishiev:2011me}, 130 GeV~\cite{Adler:2002sc}, and 200 GeV~\cite{Adamczyk:2017vfu}. At the LHC, the ALICE Collaboration has  carried out  measurement at  2.76 TeV~\cite{Adam:2015gsa} and 5.02 TeV~\cite{Acharya:2020sbc} in Pb--Pb UPCs and at 5.55 TeV in Xe--Xe UPCs~\cite{ALICE:2021jnv}.  The measured $\rho$ cross-section is seen to scale nearly linearly with the atomic number, with ALICE finding a best fit $\sigma\propto A^{0.96\pm 0.02}$.  The cross-section is smaller than is expected from a Glauber calculation (even including generalized vector meson dominance), but is consistent with a Glauber-Gribov calculation \cite{Frankfurt:2015cwa}.  The latter approach includes for high-mass intermediate state fluctuations.  These topics can be further studied by examining excited meson states.
\\

The STAR Collaboration~\cite{STAR:2009giy} also observed  the coherent production of  four charged pions~\cite{STAR:2009giy} which could be related to an excited state of the  $\rho^{0}$ vector meson but at a perturbative scale. Determining the cross-section is not possible without making assumptions about the branching ratios, but the rate seems consistent with the expectations of generalized vector meson dominance \cite{Frankfurt:2002sv}.
Both collaborations also observed an intriguing signal of an state in coherent di-pion production at a mass around 1.7 GeV/$c^2$ which could be related to another $\rho^{0}$ excited state~\cite{Klein:2016dtn,Acharya:2020sbc}.   \\

All these studies were performed at mid-rapidity in the corresponding laboratory frame. The availability of data for all these systems and energies provides an exacting challenge to theoretical descriptions of this process, whose cross section has been described more or less successfully using a variety of approaches; e.g.~\cite{Klein:1999qj,Guzey:2016piu,Cepila:2018zky}. In addition, there are at least three specific measurements to highlight: ($i$) The observation of interference effects originated in the fact that each of the two incoming projectiles can act either as a source of the photon or as a target of the interaction~\cite{Klein:1999gv,STAR:2008llz}, ($ii$) the mapping of the impact-parameter dependence of the target structure for gold ions obtained as a Fourier transform of the Mandelstam-$t$ dependence of the cross section~\cite{Adamczyk:2017vfu}, and ($iii$)  the dependence on the atomic mass number of this process at a centre-of-mass energy of the photon--nucleus system of 65 GeV per nucleon obtained from measurements off Pb~\cite{Acharya:2020sbc}, Xe~\cite{ALICE:2021jnv}, and protons~\cite{H1:2020lzc}, where a clear indication of strong shadowing was found along with the observation that the black-disc limit of QCD has not been yet reached.\\

Regarding the diffractive photoproduction of $\mathrm{J/}\psi$ the main results from UPCs have been obtained at the LHC~\cite{Contreras:2015dqa} after a first proof-of-principle measurement at RHIC~\cite{PHENIX:2009xtn}. Coherent production has been investigated by the ALICE Collaboration in two ranges of rapidity, central and forward, and two different energies $\sqrt{s_{\rm NN}} = 2.76$ TeV and 5.02 TeV~\cite{Abelev:2012ba,Abbas:2013oua,Acharya:2019vlb,ALICE:2021gpt,ALICE:2021tyx}. The CMS Collaboration published a cross section at semi-central rapidities~\cite{Khachatryan:2016qhq} at $\sqrt{s_{\rm NN}} = 2.76$ TeV, and recently the LHCb Collaboration presented results $\sqrt{s_{\rm NN}} = 5.02$ TeV covering forward rapidities~\cite{LHCb:2021bfl}. Together, these measurements allow for the study of the rapidity dependence of $\mathrm{J/}\psi$ diffractive photoproduction in a large kinematic range which corresponds to three orders of magnitude in  Bjorken $x$ from $10^{-2}$ to $ 10^{-5}$. These results provide new constraints on the evolution of the nuclear gluon distribution at large energies, see e.g.~\cite{Guzey:2020ntc,Bendova:2019psy}. In particular, the results from~\cite{ALICE:2021tyx} provide a look at the transverse structure of Pb nuclei at  the Bjorken-$x$ range $(0.3-1.4)\times10^{-3}$ which are the first step towards the mapping of the gluon  distribution in impact parameter at a perturbative scale.  A proof-of-principle measurement for incoherent production has been has been performed by the ALICE Collaboration in Pb--Pb UPCs at $\sqrt{s_{\rm NN}} = 2.76$ TeV~\cite{Abbas:2013oua}. 
\\

The coherent production of $\psi^\prime$ has also been measured by the ALICE Collaboration~\cite{Adam:2015sia,ALICE:2021gpt}; this state is interesting to understand the spin structure of the interaction, in particular to constraint the modelling of the wave function of the vector meson (see e.g.~\cite{Krelina:2018hmt}), which is a non-perturbative component of all theoretical predictions. Finally, the CMS collaboration has made an initial measurement of $\Upsilon$ photoproduction in $pA$ collisions \cite{Chudasama:2016eck}.  Another interesting related result is the measurement of coherent  $\mathrm{J/}\psi$  photoproduction in {\em peripheral} collisions, that is with a geometrical overlap of the colliding nuclei, that have been performed by the ALICE Collaboration in the Pb--Pb system~\cite{ALICE:2015mzu} and by the STAR Collaboration in the Au--Au and UU systems~\cite{STAR:2019yox}. These measurements open up interesting questions on the meaning of coherence in these quantum processes and the possibility to study new effects, e.g. the interaction of the $\mathrm{J/}\psi$ with the quark-gluon plasma created in such collisions~\cite{Klusek-Gawenda:2015hja,Shi:2017qep,Zha:2017jch}; they also offer a tool to study the energy dependence of  $\mathrm{J/}\psi$ production by offering a measurement in a different impact-parameter range than those in UPCs~\cite{Contreras:2016pkc}.

\subsection{Future measurements on diffractive vector meson photoproduction in UPCs}

In the near future the Run 3 and 4 of the LHC will provide an enormous data set of UPCs; see e.g. Table 12 of~\cite{Citron:2018lsq}. In the middle term, the Electron Ion Collider will be the experimental facility to study the QCD structure of nuclei, including diffractive vector meson production~\cite{Accardi:2012qut}. 

One of the key measurements to be performed with the new LHC data set is the Bjorken-$x$ evolution of coherent diffractive vector meson production for as many different mesons---that is, mass scales---as possible. To achieve this, the two contributions to the nucleus--nucleus cross section, one with a high-, the other with a low-energy photon have to be disentangled. In principle, this requires to perform a given measurement at a fix rapidity, but different impact-parameter ranges. There are two proposals on how to do this: using UPCs in conjunction with the peripheral collisions mentioned above~\cite{Contreras:2016pkc}, and using events where in addition to the photon exchange producing the vector meson, there is an extra nuclear dissociation process that acts as a selector of a different impact-parameter range~\cite{Baltz:2002pp,Guzey:2013jaa}. Measurements of coherent $\rho^{0}$ production at mid-rapidity in Pb--Pb~\cite{Acharya:2020sbc} and Xe--Xe~\cite{ALICE:2021jnv} UPCs accompanied by neutrons at beam rapidities, product of the nuclear dissociation, are correctly described by the NOON model~\cite{Broz:2019kpl} giving us confidence that the relevant physics is understood (at the current precision of the data) and paving the way to the application of this method. 

Another eagerly awaited measurement is the dependence on Mandelstam-$t$ of the incoherent production of vector mesons at a given rapidity. This process is sensitive to quantum fluctuations at the sub-nucleon scale~\cite{Mantysaari:2020axf}. Model predictions, e.g.~\cite{Mantysaari:2017dwh,Cepila:2017nef}, expect a one order of magnitude increase of the cross section at $|t|\sim 1$ GeV$^2$ when the sub-nuclear quantum fluctuations are taken into account with respect to the case where the relevant degrees of freedoms are the nucleons. Such a measurement will be feasible at the LHC in the near future. 

Another interesting measurement to be performed at the LHC is the study of the angular correlations of the decay products of the vector meson. Both the quasi-real photons and the gluons participating in the interaction are linearly polarised. This, and the presence of the interference effects mentioned above produce new angular correlations with a particular dependence on the transverse momentum of the vector meson. These observables are a complement to the traditional polarisation measurements and are also sensitive to the QCD structure of the target. See e.g.~\cite{Zha:2018jin,Xing:2020hwh,Zha:2020cst}.
\\

At least two other techniques can be used to measure gluon distributions using UPCs: dijets and open charm.  These measurements are theoretically cleaner than vector mesons, since they involve only single-gluon exchange, so the uncertainties involving color neutralization are much smaller.  But, the final states are more complicated, and since there is a color string connecting the mid-rapidity state and the target nucleon remnants, the reaction cannot be fully exclusive.   The ATLAS collaboration has already made the first preliminary measurements of  dijet photoproduction \cite{Angerami:2017kot}.  ATLAS explored the region where the leading jet had $p_T >20 $ GeV, and dijet mass above 35 GeV, giving them a reach down to $x\approx 3\times 10^{-3}$.  Charm photoproduction is an attractive alternative approach to reach down to lower $x$ values, since it should be possible to measure charm down to threshold, $M_{c\overline c}\approx 4$ GeV.   The rates for charm are high \cite{Klein:2002wm,Adeluyi:2011rt,Goncalves:2017zdx}, and, at the LHC, open $b\overline b$ and potentially, with $pA$ collisions or lighter ions, $t\overline t$ \cite{Klein:2000dk}. 

\subsection{The ratio of \texorpdfstring{$\Psi(2s)$}{Psi(2s)} and \texorpdfstring{$J/\Psi$}{J/Psi} photoproduction cross-sections as a tool to quantify non-linear QCD evolution}

Exclusive photoproduction of charmonium at the Large Hadron Colllider (LHC) provides an excellent testing ground for the description of the low $x$ gluon distribution, since it allows for a direct observation of
the energy dependence of the photoproduction cross-section which
directly translates into the $x$-dependence of the underlying gluon
distribution. 
photoproduction of bound states of charm quarks, {\it
  i.e.}  $J/\Psi$ and $\Psi(2s)$ vector mesons, are  of particular interest, since the charm mass provides a hard scale at the
border between soft and hard physics and the observable is therefore
expected to be particularly sensitive to the possible presence of a
semi-hard scale associated with the transition to the saturation
region, the so-called saturation scale. It is therefore ideal to search for potential deviations from linear QCD evolution.

Studies in the literature for this process, which take into account
effects due to gluon saturation, exist  both on the level of
dipole models \cite{Lappi:2013am, Armesto:2014sma,Goncalves:2014wna,Goncalves:2014swa,Kowalski:2006hc,Cox:2009ag, Cepila:2017nef}
and complete solutions to non-linear BK equation
\cite{Ducloue:2016pqr,Cepila:2018faq, ArroyoGarcia:2019cfl}. At the same time also descriptions on collinear factorization
\cite{Jones:2013eda,Jones:2013pga,Jones:2016icr,Szczurek:2017uvc, Flett:2020duk} and
 linear NLO BFKL evolution
\cite{Bautista:2016xnp, ArroyoGarcia:2019cfl}, provide an excellent description of data, see also the discussion in \cite{ArroyoGarcia:2019cfl,Hentschinski:2020yfm}. It is therefore not entirely clear, which is the appropriate description of data. While at first one might conclude that center of mass energies are simply not yet high enough to see the onset of non-linear effects, there are also indications that at least some linear frameworks turn unstably at highest center of mass energies available at LHC \cite{ArroyoGarcia:2019cfl}. \\

\begin{figure}[t]
    \centering
    \includegraphics[width=0.48\textwidth]{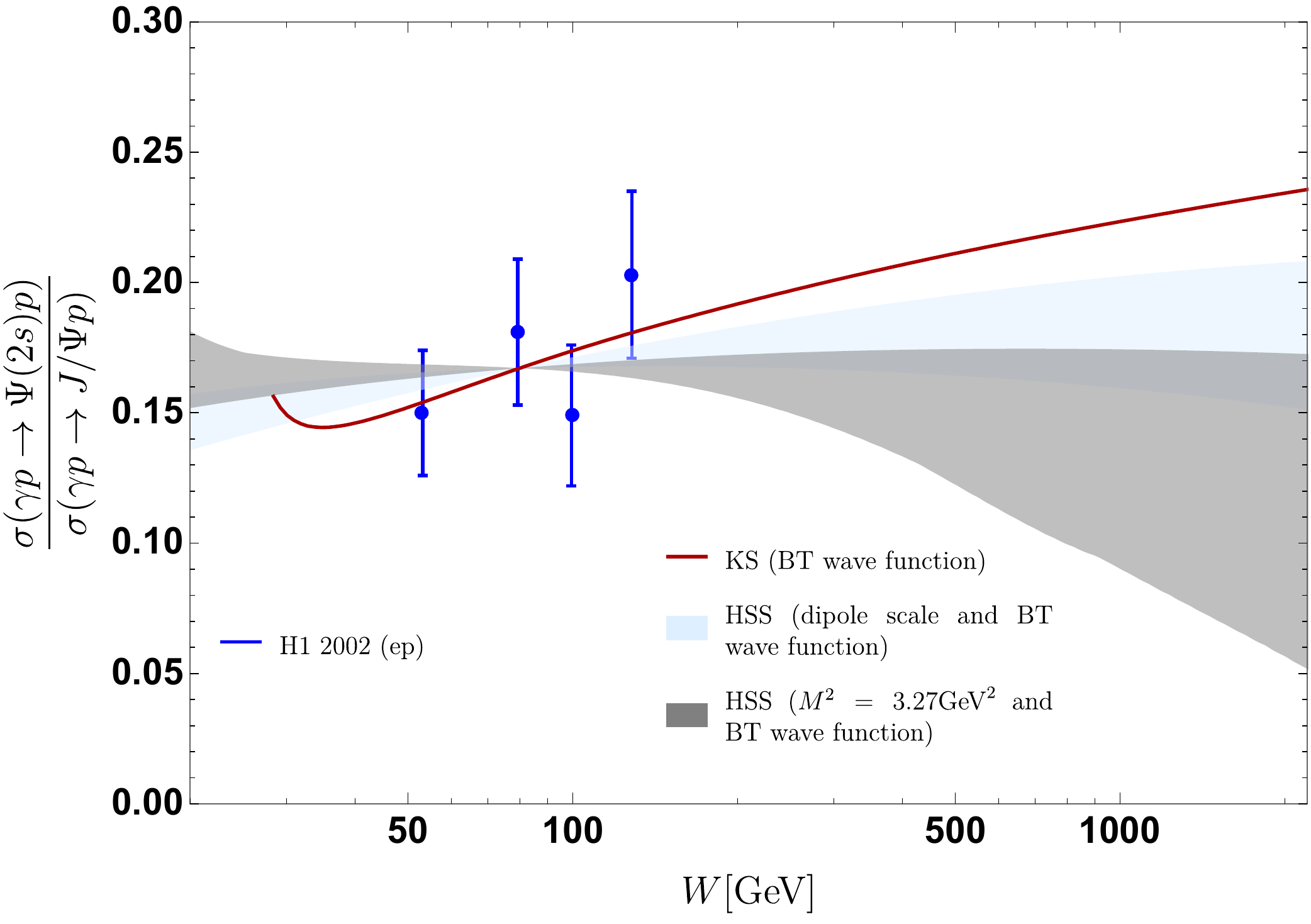}
    \includegraphics[width=0.48\textwidth]{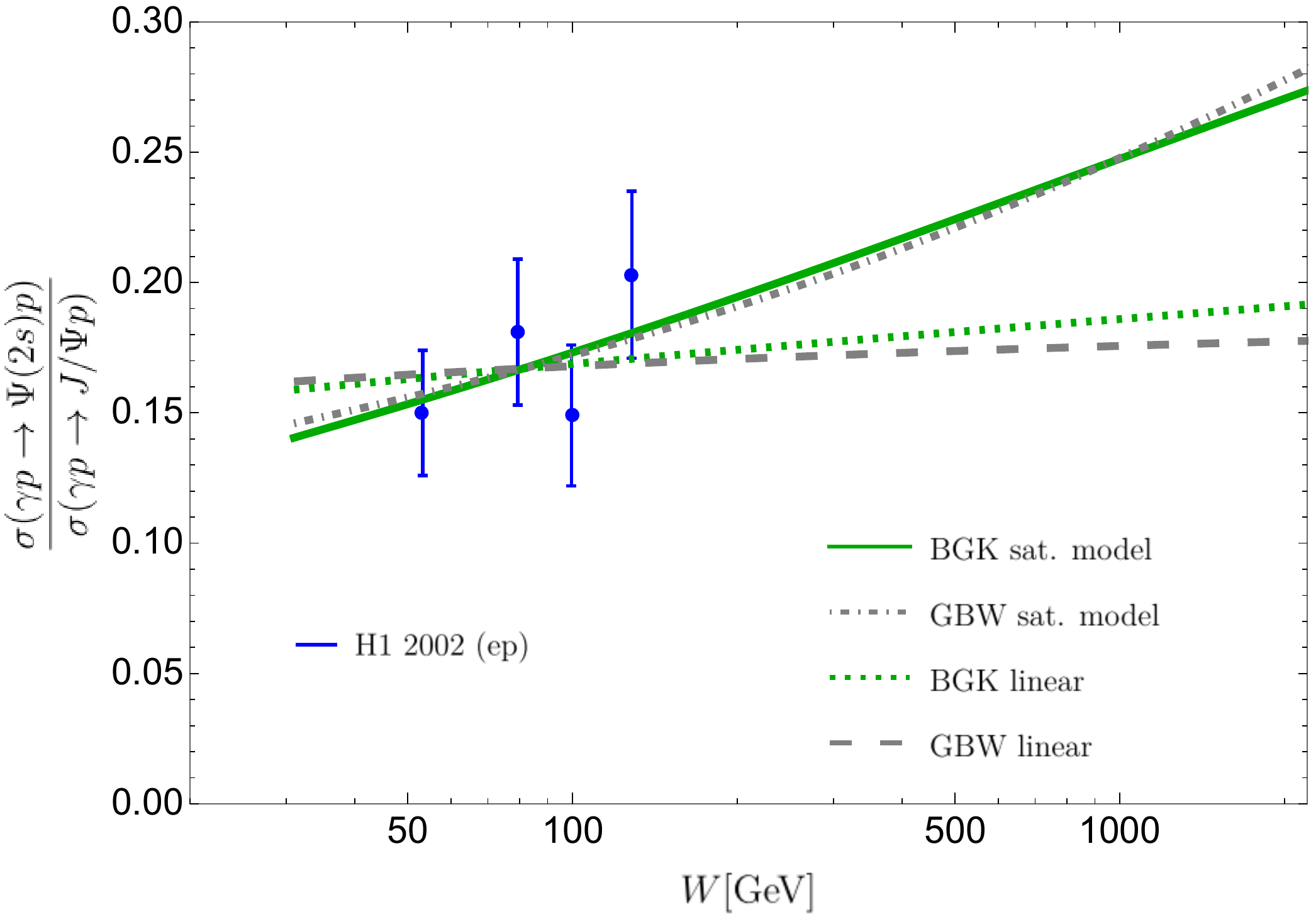}
    \caption{ Energy dependence of the ratio of $\Psi(2s)$ vs. $J/\Psi$
    photoproduction cross-section. We further
    display photoproduction data measured at HERA by the H1
    collaboration \cite{H1:2002yab}, which we further use to adjust the normalization. Left:  Implementation of the  KS and HSS gluon
    distributions for the Buchm\"uller-Tye (BT) vector meson wave functions, see \cite{Hentschinski:2021lsh} for details. The shaded regions
    correspond to a variation of the scale
    $\bar{M} \to \{\bar{M}/\sqrt{2}, \bar{M} \sqrt{2}\}$.Right: Implementation based on BKG and GBW saturation model and their corresponding linearized versions, using  numerical values found in \cite{Golec-Biernat:2017lfv} from a fit to HERA data. }
    \label{fig:ratioVM}
\end{figure}

A similar conclusion has been drawn in  \cite{Hentschinski:2020yfm}, where it  has been found that the energy dependence of the $J/\Psi$  and $\Psi(2s)$ cross-section is not able to distinguish between the non-linear (Kutak-Sapeta (KS) gluon \cite{Kutak:2012rf}, subject to non-linear Balitsky-Kovchegov (BK) evolution) and linear (Hentschinski-Salas-Sabio Vera gluon (HSS)
\cite{Hentschinski:2012kr,Hentschinski:2013id},  subject to linear NLO BFKL evolution) low $x$ evolution. On the other hand the ratio of both cross-sections was found to reveal a characteristically different energy dependence for linear and non-linear QCD evolution, see Fig.~\ref{fig:ratioVM}, left. 
Leaving large uncertainties associated with fixed scale HSS evolution aside(see  \cite{Hentschinski:2020yfm} for a detailed discussion), the stabilized "dipole scale" HSS gluon, subject to linear NLO BFKL evolution predicts a constant ratio of both photoproduction cross-section. The KS gluon, subject to non-linear BK evolution, predicts on the other hand a rise of the ratio. To understand this behavior better, it is instructive to analyze the  the same observable within a simple saturation model. The latter yields a particular simple form which allows us to gain some intuitive understanding why linear and non-linear QCD dynamics yield a different prediction for the $\Psi(2s)$ over $J/\Psi$ ratio. In the high energy limit, the relevant quantity of interest is the  imaginary part of the scattering amplitude,
 \begin{align}
  \label{eq:amp}
  \Im\text{m} \mathcal{A}_{\gamma p \to Vp}(W^2, t=0) & = \int d^2 {\bm r}  \left [\sigma_{q\bar{q}} \left(\frac{M_V^2}{W^2}, r \right)  \overline{\Sigma}_T^{(1)}( r)
 + 
\frac{d \sigma_{q\bar{q}} \left(\frac{M_V^2}{W^2}, r \right)}{dr} \overline{\Sigma}_T^{(2)}(r)
 \right],
\end{align}
which encodes the energy dependence related to the low $x$ evolution of inclusive low $x$ evolution. Here  $r = |{\bm r}|$ the transverse separation of the quark anti-quark pair and the $z$ the photon momentum fraction and   $\overline{\Sigma}_T^{(1,2)}$
describes the transition of a transverse polarized photon into a vector
meson $V$ \cite{Cepila:2019skb}; see \cite{Hentschinski:2021lsh} for details.  Within this approach the entire energy dependence is contained in the dipole cross-section $\sigma_{q\bar{q}}$, which for the 
 Golec-Biernat W\"usthoff saturation model takes the following simple form \cite{Golec-Biernat:1998zce}
\begin{align}
  \sigma_{q\bar{q}}(x, r) & = \sigma_0 \left(1-e^{-r^2 Q_s^2(x)/4} \right)
& Q_s^2(x) & =Q_0^2 (x/x_0)^\lambda
\end{align}
where $Q_s(x)$ yields within this model the saturation scale, which carries the entire energy dependence; a linearized version of this model, which yields a power-like growth with energy is then obtained through an expansion of the above expression for small saturation scales,
\begin{align}
\label{eq:GBWlin}
   \sigma_{q\bar{q}}^{\text{lin.}}(x, r) & = \sigma_0 r^2 Q_s^2(x)/4.
\end{align}
Inserting Eq.~(\ref{eq:GBWlin}) into Eq.~(\ref{eq:amp}), it is immediately clear that the saturation scale -- which carries the essential $W$ dependence  -- cancels for the ratio of $\Psi(2s)$ and $J/\Psi$ photoproduction cross-sections, up to a small logarithmic correction, related to the energy dependence of the diffractive slope of $J/\Psi$ and $\Psi(2s)$ cross-sections,  \cite{Cepila:2019skb, Hentschinski:2021lsh}. While the complete saturation model agrees with the linear approximation in the region $r \to 0$, they start to disagree for large dipole sizes  and  it is  this region where the wave-function overlap differs for the production of vector mesons $\Psi(2s)$ and $J/\Psi$, due to the presence of the node in the $2s$ wave function.  We further show results due to the  DGLAP improved saturation model, the so-called Bartels Golec-Biernat Kowalski (BGK) model, which replaces $Q_s^2(x) \to  4\pi \alpha_s(\mu(r)) x g(x, \mu(r))/3$, where $\alpha_s(\mu)$  and $xg(x, \mu)$ denotes the strong coupling constant and the collinear gluon distribution respectively, evaluated at  an $r$-dependent scale $\mu$ in the perturbative region. Similar to the HSS gluon, such a dipole size dependent saturation scale  prevents an exact cancellation of the energy dependence in the linear approximation. Nevertheless  this merely affects the perturbative region of small dipole sizes $r < 1$~GeV$^{-1}$ and therefore maintains the observed rise of the ratio for the complete saturation versus an approximately constant ratio for the linear approximation.We therefore suggest to extract from existing data and future measurements $\Psi(2s)$ and $J/\Psi$  ratio, since the energy behavior of this ratio allows to draw conclusion on the presence of non-linear QCD dynamics.  The feature is both present for unintegrated gluon distributions, which have been obtain from a numerical solution to  linear and non-linear QCD evolution, fitted to DIS data, as well as for analytic dipole models, where the distinction of linear and non-linear realization is somehow easier.  While there exist LHCb data for the energy dependence of the $J/\Psi$ and $\Psi(2s)$ photoproduction cross-section, extracted from $pp$ collisions \cite{LHCb:2018rcm} (see also \cite{ALICE:2021gpt} for  $PbPb$ data),  there exist only H1 data for the energy dependence of the ratio of both cross-sections, at relatively low values of $W$ and with still considerable uncertainties. We believe that an extraction of the ratio from both combined HERA and LHC data would be highly beneficial to pin down the size of non-linear low $x$  QCD evolution at the LHC. \\

\subsection{Planned measurements at the Electron Ion Collider}

Looking further ahead, in the early 2030s the U. S. Electron Ion Collider (EIC) should provide high-precision measurements of vector mesons over a wide range of Bjorken$-x$ and $Q^2$; the $Q^2$ of the photon can be measured independently of the rest of the reaction, allowing us to probe the nucleus using $q\overline q$ dipoles of different lengths \cite{Accardi:2012qut,AbdulKhalek:2021gbh}.  The high center-of-mass energy (up to about 140 GeV) and high luminosity will allow the EIC to study large samples of light and heavy mesons (including the three $\Upsilon$ states) \cite{Lomnitz:2018juf}.  The expected event samples range from about 50 billion $\rho^0$ per year down to about 140,000 $\Upsilon (1S)$ per year.  It will also be able to study exotic states (including the XYZ states) via Reggeon exchange reactions \cite{Klein:2019avl,Albaladejo:2020tzt}.   The high luminosity will allow for precise multi-dimensional studies, including measurements of Generalized Parton Distributions, measurements out to kinematic extremes (i. e. large $|t|$ etc.) and studies of rarely produced mesons and decays. 

The EIC will take data with a variety of different ions, so will be able to study how low$-x$ gluons evolve with nuclear size.  Light ions will be of special interest.  The EIC detectors forward spectrometers are expected to be able to detect scattered protons and light ions, allowing for a measurement of $|t|$ even if the scattered electron is not seen or poorly measured.  Light ion studies will allow for the study of neutron targets, and studies with deuterium and other very light ions will allow for measurements of the nuclear force in relatively simple systems. 

The EIC detectors are being designed to be extremely hermetic, so will be able to record vector mesons over a wide range in Bjorken$-x$ and to accurately separate coherent and incoherent production over a wide range of $|t|$ \cite{AbdulKhalek:2021gbh}; this is necessary to fully apply the Good-Walker paradigm.    The large event samples and precision detectors will allow for precise studies of the variation in gluon density and transverse position within the target.  And, the inclusion of relatively precise calorimetry that is sensitive down to low energies will allow us to study final states that include $\gamma$ and $\pi^0$, allowing for the study of a wider range of mesons.



\section{Odderon discovery and diffractive jets}
\label{sec:odder-disc-diffr}

\noindent {\bf Main contributor:} Christophe Royon \\

In this section we provide some details on the recent discovery of the Odderon as well as inclusive diffractive measurements and explorations of the soft Pomeron structure. 

\subsection{Soft diffraction and the Odderon discovery by the D0 and TOTEM experiments}

\begin{figure}[t]
\centering
\includegraphics[width=0.45\textwidth]{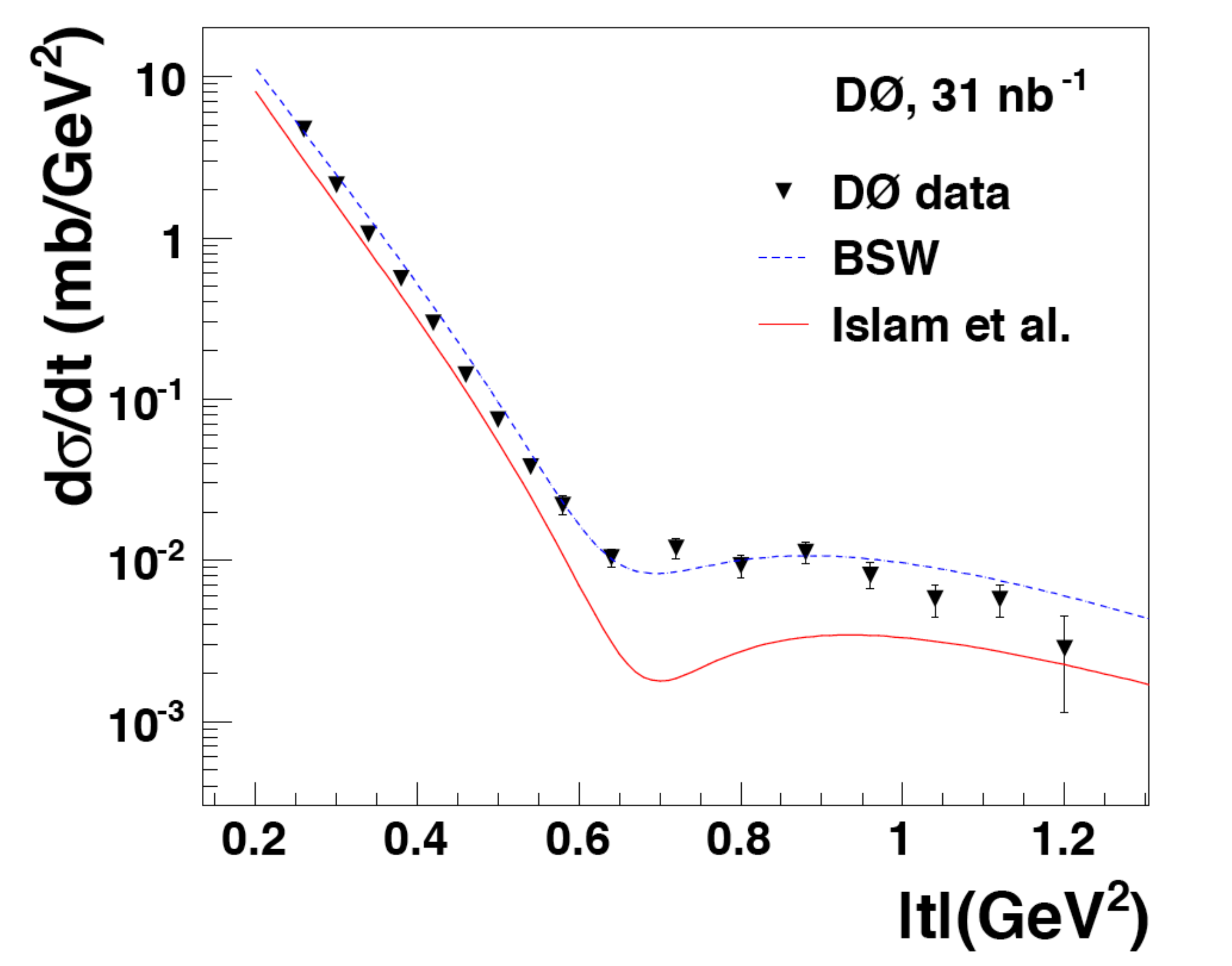}
\includegraphics[width=0.5\textwidth]{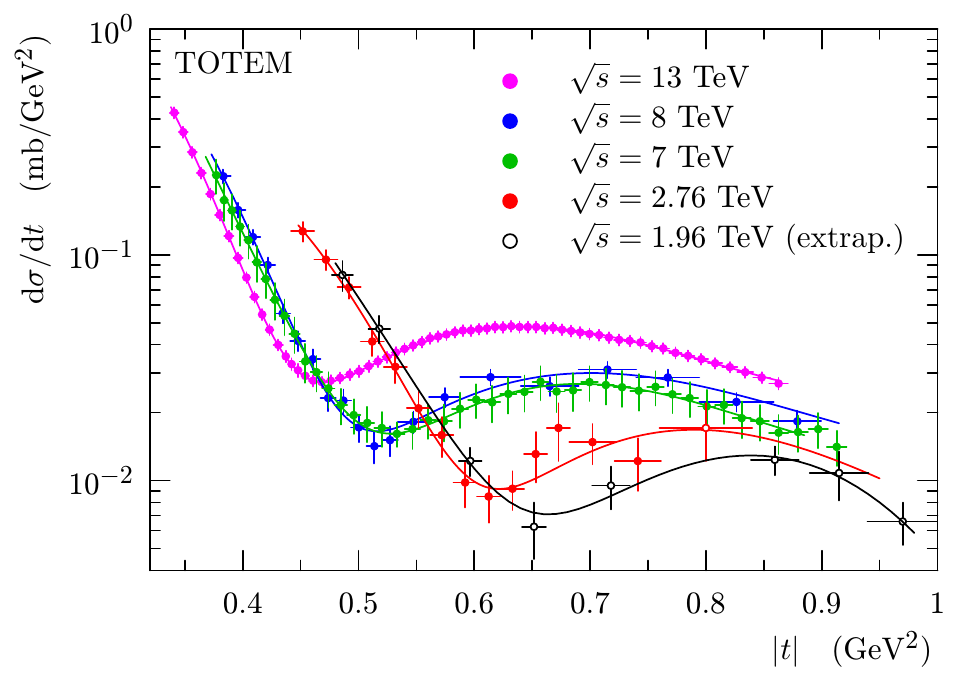}
\caption{Left: $p \bar{p}$ elastic cross section as a function of $|t|$ at 1.96 TeV from the D0 collaboration at the Tevatron. Right: $pp$ elastic cross sections as a function of $|t|$ at 2.76, 7, 8, and 13 TeV from the TOTEM collaboration at the LHC (full circles), and extrapolation to the Tevatron center-of-mass energy at 1.96 TeV (empty circles).}
\label{fig1}
\end{figure}

Soft diffraction and elastic interactions have been studied for the last 50 years at different colliders. Elastic $pp$ and $p \bar{p}$ scattering at high energies of the Tevatron and the LHC for instance corresponds to the $pp \rightarrow pp$ and $p \bar{p} \rightarrow p \bar{p}$ interactions where the protons and antiprotons are intact after interaction and scattered at very small angle, and nothing else is produced. In order to measure these events, it is necessary to detect the intact protons/antiprotons after interactions in dedicated detectors called roman pots and to veto on any additional activity in the main detector.

Many experiments have been looking for evidence of the existence of the Odderon~\cite{Lukaszuk:1973nt,Martynov:2018sga} in the last 50 years, and one may wonder why the Odderon has been so elusive. At ISR energies, at about a center-of-mass energy of 52.8 GeV~\cite{Breakstone:1985pe, Erhan:1984mv,UA4:1986cgb,UA4:1985oqn,Nagy:1978iw}, there was already some indication of a possible difference between $pp$ and $p\bar{p}$ interactions. Differences are about 3$\sigma$ but this was not considered to be a clean proof of the Odderon.  This is due to the fact that elastic scattering at low energies can be due to exchanges of additional particles to Pomeron and Odderon, namely $\rho$, $\omega$, $\phi$ mesons and Reggeons. It is not easy
to distinguish between all these possible exchanges, and it becomes quickly model dependent. This is why the observed difference at 52.8 GeV was estimated to be due to $\omega$ exchanges and not to the existence of the Odderon.
The advantage of being at higher energies (1.96 TeV for the Tevatron and 2.76, 7, 8 and 13 TeV at the LHC~\cite{D0:2012erd,TOTEM:2018psk, TOTEM:2011vxg,TOTEM:2015oop,TOTEM:2018hki} is that meson and Reggeon exchanges can be neglected. It means that a possible observation of differences between $pp$ and $p \bar{p}$ elastic interactions at high energies would be a clear signal of the Odderon. The D0 and TOTEM elastic $d\sigma/dt$ data are shown in Fig.~\ref{fig1}. The difficulty to compare between $pp$ and $p\bar{p}$ elastic scatterings is that one has to extrapolate the $pp$ measurements from TOTEM to Tevatron center-of-mass energies~\cite{TOTEM:2020zzr}.

\begin{figure}[t]
\centering
\includegraphics[width=0.5\textwidth]{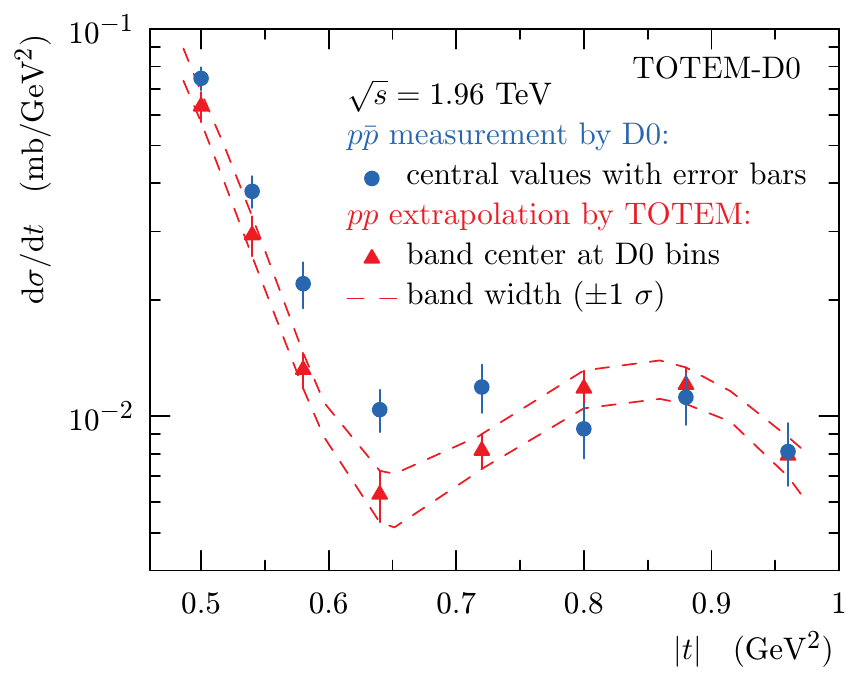}
\caption{Comparison between the D0 $p\bar{p}$ measurement
at 1.96 TeV and the extrapolated TOTEM $pp$ cross section, re-scaled to match the OP of
the D0 measurement.  The dashed lines show the
1$\sigma$ uncertainty band on the extrapolated $pp$ cross section.}
\label{fig5}
\end{figure}

The comparison between the $p \bar{p}$ elastic $d \sigma/dt$ measurement by the D0 collaboration and the extrapolation of the TOTEM $pp$ elastic $d \sigma/dt$ measurements is shown in Fig.~\ref{fig5}, including the 1$\sigma$ uncertainty band as a red dashed line~\cite{TOTEM:2020zzr}. The comparison is only made in the common $t$ domain for both $pp$ and $p \bar{p}$ measurements and show some differences in the dip and bump region between $|t|$ of  0.55 and 0.85 GeV$^2$.  Given the constraints on the optical point normalization and logarithmic slopes of the elastic cross sections, the $\chi^2$ test leads to a significance of 3.4$\sigma$. Combining this result with previous measurements of TOTEM of $\rho$~\cite{TOTEM:2017sdy} and the total cross section, the significance ranges from 5.3 to  5.7$\sigma$ (depending on the model).
Models without colorless $C$-odd gluonic compound or the Odderon are excluded by more than 5$\sigma$.

Further measurements of elastic $pp$ cross sections will happen at higher LHC energies (such as 13.6 and 14 TeV) and the  Odderon production will be performed in additional channels, such as the production of $\omega$ mesons. It is also clear that the discovery of the Odderon is likely related to the existence of glueballs, and the search for their production will happen at the LHC, RHIC and the EIC.

\subsection{Inclusive diffraction measurements at the LHC and sensitivity to the Pomeron structure}
Hard diffraction correspond to events when at least one proton is intact after interaction at the LHC and correspond to the exchange of a colorless object called the Pomeron. Many measurements at the LHC can constrain
the Pomeron structure in terms of quarks and gluons that has been
derived from QCD fits at HERA and at the Tevatron. All the
studies have been performed using the Forward Physics Monte Carlo
(FPMC), a generator
that has been designed to study forward physics, especially at the LHC~\cite{Boonekamp:2011ky,Jung:2009eq}.


One can first probe if the Pomeron is universal between
$ep$ and $pp$ colliders, or in
other words, if we are sensitive to the same object at HERA and the LHC.
Tagging both diffractive protons in
ATLAS and CMS allows to probe the QCD evolution of the gluon and quark densities
in the Pomeron  and to compare with the HERA measurements. In addition,
it is possible to assess the gluon and quark densities
using the dijet and $\gamma + jet$ productions~\cite{Marquet:2013rja,Kepka:2007nr,Marquet:2016ulz,Chuinard:2015sva}.
The different diagrams of the processes that can be studied at the LHC
are shown in Fig.~1, namely double Pomeron exchange (DPE) production of dijets (left),
of $\gamma +$jet (middle), sensitive respectively to the gluon and quark contents of the
Pomeron, and the jet gap jet events (right).

\begin{figure}[t]
\begin{center}
\includegraphics[width=7.5cm]{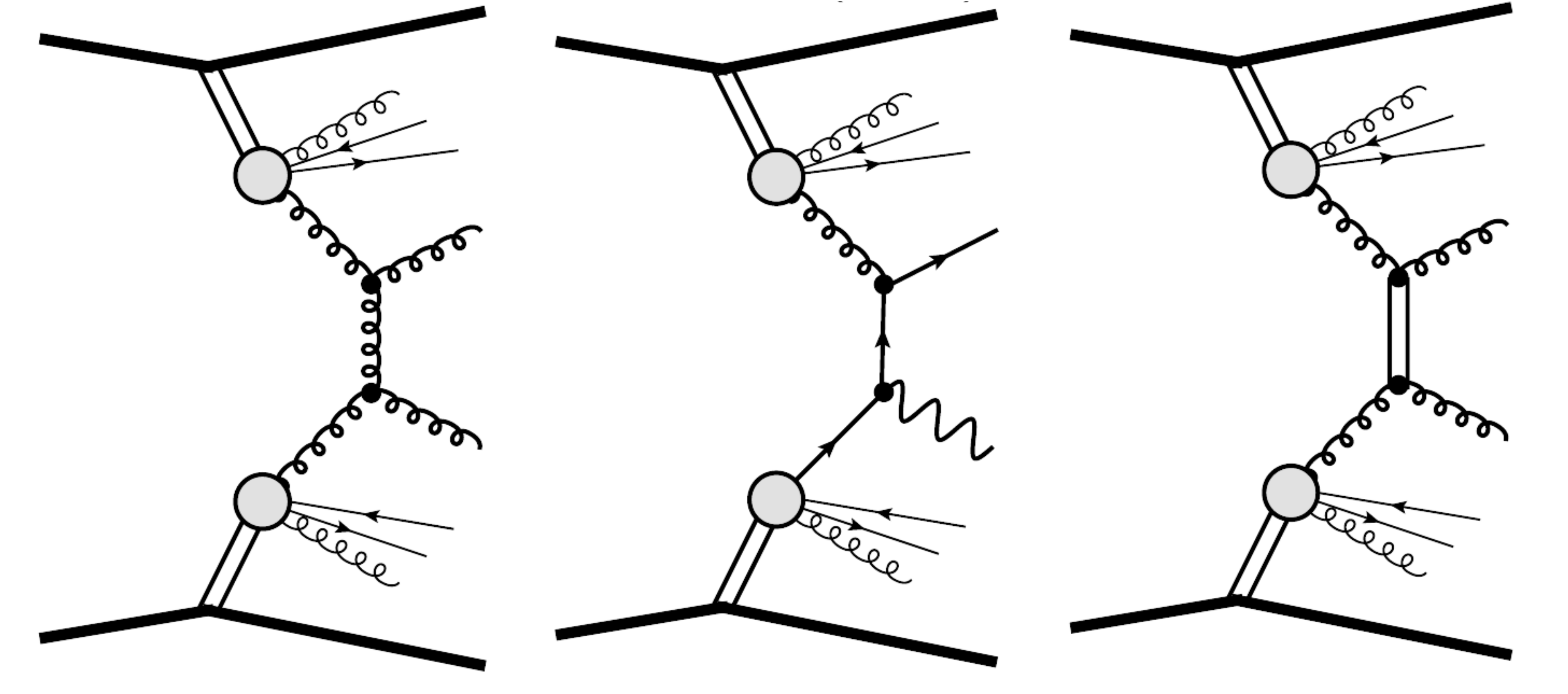}
\caption{Inclusive diffractive diagrams. From left to right: jet production and  $\gamma +$jet production in in inclusive double
Pomeron exchange, as well as  jet gap jet events involving  double
Pomeron exchange}
\label{d0b}
\end{center}
\end{figure}

The measurement of the dijet cross section is directly sensitive to the gluon density in the Pomeron and the $\gamma+$jet and $W$ asymmetry measurements~\cite{Chuinard:2015sva} 
are sensitive to the quark densities in the Pomeron.
However, diffractive measurements are also sensitive to
the survival probability which needs to be disentangled from PDF effects, and many different measurements will be needed to distinguish between them.

It is clear that understanding better diffraction and probing different models will be one of the key studies to be performed at the high luminosity LHC, the EIC and any future hadron collider.


\section{Electroweak  and Beyond the Standard Model Physics}
\label{sec:EW}

\noindent \textbf{Main Contributors:} Cristian Baldenegro, Andrea Bellora,Victor Paulo Gonçalves, Sylvain Fichet, Gero von Gersdorff, Valery Khoze,  Michael Pitt, Christophe Royon, Gustavo Gil da Silveira, Marek Tasevsky \\

While the bulk of this white paper focuses on aspects related to strong interactions in the limit of high energies and densities, there exists also an increased interest in the study of electroweak processes, which rely on dedicated forward detectors for their analysis and which are capable to contribute to searches for new physics at LHC.

\subsection{Precision Proton Spectrometer (PPS) and ATLAS Forward Proton detector (AFP) at high luminosity}
\label{sec:high_lumi}
\subsubsection{Introduction}

The CERN's Large Hadron Collider (LHC) will be restarting its operation this year at a record-breaking energy of $\sqrt{s}=13.6$~TeV. The physics run is expected to last until the end of 2025, collecting integrated luminosity of about 300 fb$^{-1}$. LHC will undergo a major upgrade following the four-year physics run, increasing its instantaneous luminosity by a factor of 5--10 larger than the nominal LHC nominal value. The High Luminosity LHC (HL-LHC) is expected to collect data corresponding to an integrated luminosity of a few ab$^{-1}$, and measure the rarest processes of the Standard Model (SM). 

\begin{figure}[!ht]
 \centering
  \includegraphics[width=0.7\linewidth]{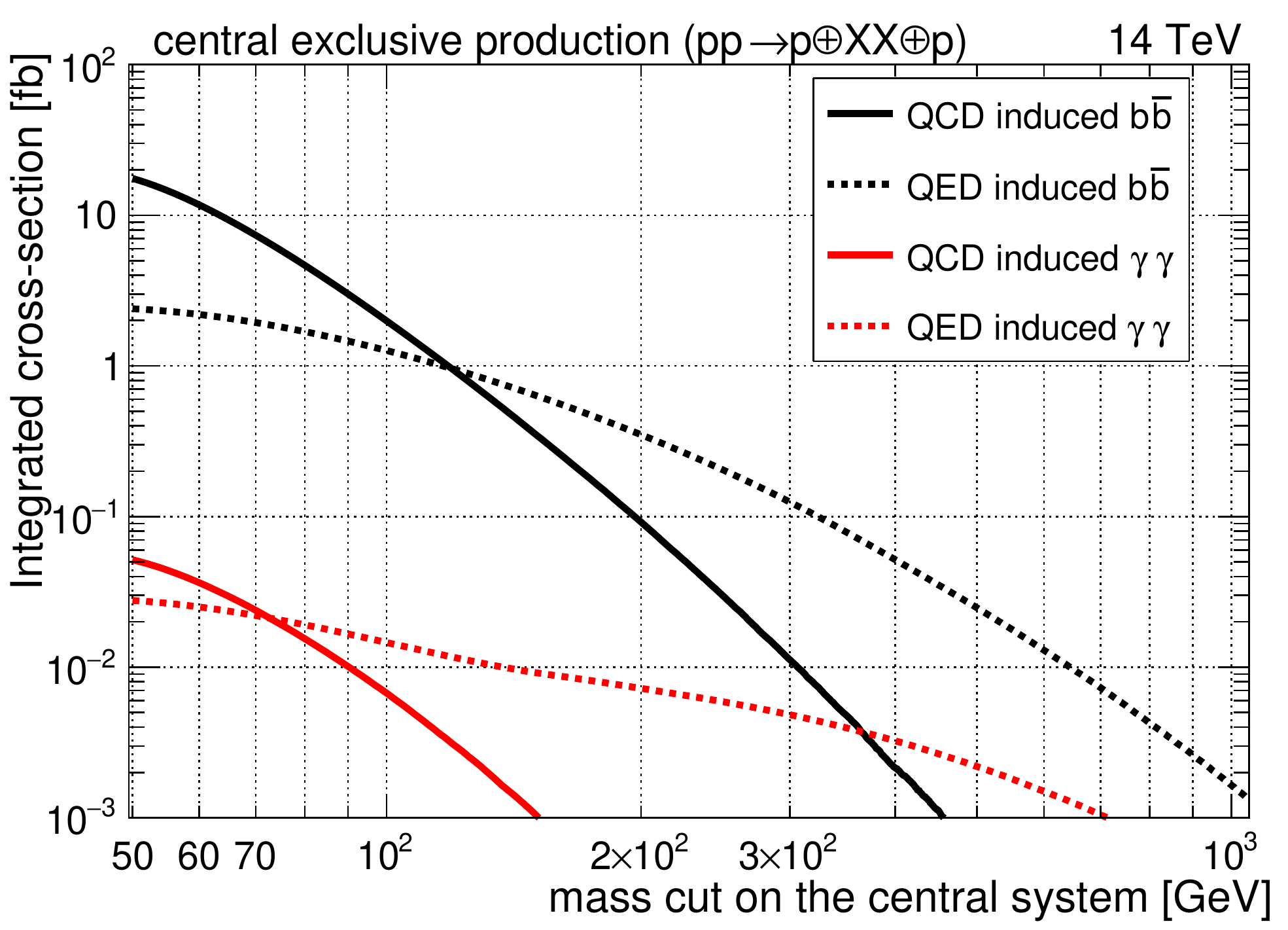}
\caption{Integrated cross sections of different exclusive processes with intact protons at $\sqrt{s} = $14\,TeV, plotted as a function of the required minimum central system mass. Taken from Ref \cite{CMS:2021ncv}.
}
\label{fig:Xsec_POM_PHO}
\end{figure}

Central Exclusive Production (CEP) is a unique process where an object $X$ is produced via $t$-channel exchange of colorless objects, photon ($\gamma$) for electromagnetic or pomeron ($\rm{I\!P}$) for strong interactions, $pp \to p \oplus X \oplus p$, where $\oplus$ stands for an absence of additional interaction between the final states. When final state particles are produced with high invariant mass, the dominant production mechanism is via photon exchange \cite{Khoze:2001xm}, in which the LHC can be considered a photon collider. Figure~\ref{fig:Xsec_POM_PHO} shows a comparison between pomeron-pomeron ($\rm{I\!P}-\rm{I\!P}$) and photon-photon ($\gamma-\gamma$) initiated processes for production cross-section of central exclusive $b\bar{b}$ and $\gamma\gamma$ events as a function of a mass, which shows the enhancement of the photon-photon scattering at high masses.

In CEP, interacting protons often emerge intact but lose a fraction of momentum and are scattered at small angles. The LHC accelerator magnets can be seen as longitudinal momentum spectrometers. The protons are deflected away from the proton bunch and can be measured by near-beam detectors installed downstream the LHC beamline, hundreds of meters from the interaction point. Such detectors, installed in movable vessels (Roman Pots) with tracking and timing capabilities, were brought online during Run~2 in 2016 by the ATLAS and CMS collaborations and were operated in standard runs.

\subsubsection{Near-beam proton spectrometers in LHC Runs 2 and 3}

The Precision Proton Spectrometer (PPS) \cite{CMS:2014sdw}, is a CMS sub-detector installed in 2016, $\sim$210 meter from the interaction point. Initially called CT-PPS (started as CMS and TOTEM project), the PPS apparatus is equipped with tracking and timing detectors. It collected more than 100 pb$^{-1}$ of integrated luminosity during LHC Run 2 and will continue to be operational with some upgrades and optimizations during LHC Run 3. During Run 2, PPS tracking detectors measured protons that have lost approximately between 2.5\% to 15\% of their initial momentum, resulting in mass acceptance between 350~GeV to 2~TeV \cite{CMS-PAS-PRO-21-001}. The data collected with PPS during 2016, with an integrated luminosity of 10 fb$^{-1}$, led to the first measurement of central exclusive di-lepton production \cite{CMS:2018uvs}, and the first search for the high mass exclusive production of photon pairs \cite{TOTEM:2021kin}, both using tagged protons. Next, using 2017 data and integrated luminosity of $\sim$30 fb$^{-1}$ a search for the exclusive production of pair of top quarks and a search for new physics in the missing mass spectrum in $pp\to p\oplus Z/\gamma + X\oplus p$ events were performed \cite{CMS-PAS-TOP-21-007,CMS-PAS-EXO-19-009}. Finally, searches for the exclusive production of di-bosons using the full Run 2 dataset were published as well \cite{CMS-PAS-SMP-21-014}.

The ATLAS Forward Proton detector (AFP) \cite{Adamczyk:2015cjy}, comprises two Roman Pot stations on each side from the interaction point with four planes of silicon pixel sensors to measure proton tracks. The far stations are additionally equipped with time-of-flight (ToF) detectors. During Run 2, ToF detectors demonstrated 20-40 ps resolution but suboptimal efficiency. AFP recorded $\sim$30 fb$^{-1}$ of integrated luminosity during Run 2, and this data was used to report on the exclusive di-lepton production \cite{ATLAS:2020mve}.

\subsubsection{Physics perspectives at HL -- LHC}

For the HL-LHC (LHC Run 4), the accelerator will be rearranged, and the current forward detectors will be dismounted. While the new detector design of forward proton spectrometers is currently under development (for example \cite{CMS:2021ncv}), the physics perspectives are presented in the following section. Two  scenarios are under consideration: 
\begin{itemize}
\item Station located in a ``warm'' region -  comprise a few stations $\sim200$~m from the interaction point, and which are suitable for the Roman Pot technology (ATLAS and CMS)
\item Station located at $420$~m in a `cold'' region - which requires a bypass cryostat and a movable detector vessel approaching the beam from between the two beam pipes, for which new developments are needed (CMS),
\end{itemize}
While QCD-induced processes are typically dominant at low masses, the photon-photon scattering is enhanced at high masses (Fig.~\ref{fig:Xsec_POM_PHO}). Fiducial cross sections for different standard model processes at $\sqrt{s} = 14$\,TeV for $\rm{I\!P}-\rm{I\!P}$ and $\gamma-\gamma$ production modes are shown 
in Table~\ref{tab:xsec-sm-processes} for different CMS PPS acceptance scenarios in HL-LHC \cite{CMS:2021ncv}.

\begin{table}[h!]
  \begin{center}
    
    \begin{tabular}{|l|c|c|c|c|}
    \hline
    \multirow{3}{*}{\textbf{Process}} & \multicolumn{4}{c|}{\textbf{fiducial cross section [fb]}} \\  
            
      & \multicolumn{2}{c}{\textbf{all stations}} & \multicolumn{2}{|c|}{\textbf{w/o 420}} \\

      & $\rm{I\!P}-\rm{I\!P}$ & $\gamma-\gamma$ & $\rm{I\!P}-\rm{I\!P}$ & $\gamma-\gamma$ \\   
        
      \hline
$\rm jj$ & $\mathcal{O}\left(10^6\right)$ &  60 & $\mathcal{O}\left(10^4\right)$  &  2 \\
$W^+W^-$ & --- &  37 & --- & 15 \\
$\mu\mu$ & ---&  46 & --- &  1.3  \\
$\rm t\bar{t}$ & --- &  0.15 & --- &  0.1 \\
H & 0.6 &  0.07 & 0 &  0 \\
$\gamma\gamma$ & --- &  0.02 & --- &  0.003 \\
    \hline
    \end{tabular}
    \caption{Fiducial cross sections of CEP of standard model processes in pp collisions at $\sqrt{s}=14$\,TeV. Two scenarios for proton tagging acceptance are shown: with and without the stations at $\pm$420\,m. (more details in \cite{CMS:2021ncv})}
    \label{tab:xsec-sm-processes}
  \end{center}
\end{table}

\paragraph*{Physics w/o 420 meter station:\\}

Standard model $\rm \gamma\gamma\rightarrow \ell^{+}\ell^{-}$ production is an important channel for both calibration and validation of the proton reconstruction, and to measure ElectroWeak contribution to Drell--Yan processes. In addition the $\gamma\gamma \rightarrow \tau^{+}\tau^{-}$ channel is of particular interest as it is sensitive to the anomalous magnetic moment (or ``$g-2$'') of the $\tau$ lepton . 
For diboson production, $\rm \gamma\gamma\rightarrow W^{+}W^{-}$ (with $\rm W^{+}W^{-} \rightarrow \mu^{+}e^{-}\nu_{\mu}\bar{\nu_{e}}$) is a particularly clean channel. The configuration of stations considered here would substantially increase the acceptance for 2-arm events, allowing a significant measurement of the SM cross section in the $\rm \mu^{+}e^{-}$ final state, which will serve as a benchmark for diboson searches in other channels and at higher masses, which provides a good means to test the interactions of photons and W bosons at high energies, and to search for Anomalous Quartic Gauge Couplings (AQGC) or other nonresonant signals of BSM physics. \\

A wide variety of BSM scenarios involving $\gamma\gamma$ production 
with forward protons have been explored in the theoretical literature (e.g \cite{deFavereaudeJeneret:2009db}). For exclusive production with intact protons, only spin-one resonances and any spin-odd states with negative parity are forbidden in $\gamma\gamma$ interactions~\cite{Landau:1948kw,Yang:1950rg}. 
This type of search is particularly 
interesting for resonances with large couplings to photons but not to gluons, which may appear in the $\rm\gamma\gamma \rightarrow X \rightarrow \gamma\gamma$ 
channel~\cite{dEnterria:2013zqi,Csaki:2015vek,Harland-Lang:2016qjy,Fichet:2016pvq,Baldenegro:2018hng}. It was shown that the expected sensitivity for axion-like particles (ALP) in CEP is expected to be competitive and complementary to other collider searches for masses above 600 GeV \cite{Baldenegro:2018hng}. Conversely, if a resonance is detected via decays to 
two photons, measuring the cross section with forward protons will help constrain its couplings to photons in a model-independent way~\cite{Fichet:2016pvq}. The use of forward protons was recently been revisited as a possible means to improve searches for pair production of supersymmetric sleptons or charginos in compressed mass scenarios~\cite{Beresford:2018pbt,Harland-Lang:2018hmi}.\\

\paragraph*{Physics including the 420 meter station:\\}

Central exclusive Higgs boson production has been extensively studied theoretically and in simulations (including the original detailed studies of the FP420 project~\cite{FP420RD:2008jqg}). In 
this case, unlike higher-mass and weakly coupled final states, gluon-gluon production 
is expected to dominate over $\gamma\gamma$ production. The cross section for CEP Higgs production in the SM has been evaluated by several groups, and the 
total cross section ranging between a few fb and a few tenths of a fb, depending on details of the survival probabilities, parton distribution functions (PDFs), Sudakov factors and other 
assumptions of the calculations. A measurement of CEP dijets at the same energy and mass range would therefore remove most of the remaining theoretical uncertainties in the Higgs cross section predictions. For the 125.4\,GeV Higgs boson production, protons could be detected in the 420\,m stations on both arms, and in the combination of the 234\,m and 420\,m stations, while the associated production with $\rm W^{+}W^{-}$ vector-boson pair has the potential for probing the Higgs sector in CEP events in the absence of the $\pm$420\,m stations. Although the exclusive production cross section is estimated to be $\sigma \approx 0.04$\,fb at tree-level, a high acceptance is expected because of the large invariant mass of the central system.\\

As discussed in \cite{Khoze:2021jkd,Khoze:2021pwd} the experiments with
forward proton spectrometers at the HL-LHC would open a promising way 
to perform a search for the QCD instantons, which are a non-trivial consequence
 of the  vacuum structure of the non-abelian  theories
(for a recent review and references see e.g. \cite {Shuryak:2021iqu}).
Instantons  describe quantum
tunneling between different vacuum sectors 
of the QCD and are arguably the best motivated yet experimentally unobserved nonperturbative
effects predicted by the Standard Model.
It is shown in \cite{Khoze:2021pwd} that for an instanton mass
$M_{inst}\geq 50$ GeV the expected central
production cross sections for the instanton-induced processes
 are of the order
of picobarns in the pure exclusive case and increase up to hundreds of pb
when the emission of spectator jets is allowed. These signal cross-sections are
encouragingly large, and under favourable background conditions 
 there is a tantalising chance that QCD  instanton effects can either
 be seen or ruled out. 
The expected experimental signature for the instanton-induced
process in the central detector
 is a large multiplicity and transverse energy
($\sum_i ET_i$) in relatively small rapidity interval ($\delta y\simeq 2-3$)
and large sphericity $S>0.8$ of the event. Note that the mean number of gluon jets radiated by 
the instanton is $\sim 1/\alpha_s$,
while the probability of the instanton creation is
$\propto \exp(-4\pi/\alpha_s)$. Therefore  to observe the clear
signature of the instanton-induced signal it is most feasible to consider
the case of the moderately
heavy instantons, $M_{inst}\geq 50-100$ GeV.
This would require measurements with the 420 m stations.


\subsection{Non-elastic contribution in photon-photon physics}
\label{sec:nonelastic}

The two-photon production of di-leptons has been largely studied at the LHC experiments in the past year \cite{CMS:2011vma,CMS:2012cve,ATLAS:2015wnx,CMS:2018uvs,ATLAS:2017sfe,ATLAS:2020mve,CMS:2013hdf,CMS:2016rtz,ATLAS:2016lse}, investigating elastic interactions at distinct colliding energies. This exclusive production presents a final state composed by the lepton pair produced at the central detector, where large rapidity gaps are present between pair and the outgoing protons in the beam line direction. Such signature differs from the usual QCD production by the absence of particle (gluon) radiation that populates the detector, largely reducing the possibility of observing this signature in the data \cite{LHCForwardPhysicsWorkingGroup:2016ote}. The interest for di-leptons comes from the fact that they can be used as luminosity monitors \cite{Khoze:2000db,Shamov:2002yi}, however the production of $W$ boson pairs via they decay channel into leptons provides a way to investigate evidences of New Physics with the use of effective theories including anomalous gauge couplings. The signal yields includes both the elastic production -- with two intact outgoing protons in the forward direction -- as well as the nonelastic production, with one or both protons dissociating into a hadronic final state, classified as semi-elastic and inelastic production, respectively (see e.g. Ref.~\cite{Harland-Lang:2020veo} for more details).\\

\begin{figure}[b]
\centering
\includegraphics[width=1.\textwidth]{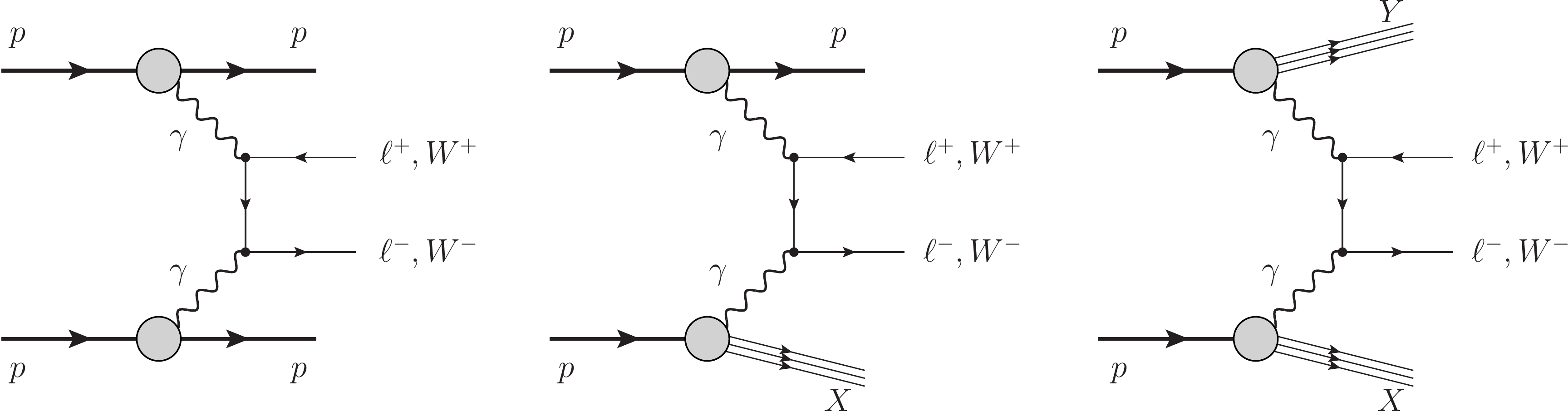}
(a) \hspace{5cm} (b) \hspace{5cm} (c)
\caption{\label{fig1-procs} Processes of particle production in two-photon interactions in hadronic collisions: (a) elastic, (b) semi-elastic, and (c) inelastic case \cite{daSilveira:2021bzs}.}
\end{figure}

\begin{figure}
\centering
\includegraphics[width=.49\textwidth]{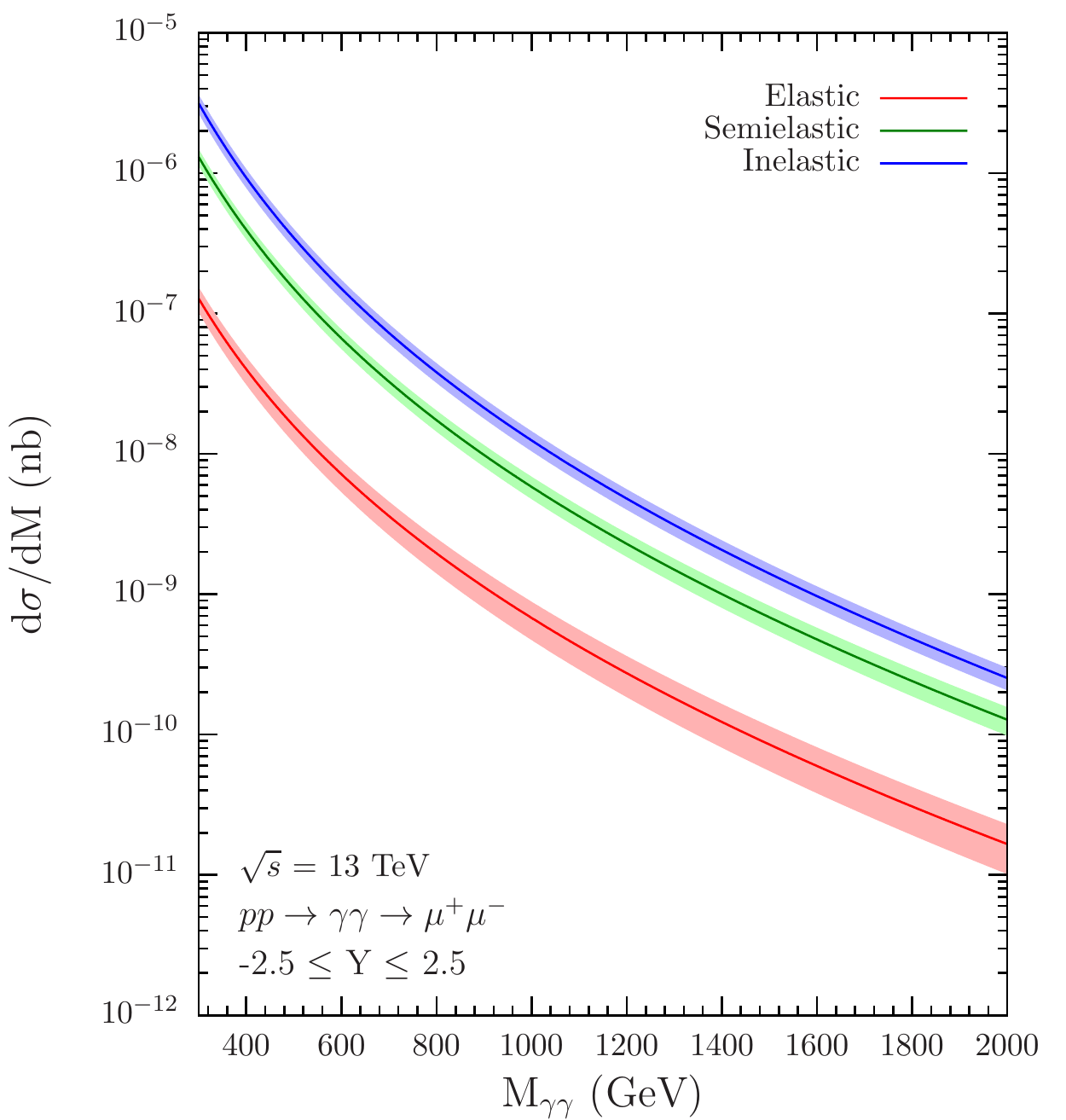}
\caption{\label{fig2-xsecs} Invariant mass distributions for the  di-muon production by $\gamma \gamma$ interactions in $pp$ collisions at $\sqrt{s} = 13$ TeV considering two distinct ranges of $M_{\gamma \gamma}$. The solid lines are the average values for the predictions and the band represent the one standard deviation uncertainty based on the different predictions \cite{daSilveira:2021bzs}.}
\end{figure}

Figure \ref{fig1-procs} illustrates the production cases. While the former is easily computed analytically with the use of photon fluxes \cite{Budnev:1975poe}, the latter is based on parton distribution functions (PDFs) with QED contribution. The typical production cross section can be expressed in terms of effective photon luminosities: $
\sigma^{i} \propto {\cal{L}}^{i}_{eff} \times \hat{\sigma}(\gamma\gamma\to\ell^{+}\ell^{-})$, where $\hat{\sigma}(\gamma\gamma\to\ell^{+}\ell^{-})$ is the tree-level cross section and ${\cal{L}}^{i}_{eff}$ is the photon luminosity for each processes:
\begin{eqnarray}
&\textrm{Fig.~1a:}& \,\,\, {\cal{L}}^{\textrm{el}}_{eff} \propto x_{1}f^{\textrm{el}}_{\gamma,1}(x_{1};Q^{2}) x_{2}f^{\textrm{el}}_{\gamma,2}(x_{2};Q^{2}), \\
&\textrm{Fig.~1b:}& \,\,\, {\cal{L}}^{\textrm{semi}}_{eff} \propto x_{1}f^{\textrm{inel}}_{\gamma,1}(x_{1};Q^{2}) x_{2}f^{\textrm{el}}_{\gamma,2}(x_{2};Q^{2}) + x_{1}f^{\textrm{el}}_{\gamma,1}(x_{1};Q^{2}) x_{2}f^{\textrm{inel}}_{\gamma,2}(x_{2};Q^{2}), \\
&\textrm{Fig.~1c:}& \,\,\, {\cal{L}}^{\textrm{inel}}_{eff} \propto x_{1}f^{\textrm{inel}}_{\gamma,1}(x_{1};Q^{2}) x_{2}f^{\textrm{inel}}_{\gamma,2}(x_{2};Q^{2}),
\end{eqnarray}
with $x_i$ is the momentum fraction of the proton carried by the photon and $Q^2$ is the photon virtuality. The non-elastic cases made use of photon PDFs based on the DGLAP evolution equations modified to include the QED parton splitting functions. Considering the different approaches used in the literature for the elastic and nonelastic contributions, an estimate for the uncertainties associated for these choices. Figure~\ref{fig2-xsecs} shows the differential cross section as function of the invariant mass of muon pairs \cite{daSilveira:2021bzs}:

\begin{eqnarray}
\frac{d \sigma^i}{d M_{\gamma\gamma}} &=& 2 M_{\gamma\gamma} \int d Y \,\, \frac{\partial^{2} {\cal{L}}^i_{eff}}{\partial M_{\gamma\gamma}^{2}\partial Y} \cdot \hat{\sigma}_{\gamma\gamma\to \mu^+ \mu^-}(M_{\gamma\gamma}^{2}=x_{1}x_{2}s) , \\
\end{eqnarray}

The curves correspond to the predictions averaged at $Q=300$~GeV among the recent parametrizations for the photon PDF: LUXqed17 \cite{Manohar:2017eqh}, MMHT2015qed \cite{Harland-Lang:2019pla}, and NNPDF31luxQED \cite{Bertone:2017bme}. 
All these parametrizations are based on the approach proposed in Ref.~\cite{Manohar:2016nzj} (See also Ref.~\cite{Luszczak:2015aoa}). The bands are evaluated as one standard deviation around the averages.

There are Monte Carlo event generators providing predictions for the elastic contribution in the two-photon di-lepton and $WW$ productions, however the nonelastic contribution is not a common feature. Given that the curves show similar shapes, it favors the possibility of obtaining a multiplicative factor that can be used to re-weight generated event samples to account for the nonelastic contributions \cite{daSilveira:2021bzs}. This prediction can be experimentally tested with forward detectors capable of observing the intact protons emerging from elastic and semi-elastic collisions, such as CMS Precision Proton Spectrometer (PPS) \cite{CMS:2014sdw} and ATLAS Atlas Forward Proton (AFP) \cite{Adamczyk:2015cjy}. A multiplicative factor has been already evaluated in previous CMS analyses \cite{CMS:2013hdf,CMS:2016rtz} in the high-mass region:
\begin{eqnarray}
F = \left. \frac{N_{\mu\mu\textrm{(data)}}-N_{\textrm{DY}}}{N_{\textrm{elastic}}} \right|_{M(\mu^{+}\mu^{-})>160\textrm{ GeV}} \end{eqnarray}
where $N_{\mu\mu\textrm{(data)}}$ is the total number of events passing the selection criteria, $N_{\textrm{DY}}$ the total number of events identified as coming from the Drell-Yan production process related to events with one or more extra tracks, and $N_{\textrm{elastic}}$ is the estimated number of elastic events from theory. In a similar fashion, theoretical predictions are used to provide a estimate of this ratio like:
\begin{eqnarray}
F_1 = \frac{\frac{d\sigma^{\textrm{el}}}{dM_{\gamma\gamma}} + \frac{d\sigma^{\textrm{semi}}}{dM_{\gamma\gamma}} + \frac{d\sigma^{\textrm{inel}}}{dM_{\gamma\gamma}}}{\frac{d\sigma^{\textrm{el}}}{dM_{\gamma\gamma}}} & \,\,\,\,\,\, \mbox{and} \,\,\,\,\,\, &
F_2 =  \frac{\frac{d\sigma^{\textrm{el}}}{dM_{\gamma\gamma}} + \frac{d\sigma^{\textrm{semi}}}{dM_{\gamma\gamma}}  }{\frac{d\sigma^{\textrm{el}}}{dM_{\gamma\gamma}}} \,\,.
\label{F_PPS}
\end{eqnarray}

\begin{figure}[t]
\centering
\includegraphics[width=.49\textwidth]{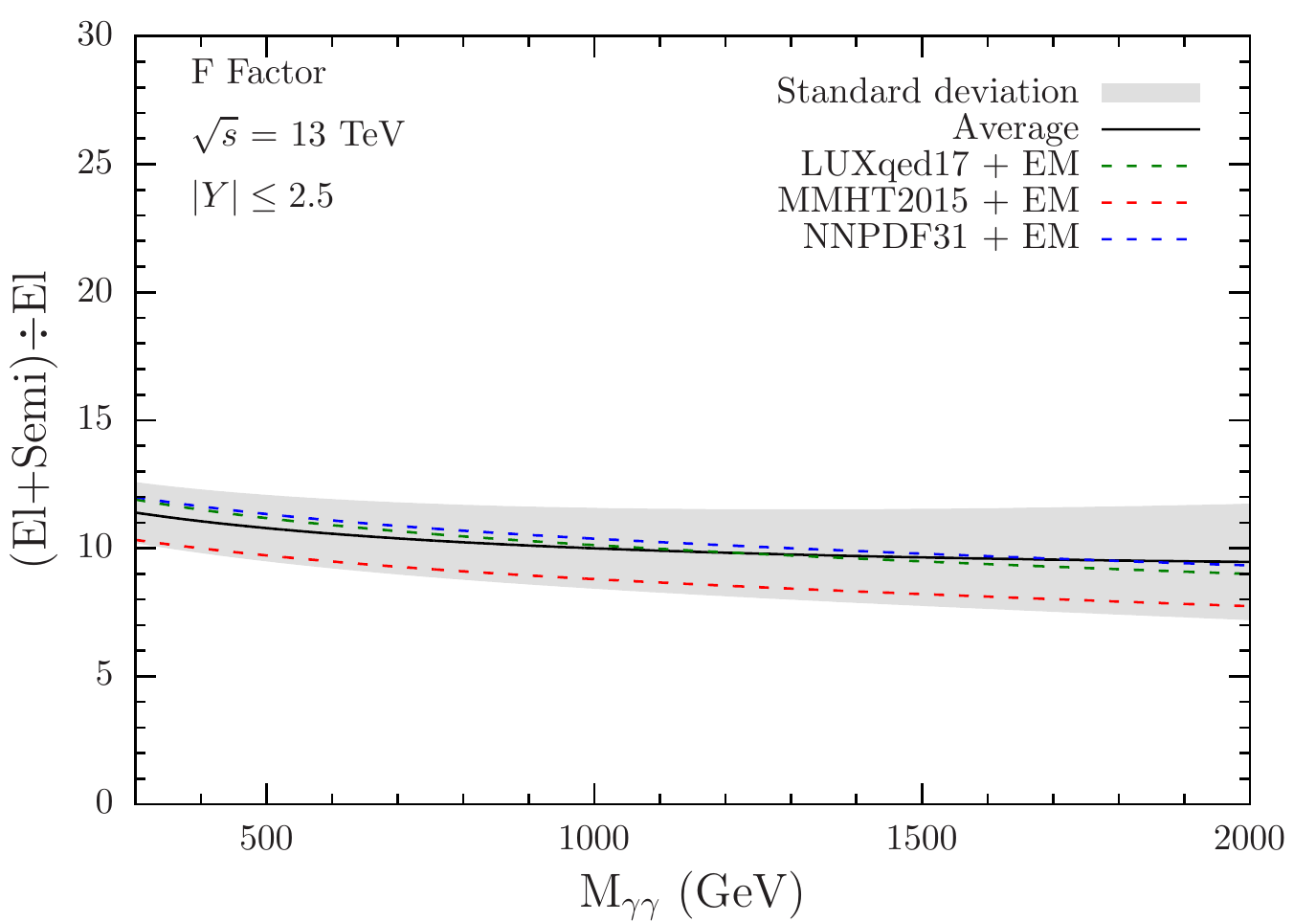}
\includegraphics[width=.49\textwidth]{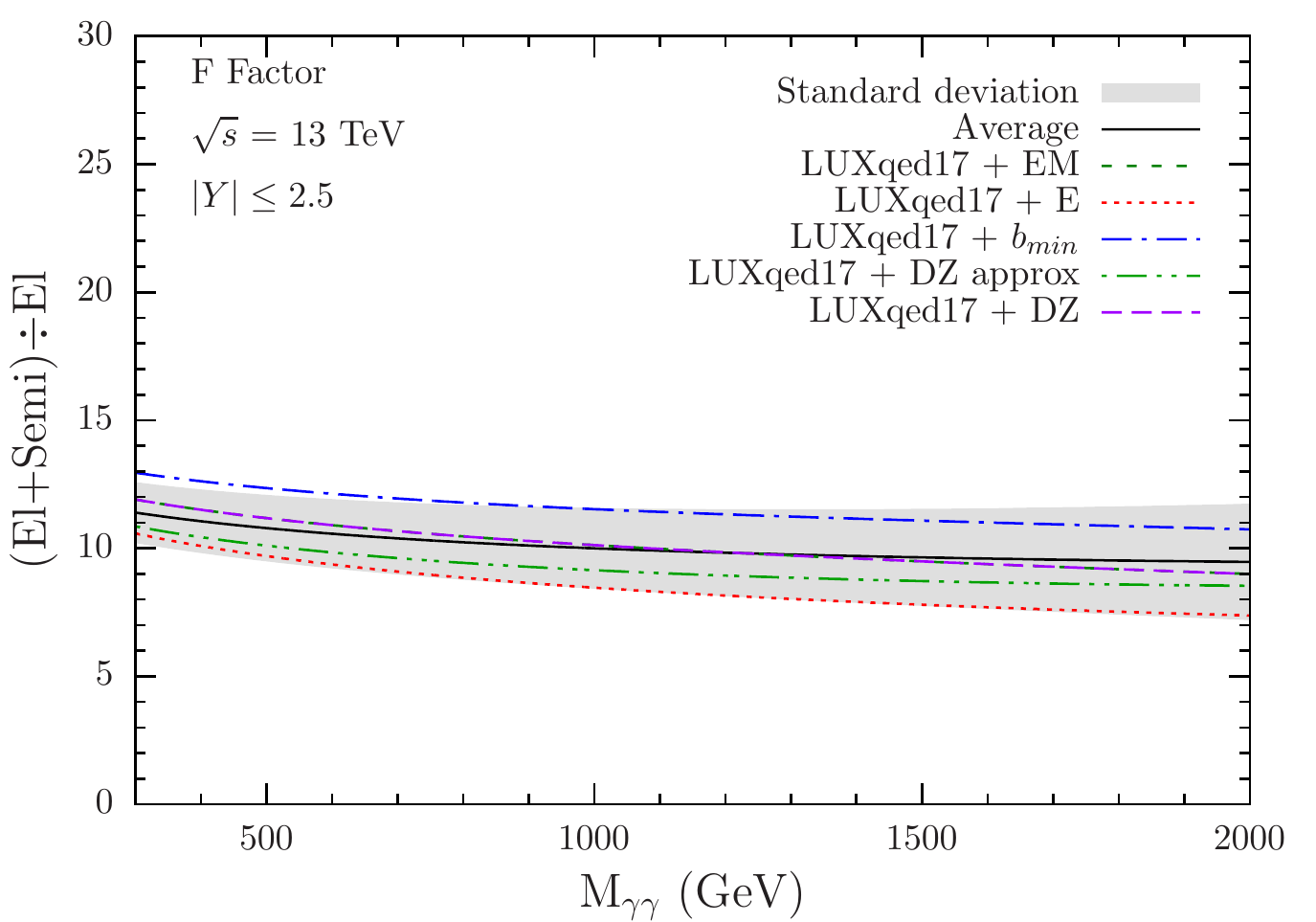}
\caption{\label{fig3-ratio} Dependence on the invariant dimuon mass of the fraction $F_2$ for  different models of the elastic (left panel) and inelastic (right panel) photon distributions \cite{daSilveira:2021bzs}.}
\end{figure}

Using the set of parametrizations for the photon PDF, one is able to evaluate these ratios in the phase-space region accessible by LHC forward detectors. Figure~\ref{fig3-ratio} presents the predictions in the mass range of 300 GeV to 2 TeV including different approaches for the elastic photon flux, see also \cite{Bailey:2022wqy} for a recent study on these effects for WW production. It shows an uncertainty of 20--40\% considering the available parametrizations. A experimental measurement of this observable would provide new insight on the parametrizations and account for a data-driven result that could be used in event generators and extend the stringency of limits for anomalous couplings. The upcoming Run3 of the LHC may provide an unique opportunity to collect enough luminosity for such measurement, opening new fronts for the investigation of photon interactions and improvement of the computational tools available in the literature (see e.g. Dark Matter searches~\cite{Harland-Lang:2018hmi} or $t\bar{t}$ production~\cite{Goncalves:2020saa,Martins:2022dfg}, both in the exclusive mode).

\subsection{Exclusive production of Higgs boson}


Central Exclusive Production (CEP) is especially attractive for three reasons: firstly, if the outgoing protons remain intact
and scatter through small angles then, to a very good approximation, the primary di-gluon system
obeys a $J_z=0$, ${\cal C}$-even, ${\cal P}$-even selection rule~\cite{Khoze:2000jm,Kaidalov:2003fw}.
Here $J_z$ is the projection of the total angular momentum along the proton beam axis. This
therefore allows
a clean determination of the quantum numbers of any observed resonance. Thus, in principle, only
a few such events are necessary to determine the quantum numbers, since the mere observation of
the process establishes that the exchanged object is in the  $0^{++}$ state. Secondly, from
precise measurements of the proton momentum losses, $\xi_1$ and $\xi_2$, and from the fact that
the process is exclusive, the mass of the central system can be measured much more precisely than
from the central detector, by the so-called missing mass method \cite{Albrow:2000na},
$M^2=\xi_1\xi_2 s$
which is independent of the decay mode. Thirdly in CEP the signal-to-background (S/B)
ratios turn out to be close to unity, if the contribution from pile-up is not considered. 
This advantageous S/B ratio is due to the combination of the $J_z=0$ selection rule, the
potentially excellent mass resolution, and the simplicity of the event signature in the
central detector.

For $pp$ collisions, the dominant contribution is expected to be from exclusive gluon-fusion production $gg \to h$ for
which the cross section predictions are still known with a limited accuracy. A similar statement applies to photon-fusion production, which is strongly enhanced in PbPb collisions  with respect to the $pp$ case, see for instance \cite{dEnterria:2019jty}.  While the $gg \to h$
is in principle calculable in perturbative QCD, a non-negligible (but conservative) spread in
cross section predictions of 0.5 -- 3.0~fb is seen due to such basic ingredients as the parton
distribution function (PDF) used and a limited control over the non-perturbative theory of soft
survival factors, $S^2$, for gluon-initiated processes in this mass
range~\cite{Harland-Lang:2014lxa}
(although these uncertainties cancel in the S/B ratio for many backgrounds). Existing experimental
data from CDF exclusive di--photon~\cite{CDF:2011unh} or LHCb $J/\Psi$ pair~\cite{LHCb:2014zwa}
or quarkonia~\cite{LHCb:2011dra} analyses rather prefer values towards the higher end of the
spread (see discussions in Refs.~\cite{Harland-Lang:2014efa,Harland-Lang:2015cta}) nevertheless
direct measurements of the exclusive Higgs production would undoubtedly allow its production rate
to be directly constrained (or for example by monitoring rates of CEP dijets or di-photons, since
the same PDFs and $S^2$ enter the respective production cross sections at the same central system
mass).

The exclusive production of Higgs boson was a flagship topic of the project FP420 (see e.g.
the title of the main document, ``The FP420 R\&D Project: Higgs and New Physics with forward
protons at the LHC''~\cite{FP420RD:2008jqg}) whose main goal was to install forward proton
detectors (FPDs) at 420~m
from the interaction point of ATLAS and CMS experiments to detect forward protons coming from
diffractive proton-induced or photon-induced interactions.

Another important feature of forward proton tagging
in the case of the Higgs boson is the fact that it enables the dominant decay modes, namely
$b\bar b$, $WW^{(*)}$, $ZZ^{(*)}$ and $\tau\tau$ to be observed in one process. In this way, it may
be possible to access the Higgs boson coupling to bottom quarks. This is challenging in
conventional search channels at LHC due to large QCD backgrounds, even though $h \to b\bar{b}$ is the
dominant decay mode for a light SM Higgs boson.
The $b\bar b$, $WW^{(*)}$ and $\tau\tau$ decay modes were studied in detail and are documented in
literature ($b\bar b$ in Refs.~\cite{CMS:2006exu,Cox:2007sw,Heinemeyer:2007tu,Heinemeyer:2010gs,Tasevsky:2013iea,Boonekamp:2005up,Tasevsky:2005ue,ATLASnote}, $WW^{(*)}$ in Refs.~\cite{CMS:2006exu,Heinemeyer:2007tu,Tasevsky:2005ue,ATLASnote,Cox:2005if,Khoze:2005hc} and $\tau\tau$ in Ref.~\cite{Heinemeyer:2007tu,Heinemeyer:2010gs} and in an unpublished diploma thesis~\cite{Vlasta}).
It was the $b\bar b$ mode that was studied in greatest detail --- thanks to advantages enumerated
above and also thanks to the most favourable prospects for this decay mode in enhancing the
production cross section in Minimal SuperSymmetric SM (MSSM), the most popular model of BSM of
those days. Prospects for other extensions were outlined in Ref.~\cite{Forshaw:2007ra} for NMSSM
(Next-to-Minimal SuperSymmetric SM) and in Ref.~\cite{Chaichian:2009ts} for a possible triplet
Higgs sector. 
Results of the above studies, including SM and BSM Higgs bosons, were reviewed in 2014 in
Ref.~\cite{Tasevsky:2014cpa} and can be summarized in the following way, noting especially the
fact that all were performed prior to the Higgs boson discovery. 

Although studies of properties of the Higgs boson with mass close to 125.5~GeV discovered by the
ATLAS \cite{ATLAS:2012yve} and CMS \cite{CMS:2012qbp} (see for example a global analysis in
Ref.~\cite{Bechtle:2014ewa}) suggest that the Higgs boson is compatible with the Standard Model,
there is still room for models of New Physics, e.g. at lower or higher masses than 125.5~GeV,
and the central exclusive production of the Higgs boson still represents a powerful tool to
complement the standard strategies at LHC. A striking feature of the CEP Higgs-boson is that this
channel provides valuable additional information on the spin and the coupling structure of Higgs
candidates at the LHC. We emphasize that the $J_z = 0$, ${\cal C}$-even, ${\cal P}$-even selection rule of
the CEP process enables us to estimate very precisely (and event-by-event) the quantum numbers of
any resonance produced via CEP. 

Signal selection and background rejection cuts are based on requiring a match between
measurements in the central detector and FPD within assumed subdetector resolutions. In addition,
pile-up backgrounds are suppressed by using Time-of-Flight (ToF) detectors, a natural part of FPD
whose utilization necessitates protons to be tagged on both sides from the interaction point
(see a recent ToF performance study in Ref.~\cite{Cerny:2020rvp}). 
The significances for the CEP Higgs boson decaying into $b\bar{b}$, $WW$ or $\tau\tau$ pairs in SM
are moderate but $3\,\sigma$ can surely be reached if the analysis tools, ToF measurement resolution
or L1 trigger strategies are improved, among others by knowing the Higgs boson mass precisely, as
discussed in Ref.~\cite{Tasevsky:2014cpa}. For example we can surely expect improvements in the
gluon-jet/$b$-jet mis-identification probability $P_{g/b}$. In the original analyses in
Refs.~\cite{Cox:2007sw,Heinemeyer:2007tu,Heinemeyer:2010gs,Tasevsky:2013iea,ATLASnote} a
conservative approach has been followed by taking the maximum of two values available at that
time in ATLAS and CMS. Meanwhile new developments were reported in reducing the light-quark-b
mis-identification probabilities in ATLAS \cite{ATLAS:2011hfa} and CMS \cite{CMS:2012feb}. Other
possibilities to improve the significances in searching for the SM Higgs in CEP are a possible
sub-10 ps resolution or finer granularity of timing detectors, the use of multivariate techniques
or a further fine-tuning or optimization of the signal selection and background rejection cuts,
thanks to the fact that the mass of the SM-like Higgs boson is already known with a relatively
high precision. The known Higgs boson mass can also greatly facilitate proposals for a dedicated
L1 trigger to efficiently save events with the CEP $H \to b\bar{b}$ candidates. Proposals made in
Ref.~\cite{Brown:2009mda}, well before the SM-like Higgs boson discovery, can thus be further
optimized.

Studying properties of Higgs bosons born exclusively with a mass around 125~GeV would require
building FPDs in the region 420~m from the interaction point. Such a possibility, as a possible
upgrade of FPDs at HL-LHC, is considered by the CMS collaboration (see e.g.
Ref.~\cite{CMS:2021ncv}). Equipping that region of the LHC beam pipe (so called ``cold region'')
by Roman Pots or Hamburg Beampipe devices was thoroughly discussed in the framework of the FP420
collaboration and all the know-how has been then put in the R\&D document~\cite{FP420RD:2008jqg}.
The constraints coming from experimental data exclude the heavy Higgs boson mass region below
400~GeV, although in special MSSM scenarios, for example Mh125 alignment
scenario~\cite{Bagnaschi:2018ofa}, masses lower than 400~GeV would still be possible, but for
``fine-tuned'' points rather than larger areas. Other extreme scenarios that are still possible
are represented by the $M_H^{125}$ scenario~\cite{Bagnaschi:2018ofa}, in which the light CP-even
Higgs is lighter than 125~GeV, and the discovered Higgs boson corresponds to the heavy CP-even
MSSM Higgs boson. The development of the $M_H^{125}$ scenario was triggered by the observation of
a local excess of 3$\sigma$ at about 96~GeV in the diphoton final state, based on the CMS Run~2
data \cite{CMS:2018cyk}. First Run~2 results from ATLAS with 80 fb$^{-1}$ in the
$\gamma\gamma$ final state (see e.g. Ref~\cite{Heinemeyer:2018wzl}) or full Run~2 ATLAS results
in the $\tau^+\tau^-$ final state \cite{ATLAS:2020zms} turned out to be weaker, but a full Run~2
analysis of the CMS data is still awaited.

\subsection{Anomalous quartic couplings with proton tagging}
High-energy photon-photon fusion processes can be studied at the CERN LHC in proton-proton collisions. In comparison to the ultraperipheral heavy-ion collisions, the impact parameter range is much smaller in pp collisions for photon exchange. The quasi-real photon energy spectrum can easily reach the TeV scale for 14 TeV pp collisions, although with a much smaller photon flux since one does not have the same $Z^4$ enhancement factor as in heavy-ion collisions. One of the main interests for studying photon-fusion processes in proton-proton collisions is its potential for discovering physics beyond the standard model (BSM). Such prospects for discovering new physics are complementary to the standard searches at the LHC, which rely on quark- and gluon-initiated processes.

In a fraction of the quasi-real photon exchange processes, the colliding protons may remain intact. In these central exclusive production processes, the photon exchange can be modelled within the equivalent photon approximation, which is based on the parametrization of the electromagnetic form factors of the proton from elastic photon-proton precision data. Non-perturbative corrections related to the underlying event activity or QCD initial-state radiation effects are absent in this case. The survival probability, which quantifies the probability that the protons remain intact after the photon exchange, has been calculated and measured to be on the order of 70-90\% (depends on the invariant mass of the central system).

The intact protons retain most of the original beam momentum, and are thus deflected at small angles with respect to the beam line. The magnetic lattice of the LHC can be used to separate these intact protons from the beam protons that did not collide. Then, these intact protons can be detected with the Roman pot detectors located at about 200 m with respect to the interaction point. If these two protons are detected together with a hard, central system at central pseudorapidities, then all the decay products of the collision have been successfully measured. The PPS and AFP detectors of CMS and ATLAS have such setups for the detection of protons at the nominal instantaneous luminosities.

The mass $m_X$ and rapidity $y_X$ of the central system are directly related to the fractional momentum loss of the scattered protons $\xi_{1,2} = \Delta p_{1,2}/p_{1,2}^\text{beam}$ via,

\begin{equation}
m_{X} = \sqrt{\xi_1\xi_2 s} \hspace{3cm} y_{X} = \frac{1}{2}\ln(\xi_1/\xi_2)
\end{equation}

This kinematical correlation is used to suppress the contributions from pileup interactions, which is the largest source of background for these measurements. The pileup contributions are such that a hard scale process (e.g., QCD production of a photon pair, jets) is paired with uncorrelated forward protons from diffractive pileup interactions. The signature would be similar to that of central exclusive production: two protons and a hard scale system at central rapidities. The cross section for soft diffractive interactions is large (on the order of 20 mb at 13 TeV). Together with the high pileup multiplicities at the LHC and at the future HL-LHC, it becomes more important to control this background. The aforementioned kinematical correlation between the forward and central system mitigate pileup. Pileup is further mitigated with time-of-flight measurements.

\begin{figure}
\centering
\includegraphics[scale=0.15]{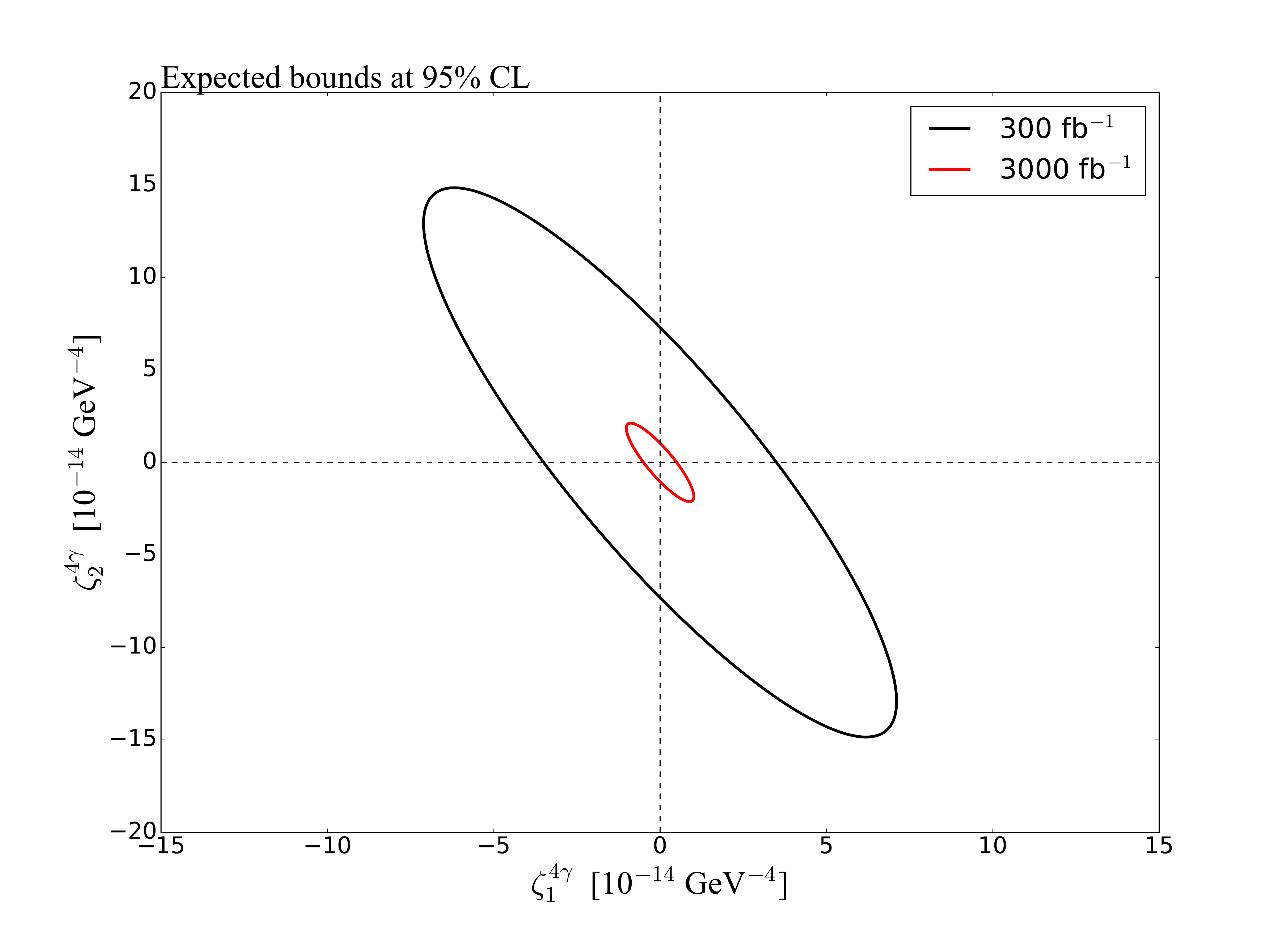}
\includegraphics[scale=0.135]{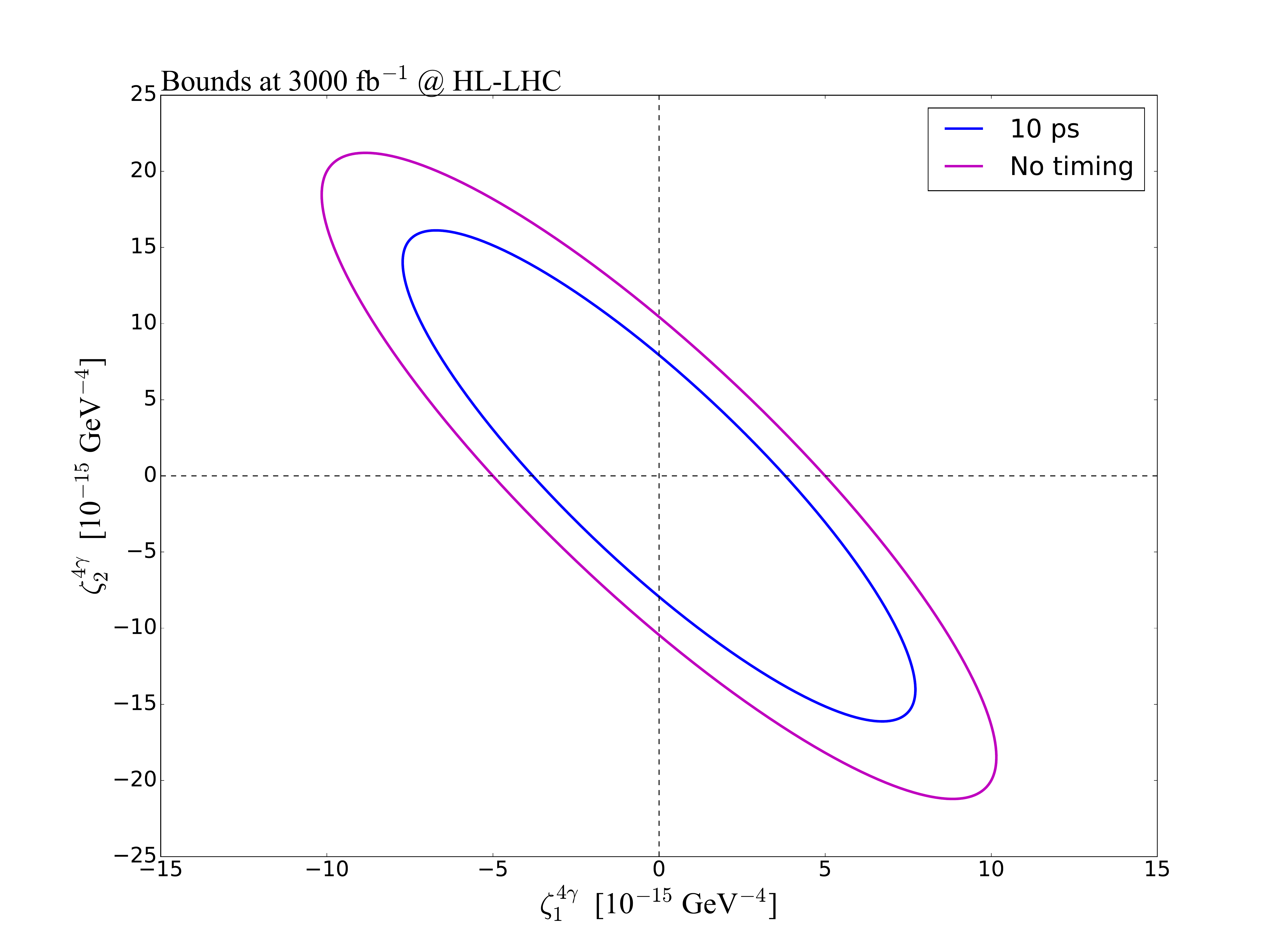}
\caption{\label{fig:sensitivity_4photon} Expected bounds at 95\% CL on the anomalous quartic coupling for 300 fb$^{-1}$ and at the HL-LHC with 3000 fb$^{-1}$ (no time-of-flight measurement) (left). Expected bounds at 95\% CL on the anomalous couplings at the HL-LHC with time-of-flight measurement with precision of 10 ps and without time-of-flight measurement (right). Figure extracted from Ref.~\cite{Azzi:2019yne}.}
\end{figure}

We now discuss a number of examples of new physics searches using proton tagging at the LHC. The scattering of light-by-light ($\gamma\gamma\to \gamma \gamma)$ is induced via box diagrams in the SM at the lowest order in perturbation theory. The experimental signature would be two photons back-to-back, with no hadronic activity, and two scattered protons. Exotic particles can contribute to light-by-light scattering via virtual exchanges at high-mass\cite{Fichet:2013gsa,Fichet:2014uka}. Generic manifestations of physics beyond the SM can be modelled within the effective field theory (EFT) formalism, under the assumption that the invariant mass of the diphoton system is much smaller than the energy scale where new physics manifests.  Among these operators, the pure photon dimension-eight operators $\mathcal{L}_{4\gamma}= 
\zeta_1^{4\gamma} F_{\mu\nu}F^{\mu\nu}F_{\rho\sigma}F^{\rho\sigma}
+\zeta_2^{4\gamma} F_{\mu\nu}F^{\nu\rho}F_{\rho\lambda}F^{\lambda\mu}$ induce the $\gamma\gamma\gamma\gamma$ interaction. The quartic photon couplings have been constrained at the CERN LHC by the CMS Collaboration with values of $|\zeta_1^{4\gamma}| ( |\zeta_2^{4\gamma}|) < 2.88 (6.02) \times 10^{-13}$ GeV$^{-4}$ at 95 \% CL \cite{TOTEM:2021kin}. At the HL-LHC, these bounds can in principle be improved down to $|\zeta_{1}| \approx 4 (8) \times 10^{-14}$ GeV $^{-4}$~\cite{Azzi:2019yne}. Time-of-flight measurements will be very important to suppress the larger amount of pileup interactions. Projections for HL-LHC conditions are shown in Fig.~\ref{fig:sensitivity_4photon}.

\begin{figure}
\centering
\includegraphics[scale=0.15]{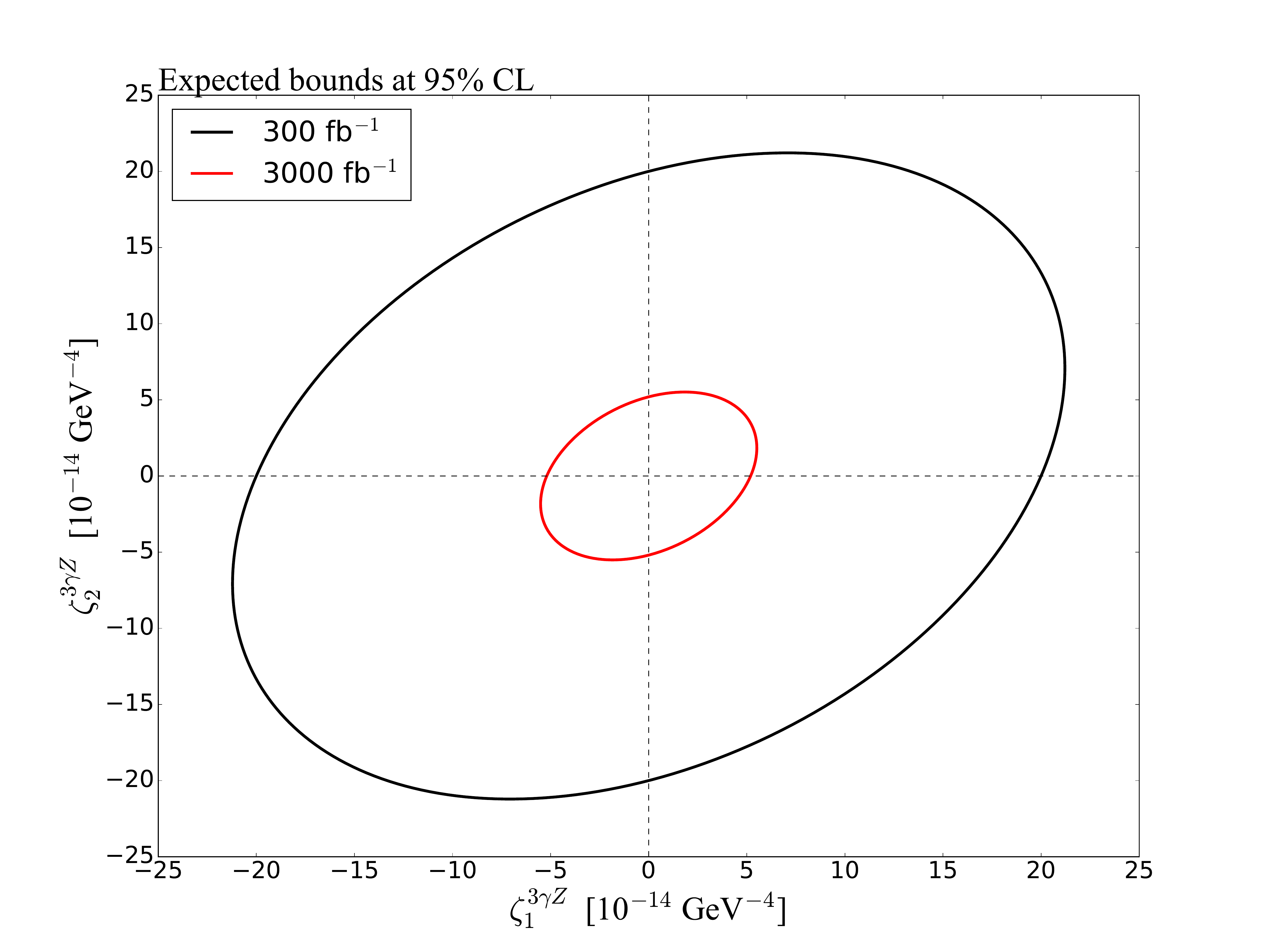}
\includegraphics[scale=0.15]{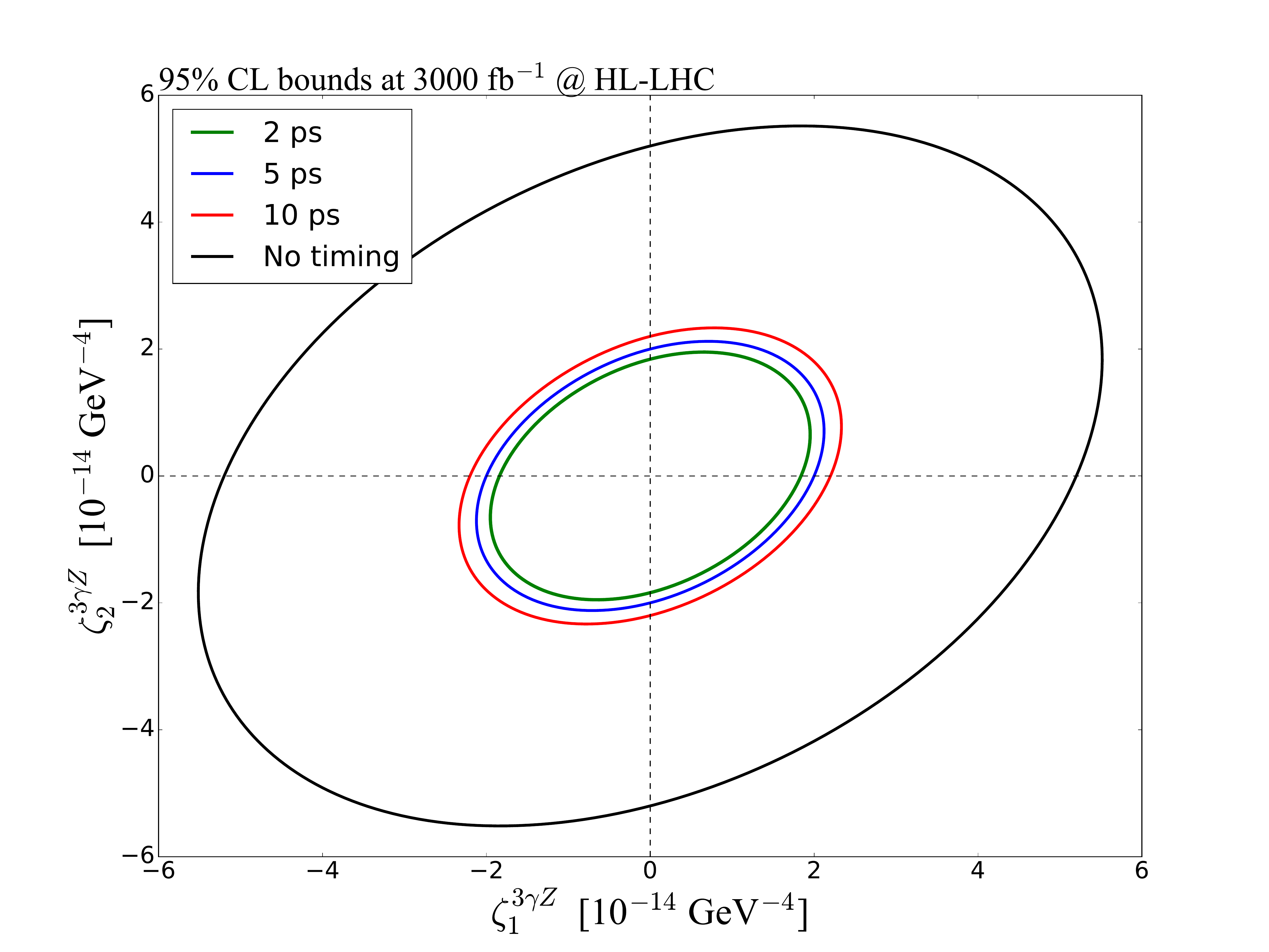}
\caption{\label{fig:sensitivity_3gammaZ} Expected bounds on the anomalous couplings at 95\% CL with 300 fb$^{-1}$ and 3000 fb$^{-1}$ at the HL-LHC (no time-of-flight measurement) (left). Expected bounds at 95\%CL for timing precision of $\delta t = 2,\,5,\,10$ ps at the HL-LHC (right). Figure extracted from Ref.~\cite{Azzi:2019yne}.}
\end{figure}

The $\gamma\gamma\to \gamma Z$ scattering process can be probed with proton tagging as well~\cite{Baldenegro:2017aen}. This process is induced at the lowest order in perturbation theory via box diagrams of particles charged under hypercharge, analogous to the SM light-by-light scattering box diagram. In the leptonic decay channel, the background can be controlled to a similar degree as the one in light-by-light scattering. New physics manifestations can be modelled using dimension-eight effective operators  $\mathcal{L}_{\gamma\gamma\gamma Z}=\zeta_1^{3\gamma Z} F^{\mu\nu}F_{\mu\nu}F^{\rho\sigma}Z_{\rho\sigma}+\zeta_2^{3\gamma Z} F^{\mu\nu} \tilde{F}_{\mu\nu}F^{\rho\sigma}\tilde{Z}_{\rho\sigma}$. The quartic $\zeta_1, \zeta_2$ couplings can be constrained down to $\approx 2 \times 10^{-13}$ GeV$^{-4}$ in Run-3 conditions~\cite{Baldenegro:2017aen}. This constrain surpasses projections based on measurements of the branching fraction of the rare $Z\to \gamma\gamma\gamma$ decay at the HL-LHC by about two orders of magnitude. The channel is experimentally very clean (an isolated photon recoiling back-to-back against a reconstructed $Z$ boson with no soft hadronic activity associated to the primary vertex). Competitive limits can already be extracted with existing data collected by ATLAS and CMS. At the future HL-LHC, the search can be expanded by considering boosted topologies of the $Z$ boson. This could help populate the region of phase-space at large $\gamma Z$ invariant masses, complementing the reach with the (cleaner) fully leptonic decay channel. Projections for the HL-LHC conditions are shown in Fig.~\ref{fig:sensitivity_3gammaZ}.

\begin{figure}[ht]
     \centering
     \includegraphics[width=0.7\textwidth]{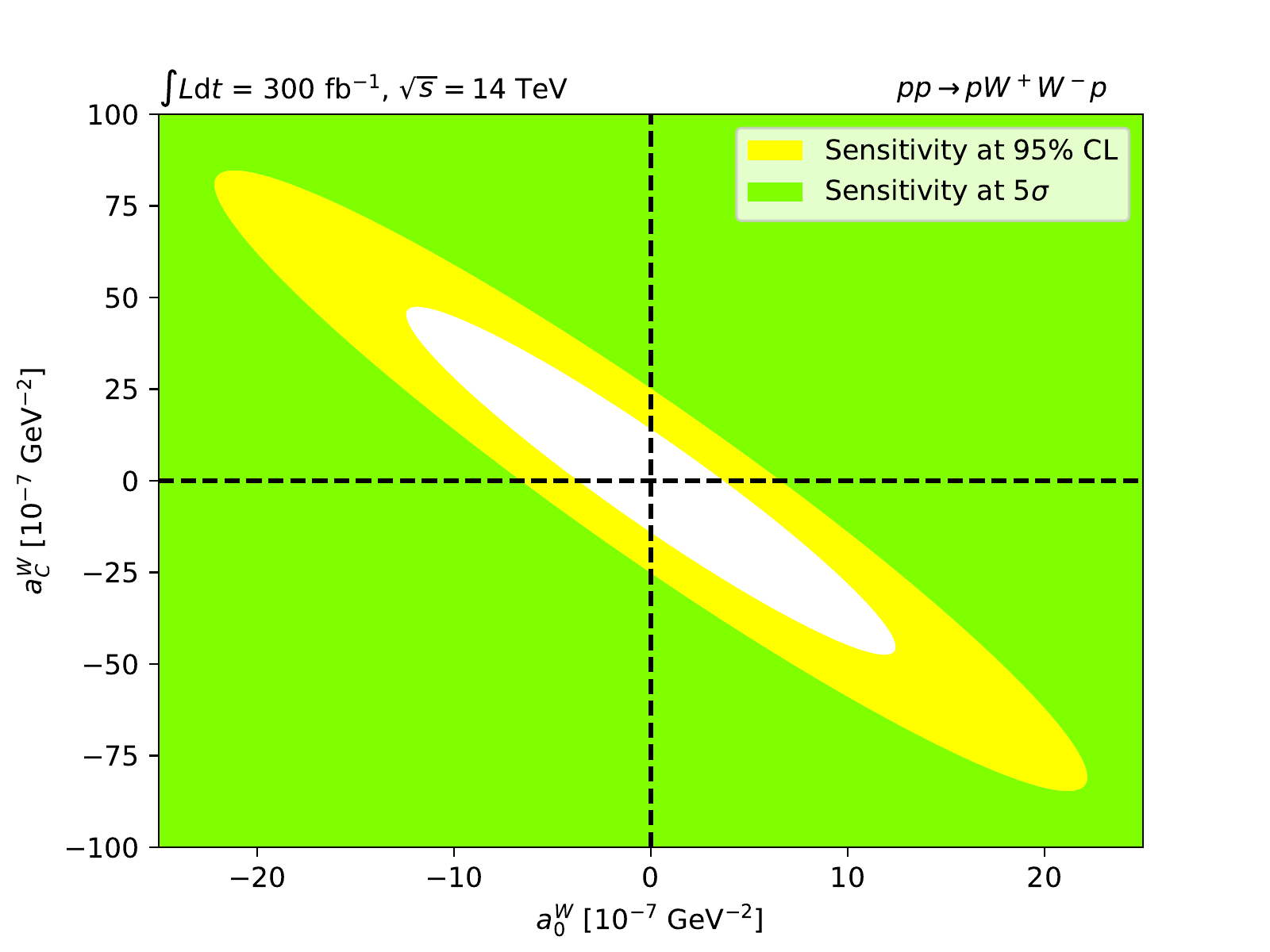}
     \caption{Projected sensitivities on the anomalous coupling parameters $a_0^W$ and $a_C^W$ without form factors. The projections are shown for $pp\rightarrow p W^+W^- p$ at 14 TeV assuming an integrated luminosity of 300 fb$^{-1}$. The yellow and green areas represent respectively the projected sensitivities at 95\% CL and $5\sigma$ combining the hadronic, semi-leptonic, and leptonic decay channels of the $W^+W^-$ system. The blank area in the center represents the region where we do not expect sensitivity to the anomalous coupling parameter. Time-of-flight measurements with 20 ps precision is assumed. Figure extracted from Ref.~\cite{Baldenegro:2020qut}.  \label{fig:a0_limit_noff}}
\end{figure}

Another process of interest is the electroweak gauge boson scattering $\gamma\gamma\to W^+W^-$. Unlike the two previous instances, the $\gamma\gamma \to W^+W^-$ process is induced at tree-level in the SM via the triple $\gamma WW$ and quartic couplings $\gamma\gamma W W$ in the electroweak sector~\cite{Chapon:2009hh}. The process has been observed already by the ATLAS Collaboration without the use of proton tagging by focusing on the purely leptonic decay channel~\cite{ATLAS:2020iwi}. However, in order to probe a region of phase-space that is sensitive to modifications of the SM interactions (high-mass modifications specifically), the proton tagging technique is necessary~\cite{Baldenegro:2020qut}. At high diboson invariant masses and high boson $p_T$, boosted topologies are kinematically favorable. The fully hadronic channel, where each of the hadronically decaying $W$ bosons are reconstructed as large radius jets, provide the best sensitivity to new physics manifestations~\cite{Baldenegro:2020qut}. Modifications to the SM can be modelled with a dimension-six interaction Lagrangian density, $\mathcal{L}_6^\text{eff} = -\frac{e^2}{8} a_0^W F_{\mu\nu}F^{\mu\nu}W^{+\alpha}W^{-}_{\alpha}-\frac{e^2}{16} a_C^W F_{\mu\alpha}F^{\mu\beta} \Big( W^{+\alpha}W^{-}_{\beta}+W^{-\alpha}W^{+}_{\beta} \Big)$. These are the only operators allowed after imposing U(1)$_\text{em}$ and global custodial SU(2)$_\text{C}$ symmetries. The expected limit on the anomalous $a_0^W$ and $a_C^W$ couplings would be at least one order of magnitude larger in the hadronic channel than in the semi-leptonic or leptonic channel combined. The projections for 14 TeV Run-3 combining all channels is shown in Fig.~\ref{fig:a0_limit_noff}. However, the use of jet substructure variables that are sensitive to the number of hard prongs in the jet are necessary (for example, $N$-subjettiness ratios) in order to tame the large QCD jet background. The sensitivity can be further expanded by considering ungroomed jet substructure variables; the ungroomed jet mass and jet shapes for central exclusive $W$ boson jets should render similar resemblance to the jet substructure of a groomed $W$ boson jet from a typical QCD interaction. The SM $\gamma\gamma \to WW$ scattering can be probed in the semi-leptonic channel at high $WW$ invariant masses, in a way such that it complements the phase-space covered by the fully leptonic channel.

\begin{figure}[ht]
     \centering
     \includegraphics[width=0.7\textwidth]{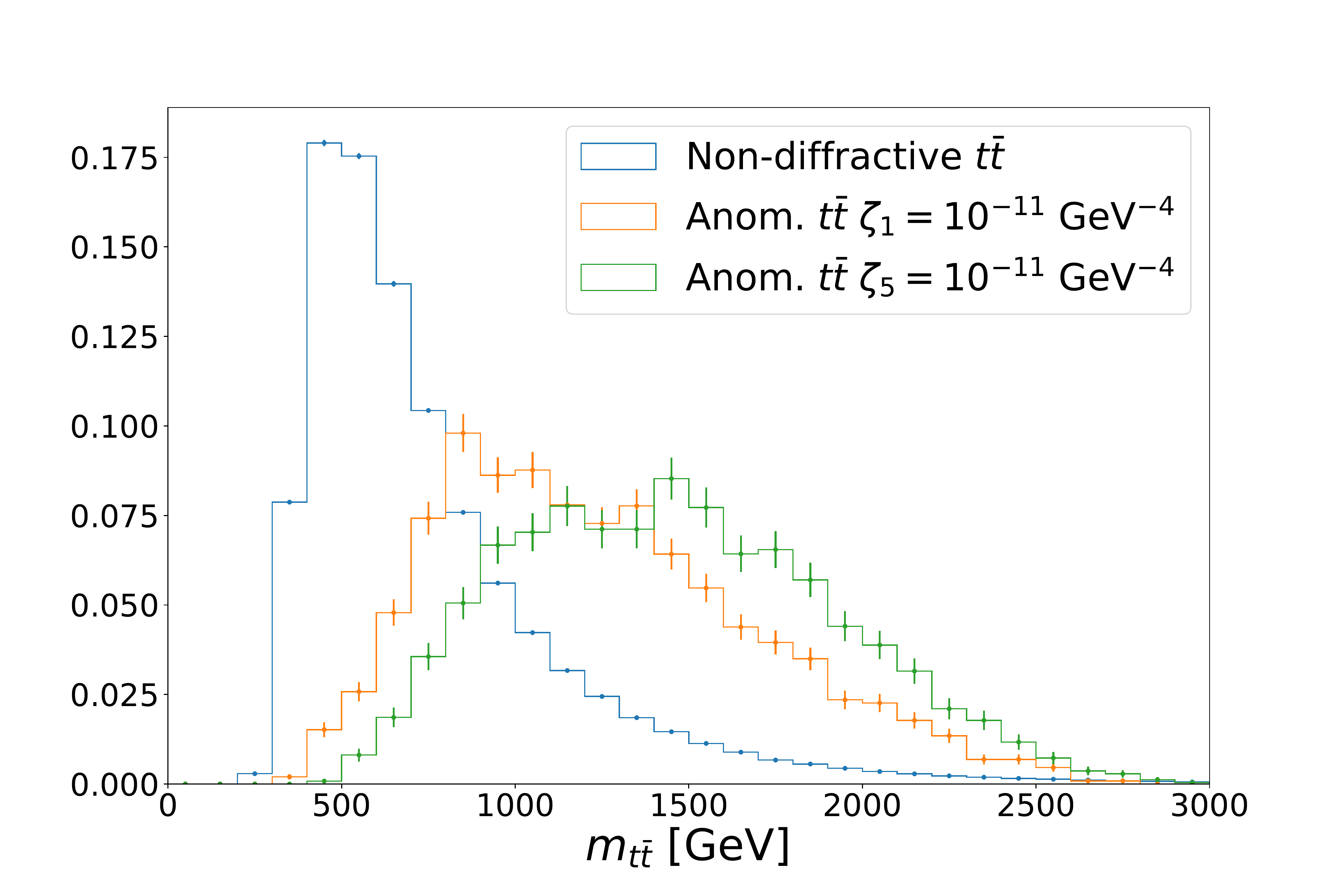}
     \caption{Mass spectrum of the $t\bar{t}$ system as reconstructed from its decay products. Only events passing the pre-selection are shown. Histograms are normalized to unity. The corresponding publication is in preparation.}
     \label{fig:mtt}
\end{figure}

In addition to pure gauge boson scattering, one can probe electromagnetic interactions in other processes such as in  $\gamma\gamma\to t\bar{t}$ scattering, which is induced at tree-level in the SM with the elementary QED vertices of the top quark and the photon. The SM process has not yet been observed. The CMS Collaboration has set an upper limit on the cross section of 0.59 pb at 95\% CL~\cite{CMS-PAS-TOP-21-007}. Although the process is induced at tree-level, the cross section is on the order of $10^{-1}$ fb before branching fraction corrections and for a typical RP acceptance in $\xi$. It is likely that evidence could be established considering the full HL-LHC luminosity. For BSM physics, we considered six different operators (four dimension-six and two dimension-eight) with $\gamma\gamma t\bar{t}$ quartic couplings. We embedded the corresponding amplitudes for six different operators, each representative of different underlying symmetries of the BSM scenarios at high masses. The constraints we expect for a typical Run-3 scenario is about $\zeta_i^{\gamma\gamma t\bar{t}} \approx 10^{-12}$ GeV$^{-4}$, for $i = 1, \dots, 6$, where $\zeta_i$ represent the anomalous quartic couplings. Focusing on high-mass back-to-back top quark pairs with proton tagging, one expects a residual QCD $t\bar{t}$ background of the order of 100 counts for 300 fb$^{-1}$ at 14 TeV. The mass distributions at particle-level for QCD $t\bar{t}$ production and predictions for anomalous couplings are shown in Fig.~\ref{fig:mtt}. The search could be expanded to include the fully-hadronic case at the HL-LHC, where the larger statistical sample allows the coverage of the region of phase-space of highly boosted top quarks.

To summarize this section, there are good prospects for expanding the search for new physics at the LHC in photon-fusion processes such that is complementary to the existing program of the CERN LHC. Other prospects for the HL-LHC era can be read in Ref.~\cite{CMS:2021ncv}.

\section{Conclusions}
\label{sec:conclusion}

Forward physics allows to address fundamental research questions related to the growth of gluon distributions in the perturbative high energy limit and their potential saturation due to the onset of unitarity corrections. It allows searching for imprints of such effects in both parton distribution functions of colliding hadrons and directly in the final state of events. Carrying out this physics program is essential for two reasons:  preparation for the future Electron Ion Collider (EIC) and the potential to answer central research questions already at LHC runs. In comparison to the LHC forward physics program, the future EIC will allow to probe the dense nuclear matter with an electron beam, ideal for the investigation and characterization of hadronic structure. \\

Identifying suitable probes at the LHC is  on the other hand far more cumbersome. Nevertheless this is worthwhile effort: due to its high center of mass energy, the LHC allows to probe hadronic matter at unprecedented values of $x$, which are several orders of magnitude below the values to be reached at the Electron Ion Collider. This is particularly true when using dedicated events in the forward region. It therefore covers regions of phase space which are completely inaccessible at the EIC and  allows for a direct comparison between high parton densities generated through low $x$ evolution and those present in large nuclei. \\

A related topic addresses the direct analysis of emission patterns, related to low $x$  -- in that case BFKL -- evolution, which can be studied using multi jet events. While challenging at the LHC, study of such evolution effects is clearly limited at an Electron Ion Collider to the  limitations in available phase space. Within the foreseeable future, such questions will be either studied at the LHC within the Forward Physics program, or they will not be studied at all. \\

While somewhat orthogonal from the point of view of the physics program, it is natural to employ forward detectors not only for the exploration of strong interactions but also for new physics searches and the study of electroweak dynamics. In particular photon-photon reactions, and related Pomeron-Pomeron fusion processes allow  for the observation of very clean events at the LHC, due to the detection of intact scattered protons and/or large rapidity gaps between the centrally produced object and the scattered proton. While their exploration is of high interest by itself, such events have further the potential to  improve existing bounds on new degrees of freedom and  to contribute to searches for new physics at the LHC. \\

Forward Physics allows therefore to address central physics questions of both nuclear and particle physics. Its physics program is strongly related to the  physics at the future EIC as well as searches for new physics at the LHC. The region of phase space explored by LHC forward physics is unique and therefore allows us to address research questions which are not accessible anywhere else.

\subsection*{Acknowledgements} This project has received funding from the European Union’s Horizon 2020 research and innovation programme under grant agreement STRONG–2020 No 824093 in order to contribute to the EU Virtual Access {\sc NLOAccess} (VA1-WG10) \& to the Joint Research activity ``Fixed Target Experiments at the LHC'' (JRA2), from the Agence Nationale de la Recherche (ANR) via the grant ANR-20-CE31-0015 (``PrecisOnium'') and via the IDEX Paris-Saclay ``Investissements d’Avenir'' (ANR-11-IDEX-0003-01) through the GLUODYNAMICS project funded by the ``P2IO LabEx (ANR-10-LABX-0038)''.
This work  was also partly supported by the French CNRS via the GDR QCD, via the IN2P3 project GLUE@NLO, via the Franco-Polish EIA (Gluegraph).
Francesco Giovanni Celiberto acknowledges support from the INFN/NINPHA project and thanks the Universit\`a degli Studi di Pavia for the warm hospitality.
G. Chachamis acknowledges support by the Funda\c{c}{\~ a}o para a Ci{\^ e}ncia e a Tecnologia (Portugal) under project CERN/FIS-PAR/0024/2019 and contract 'Investigador auxiliar FCT - Individual Call/03216/2017' and from the European Union's Horizon 2020 research and innovation programme under grant agreement No. 824093.
Michael Fucilla, Mohammed M.A Mohammed and Alessandro Papa acknowledge support from the INFN/QFT@COL\-LI\-DERS project.
Krzysztof Kutak acknowledges the support by Polish National Science Centre grant no.\ DEC-2017/27/B/ST2/01985. 
Andreas van Hameren acknowledges the support by Polish National Science Centre grant no.\ 2019/35/B/ST2/03531. 
M.A.O.’s work was partly supported by the ERC grant 637019 ``MathAm''.
The work of SRK is supported in part by the U.S. Department of Energy, Office of Science, Office of Nuclear Physics, under contract number DE- AC02-05CH11231.
M.~Hentschinski acknowledges support  by Consejo Nacional de Ciencia y Tecnolog{\'\i}a grant number
A1 S-43940 (CONACYT-SEP Ciencias B{\'a}sicas). 
The work of Lech Szymanowski is supported  by the grant  2017/26/M/ST2/01074 of the National Science Center in Poland.
J. Jalilian-Marian acknowledges support by the US Department of Energy's Office of Nuclear Physics through Grant No. DE-SC0002307.
The work of Lech Szymanowski is supported respectively by the grant  2017/26/M/ST2/01074 of the National Science Center in Poland.
The research of M.S. was supported by the U.S. Department of Energy, Office of Science, Office of Nuclear Physics, under Award No. DE-FG02-93ER40771.
M.T. supported by MEYS of Czech Republic within the project LTT17018.

\bibliography{references}

\end{document}